\documentclass[aps,twocolumn,nofootinbib]{revtex4}
\usepackage{graphicx}
\usepackage{bm}

\begin{document} 

\title{\bf The compositional and evolutionary logic of metabolism}

\author{Rogier Braakman}

\affiliation{Santa Fe Institute, 1399 Hyde Park Road, Santa Fe, NM
87501, USA}

\author{Eric Smith}

\affiliation{Santa Fe Institute, 1399 Hyde Park Road, Santa Fe, NM
87501, USA}

\date{\today}
\begin{abstract}

  Metabolism is built on a foundation of organic chemistry, and
  employs structures and interactions at many scales. Despite these
  sources of complexity, metabolism also displays striking and robust
  regularities in the forms of modularity and hierarchy, which may be
  described compactly in terms of relatively few principles of
  composition. These regularities render metabolic architecture
  comprehensible as a system, and also suggests the order in which
  layers of that system came into existence. In addition metabolism
  also serves as a foundational layer in other hierarchies, up to at
  least the levels of cellular integration including bioenergetics and
  molecular replication, and trophic ecology. The recapitulation of
  patterns first seen in metabolism, in these higher levels, motivates
  us to interpret metabolism as a source of causation or constraint on
  many forms of organization in the biosphere. Many of the forms of
  modularity and hierarchy exhibited by metabolism are readily
  interpreted as stages in the emergence of catalytic control by
  living systems over organic chemistry, sometimes recapitulating or
  incorporating geochemical mechanisms.

  We identify as modules, either subsets of chemicals and reactions,
  or subsets of functions, that are re-used in many contexts with a
  conserved internal structure. At the small molecule substrate level,
  module boundaries are often associated with the most complex
  reaction mechanisms, catalyzed by highly conserved enzymes.
  Cofactors form a biosynthetically and functionally distinctive
  control layer over the small-molecule substrate. The most complex
  members among the cofactors are often associated with the reactions
  at module boundaries in the substrate networks, while simpler
  cofactors participate in widely generalized reactions. The highly
  tuned chemical structures of cofactors (sometimes exploiting
  distinctive properties of the elements of the periodic table)
  thereby act as ``keys" that incorporate classes of organic reactions
  within biochemistry.

  Module boundaries provide the interfaces where change is
  concentrated, when we catalogue extant diversity of metabolic
  phenotypes. The same modules that organize the compositional
  diversity of metabolism are argued, with many explicit examples, to
  have governed long-term evolution. Early evolution of core
  metabolism, and especially of carbon-fixation, appears to have
  required very few innovations, and to have used few rules of
  composition of conserved modules, to produce adaptations to simple
  chemical or energetic differences of environment without diverse
  solutions and without historical contingency. We demonstrate these
  features of metabolism at each of several levels of hierarchy,
  beginning with the small-molecule metabolic substrate and network
  architecture, continuing with cofactors and key conserved reactions,
  and culminating in the aggregation of multiple diverse physical and
  biochemical processes in cells.

\end{abstract}

\maketitle

\section{Introduction}

The chemistry of life is distinguished both by its high degree of
order and by its essential dependence on a number of
far-from-equilibrium reactions~\cite{Schrodinger:WIL:92}.  While in
some cases reactions may be treated as isolated subsystems with
equilibrium
approximations~\cite{Smith:NS_thermo_I:08,Smith:NS_thermo_II:08}, such
isolations are themselves cumulative deviations far from equilibrium,
reflecting the system-level properties of life as a whole. The
dynamical order of life's chemistry is maintained by the
non-equilibrium transfer of electrons through the biosphere. Free
energy from potential differences between electron donors and
acceptors can be derived from a variety of biogeochemical
cycles~\cite{Falkowski:biogeo_cycles:08}, but within cells electron
transfer is mediated by a small number of universal electron carriers
which drive a limited array of organic
reactions~\cite{Lengeler:BP:99}.  Together these reactions make up
metabolism, which governs the chemical dynamics both within organisms
and across ecosystems.  The universal and apparently conserved
metabolic network transcends all known species diversification and
evolutionary change~\cite{Morowitz:BCL:92,Smith:universality:04}, and
distinguishes the biosphere within the major classes of planetary
processes~\cite{Rankama:GC:50}.  We identify metabolism with the quite
specific substrate architecture and hierarchical control flow of this
network, which provide the most essential characterization of the
chemical nature of the living state.

Understanding the structure of metabolism is central to understanding
how physics and chemistry constrain life and evolution.  The
polymerization of monomers into selected functional macromolecules,
and the even more complex integration and replication of complete
cells, form a well-recognized hierarchy of coordination and
information-carrying processes.  However, in the sequence of
biosynthesis these processes come late, and they involve a much
smaller and simpler set of chemical reactions than core metabolism,
the network in which all basic monomer components of biomass are
created from environmental inputs.  

Because the core is the origin of all biomass, its flux is perforce
higher than that in any secondary process; only membrane electron
transport (reviewed in Ref.~\cite{Falkowski:biogeo_cycles:08}) has
higher energy flux. For example, Ref.~\cite{Bar_Even:enzyme_parms:11}
notes that, over a broad sample of enzymes collected from the
literature, those for secondary metabolic reactions have rates $\sim 1
/ 30$ the typical rates of enzymes for core reactions. 

The combined effects of a higher diversity of constraints from
chemistry and physics and a higher density of mass flux within core
metabolism relative to other processes in living systems have major
impacts on the large-scale structure of evolution, as we will
show. Metabolism is the sub-space of organic chemistry over which life
has gained catalytic control, and because in the construction and
optimization of biological phenotypes all matter flows through this
sub-space, its internal structure imposes a very strong filter on
evolution.

In this review we identify a number of organizing principles behind
the major universal structures and functions of metabolism.  They
provide a simple characterization of metabolic architecture,
particularly in relation to microbial metabolism, ecology, and
phylogeny, and the major (biogeochemical) transitions in evolution.
We often find the same patterns of organization recapitulated at
multiple scales of time, size, or complexity, and can trace these to
specific underlying chemistry, network topology, or robustness
mechanisms.  Acting as constraints and sources of adaptive variation,
they have governed the evolution of metabolism since the earliest
cells, and some of them may have governed its emergence.  They allow
us to make plausible reconstructions of the history of metabolic
innovations and also to explain certain strong evolutionary
convergences and the long-term persistence of the core components of
metabolic architecture.

Many structural motifs in both the substrate and control levels of
metabolism may be interpreted as functional modules.  By isolating
effects of perturbation and error, modularity can both facilitate
emergence, and support robust function, of hierarchical complex
systems~\cite{Simon:arch_compl:62,Simon:Org_Comp_Sys:73}.  It may also
affect the large-scale structure of evolution by favoring variation in
the regulation and linkage between modules, while conserving and
thereby minimizing disruption of their internal architecture and
stability~\cite{Ancel:mod_RNA:00,Fontana:evo_devo_RNA:02}.  This can
enhance evolvability through two separate effects. An increased
\emph{phenotypic} (i.e. structural or functional, as opposed to
\emph{genotypic} or sequence) robustness of individual modules gives
access to larger genetic neutral spaces and thus a greater number of
novel phenotypes at the boundaries of these spaces~\cite{Wagner08}.
At the same time, concentrating change at module interfaces, and
allowing combinatorial variation at the module level, can decrease the
amount of genetic variation needed to generate heritable changes in
aggregate phenotypes~\cite{Wagner96,Kirschner98}. It has been argued
that asymmetries in evolutionary constraints can be amplified through
direct selection for evolvability, and that this is a central source
of modularity and hierarchy within biological
systems~\cite{Wagner96,Kirschner98,Gerhart:Cells:97,Gerhart:facil_vary:07}.

These functional consequences of modularity lead us to expect that
metabolism will be modular as a reflection of the requirements of
emergence and internal stability.  Certainly we observe this
empirically; many topological analyses of metabolic networks find a
modular and hierarchical structure~\cite{Ravasz02,Guimera05,Newman06}.
Because of the higher mass flux and more diverse chemistry in core
pathways, we also expect that modularity in their subnetwork will have
the greatest influence on evolutionary dynamics.  In
Sec.~\ref{sec:core_carbon} we will review a range of evidence
supporting this expectation, which suggests that innovations in core
${\mbox{CO}}_2$ fixation were a large part of the cause for major
divergences in the deep tree of
life~\cite{Braakman:carbon_fixation:12}.

To understand the origin and evolutionary consequences of modularity
in metabolism, however, we will need system-level representations that
go beyond topology, to include sometimes quite particular distinctions
of function.  Details of substrate chemistry, enzyme grouping and
conservation, and phylogenies of metabolic modules, in particular, are
rich sources of functional information and context.  While we will
find it significant that some module boundaries are recapitulated at
many levels, differences between levels will also help to distinguish
modularity originating in reaction mechanisms and network topology of
the small-molecule metabolic substrate, from possibly independent
higher-level forms of modularity in the regulation of flux rates or
phenotypic expression by the macromolecular components of cells.  As
an example of the second kind of regulatory control, it has been
argued that the modular constraints observed in amino acid
biosynthetic pathways are due to evolutionary optimization of the
overall kinetics and dynamic responses of these
pathways~\cite{Monod63,Savageau74,Savageau75}.  These forms of
modularity arise from mechanisms such as allosteric inhibition of
enzymes and the tuning of enzyme specific activities, which are
brought into existence by the underlying network topology and
molecular inventory of metabolites. We will return in
Sec.~\ref{sec:rate_regulation} to ways in which regulation of networks
may have been essential to the stability of their underlying
architecture. Recognizing the distinct character of architectural
motifs and control mechanisms at different levels will enable us to
reconstruct steps in metabolic evolution and identify their
environmental drivers.

\subsection{Hierarchy in metabolism, and the role of individuals and
  ecosystems} 

While most metabolic conversions are performed within cells, there is still a significant number that take place at the cell population level (for example those involving siderophores and secreted enzymes).  In general it is important to appreciate
  that a complete accounting of biochemical fluxes not
  only will span many levels of biological organization, but also may
  incorporate multiple distinct internal modes of organization.  In
  addition to the standard ecological distinction between autotrophy
  and heterotrophy, scientists working in the area of bioremediation,
  for example, have coined the term \emph{epi-metabolome} to refer to
  those compounds that due to their slow degradation are freely
  diffusible across microbial communities~\cite{deLorenzo08}. Thus,
the causes and roles of evolutionary changes, even though they arise
within cellular lineages, may be only partly explained by organization
at the cellular or species level. Other levels that must also be
considered include the meta-metabolome of trophic
ecosystems~\cite{Elser:genes_ecosystems:00,Venter:shotgun_sargasso:04,%
  Borenstein:metabolome:08,Klitgord11}, and the links to
geochemistry~\cite{Amend:Energetics:01,Reysenbach:vents:02,%
  Martin:vents:08,Shock:mineral_energy:09,Erwin:Cambrian:11,%
  Erwin:ecol_metazoa:12,Erwin:radiation:12,Erwin:macroevol:12}.  The
great biogeochemical cycles -- of carbon, nitrogen, phosphorus, or
many metals -- combine physiological, ecological, and even geochemical
links such as mantel convection or continental weathering~\cite{Redfield34,Redfield58}.

An important additional empirical observation is that the deepest
universal features of metabolism are reliably seen not at the
individual, but at the ecosystem
level~\cite{Smith:universality:04,Morowitz:EFOoL:07}.  The
single-organism metabolisms among members of complex ecosystems may
vary extremely widely~\cite{Rodrigues:met_nets:09}, because different
organisms perform different segments of biosynthetic or degradative
pathways, using trophic links (predation, parasitism, symbioisis,
syntrophy, saprophyty) to obtain what they do not make.  The
aggregate, or net, pathways to which these individual metabolisms
  contribute, once assembled through their trophic links, mostly
remain within standard networks as reflected in databases such as
KEGG~\cite{Kanehisa:KEGG_update:06} or
UniProt~\cite{Uniprot:uniprot_update:11}.  Appreciating that the
redundancy in metabolism would permit the assembly of a comprehensible
metabolic chart despite the bewildering variety of species was the
major contribution of Donald Nicholson~\cite{Dagley:pathways:70}.  A
dynamical interpretation of the universality of metabolism that may be
more important for understanding evolution is that ecosystems have
dynamics inherent to their own level of aggregation that is not
captured in their descriptions merely as assembled communities of
species. Such dynamics are expressed as limits on, or long-range
evolutionary convergences of, innovations within metabolism.

The corollary, that individuality is a derived characteristic of
living systems within a larger framework of metabolic regularity, fits
well with the modern understanding that individuality takes many forms
which must be explained within their contexts~\cite{Buss:EvolInd:07}.
Alternatively, in more conventional genetic descriptions of
evolution~\cite{Fisher:gen_theory:30,Ewens:pop_gen:04}, metabolic
completeness, trophic as well as physiological flux balance, and
network-level response to fluctuations are explicit features
contributing to an organism's fitness within a co-evolving or
constructed environment~\cite{OdlingSmee:NC:03}.

We can to a considerable extent disentangle the inherent chemical
hierarchy of metabolism from the evolutionary hierarchy of species by
studying variations in the anabolic (biosynthetic) versus catabolic
(degradative) pathways within organisms, along with the relations of
autotrophy (self-feeding) versus heterotrophy (feeding from others) in
the ecological roles of species.  We can argue for the existence of a
universal anabolic, autotrophic
network~\cite{Srinivasan:aquifex_chart:09,Srinivasan:aquifex_analysis:09}
that comprises the chemistry essential to life.  We can then separate
the structural requirements and evolutionary history of the universal
network from secondary complexities, which we will argue originate in
the diversification of species and the concurrent processes of
assembly of ecological communities.

Within the universal (and apparently essential) network we may
identify further layers, with distinct functions and plausibly
distinct origins. A functioning metabolism is both a network of fluxes
through substrate molecules, and a set of hierarchical relations in
which some of the more complex structures control the kinetics of
flows within the network.  Within the substrate network,
distinguishable subnetworks include the core network to synthesize CHO
backbones, networks radiating from the core that incorporate N, S, P,
or metals, higher-order networks that assemble complex organics from
``building blocks'', and still others that synthesize all forms of
polymers from small organic monomers.  Within the control hierarchy,
the layers of cofactors, oligomer catalysts, and integrated cellular
energetic and biosynthetic subsystems are qualitatively distinct.

\begin{center}
\begin{figure}[t!]
   \includegraphics[scale=0.68]{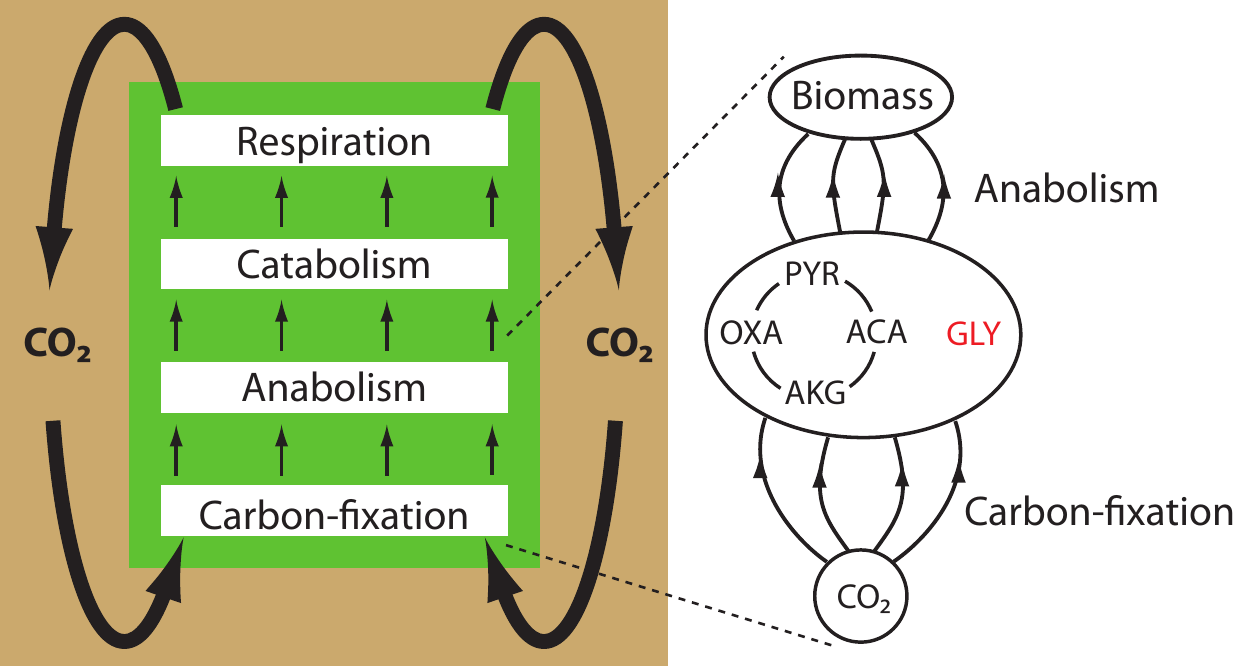}
  \caption{
  The metabolic structure of the biosphere. The biosphere as a whole (green)
  can be described as implementing a global biological carbon cycle
  based on ${\mbox{CO}}_2$, with carbon-fixation as the metabolic
  foundation that anchors it within geochemistry (brown). The small
  organic molecules produced during fixation of 
  ${\mbox{CO}}_2$ are subsequently transformed and built up into the
  full diversity of known biomolecules through the process of
  anabolism, before ultimately being broken back down through
  catabolism and re-released as ${\mbox{CO}}_2$ through respiration.
  The striking modularity of metabolism is expressed in the fact that
  the interface between carbon-fixation and anabolism consists of a
  very small number of small organic molecules (shown schematically at
  right-center). The key observation that in addition to intermediates
  in the citric acid cycle -- from which nearly all anabolic pathways
  emanate ~\cite{Smith:universality:04} -- glycine (red) should be included in
  this set allows a complete reconstruction of the evolutionary
  history of carbon-fixation~\cite{Braakman:carbon_fixation:12}. These
  points are explained in detail in
  Sec.~\ref{sec:core_carbon}. Abbreviations: Acetyl-CoA (ACA);
  Pyruvate (PYR); Oxaloacetate (OXA); $\alpha$-Ketoglutarate (AKG);
  Glycine (GLY). 
    \label{fig:Biosphere_global} 
  }
\end{figure}
\end{center}

The foundation of autotrophy -- and more generally the anchor that
embeds the biosphere within geochemistry -- is \emph{carbon-fixation},
the transformation of ${\mbox{CO}}_2$ into small organic molecules
(see Fig.~\ref{fig:Biosphere_global}).  A recent
study~\cite{Braakman:carbon_fixation:12} combining evidence from
phylogeny and metabolic network reconstruction -- an approach we refer
to as ``phylometabolic" reconstruction -- showed that all carbon
fixation phenotypes may be related by an evolutionary tree with very
high (nearly perfect) parsimony, and a novel but sensible phenotype at
the root.  The branches representing innovations in carbon fixation
were found to trace the standard deep divergences of bacteria and
archaea.  More striking, this work showed that likely environmental
drivers could be identified for most divergences, suggesting that deep
evolution reflects first incursions into novel geochemical
environments~\cite{Braakman:carbon_fixation:12}. The tight coupling of
the reconstructed phylogeny to geochemical variety suggests that
constraints from chemistry and energetics drove early evolution in
predictable ways, leaving little need to invoke historical
contingency. We will discuss these points, placing them in context, in
detail in Sec.~\ref{sec:core_carbon}.

\subsection{Catalytic control and origins of modularity in metabolism}

While carbon-fixation draws on all levels of biological organization
(requiring integration and control of many cellular components),
evolution in the network of its small-molecule substrate has consisted
only of changes in the use of a small number of clearly defined
reaction sequences.  The disruption, disconnection, or reversal of
these modules accounts for the full diversity of modern
carbon-fixation.  As we will show below, the module structure is
further defined by a distinction between two types of chemistry.
Within modules, the reactions are mainly (de-)hydration or
(de-)hydrogenation reactions, catalyzed by enzymes from common and
highly-diversified families.  The interfaces between modules are
created (and distinguished) by key carboxylation reactions, catalyzed
by highly conserved enzymes, often involving special metal centers
and/or complex organic cofactors.  The congruence of phylogenetic
branching with topological and chemical module boundaries suggests
that a very small number of catalytic innovations were the key
bottlenecks to evolutionary diversification, against a background of
facile and readily re-used organic chemistry.

Topological modularity in the small-molecule substrate network is
often associated with functional divisions in the more complex
molecules that control metabolism, particularly the cofactors, showing
that their metabolic role is also an evolutionary role.  As carriers
of electrons or essential functional groups, cofactors regulate
kinetic bottlenecks in metabolic networks.  Here again we wish to
distinguish the chemical complexity of metabolic reaction mechanisms,
and the role of cofactors in those reactions, from higher level
control of kinetics through regulation of concentrations, which may
optimize pathway robustness against perturbations~\cite{Coelho09}.
The structurally most elaborate cofactors tend to facilitate the
chemically most complex reaction mechanisms. As a result, the
distinction between presence and absence of these cofactors is
effectively the absolute presence or absence of reaction mechanisms (a
``topological'' network property), as contrasted with a finite rate
adjustment.  Thus, the appearance and diversification of families of
biosynthetically related cofactors introduced functions which served
as ``keys'' to domains in organic chemistry, incorporating these
within biochemistry.  As a result we may often map biosynthetic
pathway diversification of cofactors onto particular lineage
divergences in the tree of life.  We will show examples of this in
Sec.~\ref{sec:cofactors}. Cofactor biosynthetic networks are
themselves often modular, with multiple biosynthetic pathways in a
family using closely related enzymes that enable structures
characteristic of the cofactor class.

The quite sharply defined roles of many modules enable us to
understand strong evolutionary convergences that have occurred within
fundamental biochemistry, and in some cases we can relate the
functioning of an entire class of substrate or control molecules to
specific chemical properties of elements or small chemical groups.
Several important module boundaries are aligned at the same points in
their substrate networks and their control layers.  This suggests to
us that lower-level substrate-reaction networks introduced constraints
on the accessible or robust forms of catalysis and aggregation that it
was later possible to build up over them.  From repeated motifs within
the substructure of modules, and from patterns of re-use or
convergence, we may identify chemical constraints on major transitions
in metabolic evolution, and we may separate the early functions of
promiscuous catalysts as enablers of chemistry, from later
restrictions of reactants as catalysts were made more specific.  The
remarkable fact that such low-level chemical distinctions (in
elements, reactions, or small-molecule networks) should have created
constraints on innovation well into the Darwinian era of modern cells
suggests these as relevant constraints also in the pre-cellular era.

\subsection{Manuscript outline}

Our main message is twofold: 1) that the structure of biosynthetic
networks and their observed variation, even though the networks are
elaborate, has a compact representation in terms of a small collection
of rules for composition, and 2) that the same rules we abstract from
composition have a natural interpretation as constraints on
evolutionary dynamics, which as a generating process has produced the
observed variants.  We intend the expression ``logic of metabolism''
to refer to the collection of architectural motifs and functions that
have apparently been necessary for persistence of the biosphere, that
have led to modularity in the physics and chemistry of life, and that
have determined its major evolutionary contingencies and convergences.

After a short description of the important global features of
metabolism in Sec.~\ref{sec:metab_global}, we will construct these at
ascending levels in the hierarchy, beginning in
Sec.~\ref{sec:core_carbon} with the networks of core carbon fixation
and the lowest levels of intermediary metabolism.  We will then, in
Sec.~\ref{sec:cofactors}, consider cofactors as the intermediate level
of structure and the first level of explicit control in biochemistry,
illustrating how key cofactor classes govern the fixation and transfer
of elementary carbon units, and introduce control over reductants and
redox state.  Both in the metabolic substrate and in the cofactor
domain, it will be possible to suggest a specific historical order for
many major innovations.  For the substrate network this will capture
conditional dependencies in the innovation of carbon fixation
strategies.  For cofactors it will allow us to approximately place the
emergence of specific cofactor functionalities within the expansion of
metabolic networks from inorganic inputs.

In Sec.~\ref{sec:innovation} we consider the processes by which
innovation occurs, specifically interplay of the introduction of
general reaction mechanisms versus selectivity over substrates.  The
modular substructure and evolutionary sequence of many of our
reconstructed innovations favors an early role for non-specific
catalysts, with substrate selectivity appearing later.  In
Sec.~\ref{sec:integration} we then list candidates for the major
organizing constraints on integration of metabolism within cells.
These include the role of compartments in linking energy systems, as
well as the coupling of physiological and genetic individuality, which
permit species differentiation, and complementary specialization
within ecological assemblies. Finally, in Sec.~\ref{sec:origins} we
discuss how the various observation made throughout the manuscript may
be used to provide context in assessing scenarios for the emergence of
life. Because we draw from several areas of research which do not
have fully-shared vocabulary, a glossary with some terms used
frequently is provided in App.~\ref{sec:glossary}.

\section{An overview of the architecture of metabolism}
\label{sec:metab_global}

\subsection{Anabolism and catabolism in individuals and ecosystems}
\label{sec:ana_cata}

Metabolic networks within organisms are commonly characterized as
having three classes of pathways: 1) catabolic pathways that break
down organic food to provide chemical ``building blocks'' or energy;
2) core pathways through which nearly all small metabolites pass
during primary synthesis or ultimate breakdown, and 3) anabolic
pathways that build up all complex chemicals from components
originating in the core.  The motif of three-stage pathways --
  catabolic, core, anabolic -- between typical pairs of metabolites
  has been abstracted into a paradigm of ``bowtie'' architecture for
  metabolism~\cite{Csete:bowties:04,Zhao:bowties:07,Riehl:coli_paths:10}.
This qualitative characterization (which may be complicated by salvage
pathways and other cross-linkages) is supported by a strong
statistical observation that most minimal pathways connecting pairs of
metabolites consist of a catabolic and an anabolic segment connected
through the core~\cite{Riehl:coli_paths:10}.  Thus, relatively
speaking, the catabolic and anabolic pathways are less densely
crosslinked than pathways within the core, from which they radiate.

The reason for this lack of cross-linking can be understood from the
explanation of path lengths in terms of number theory and string
chemistries in Ref.~\cite{Riehl:coli_paths:10}. Lengths of typical
optimal paths between pairs of metabolites in \emph{E.~coli} are
logarithmic in carbon count, because they decompose molecules into
small prime ``factors'' in the core which are then modified by single
carbons to other prime factors and re-assembled. Thus, optimal
conversions within the bowtie consists of finding common molecular
``divisors" of input and output metabolites, which in actual metabolic
chemistry are familiar 2-, 3-, and 5-carbon groups.  We will
  argue that, when other chemical and phylogenetic evidence is taken
  into account, the fact that short paths exist from most metabolites
  to a small set of building blocks is more likely a reflection of the
  prior role of the core (where building blocks are created) in
  defining the possibilities for later anabolism and thus the
  metabolites reached by the bowtie.
    
\begin{center}
\begin{figure}[ht]
  \includegraphics[scale=0.75]{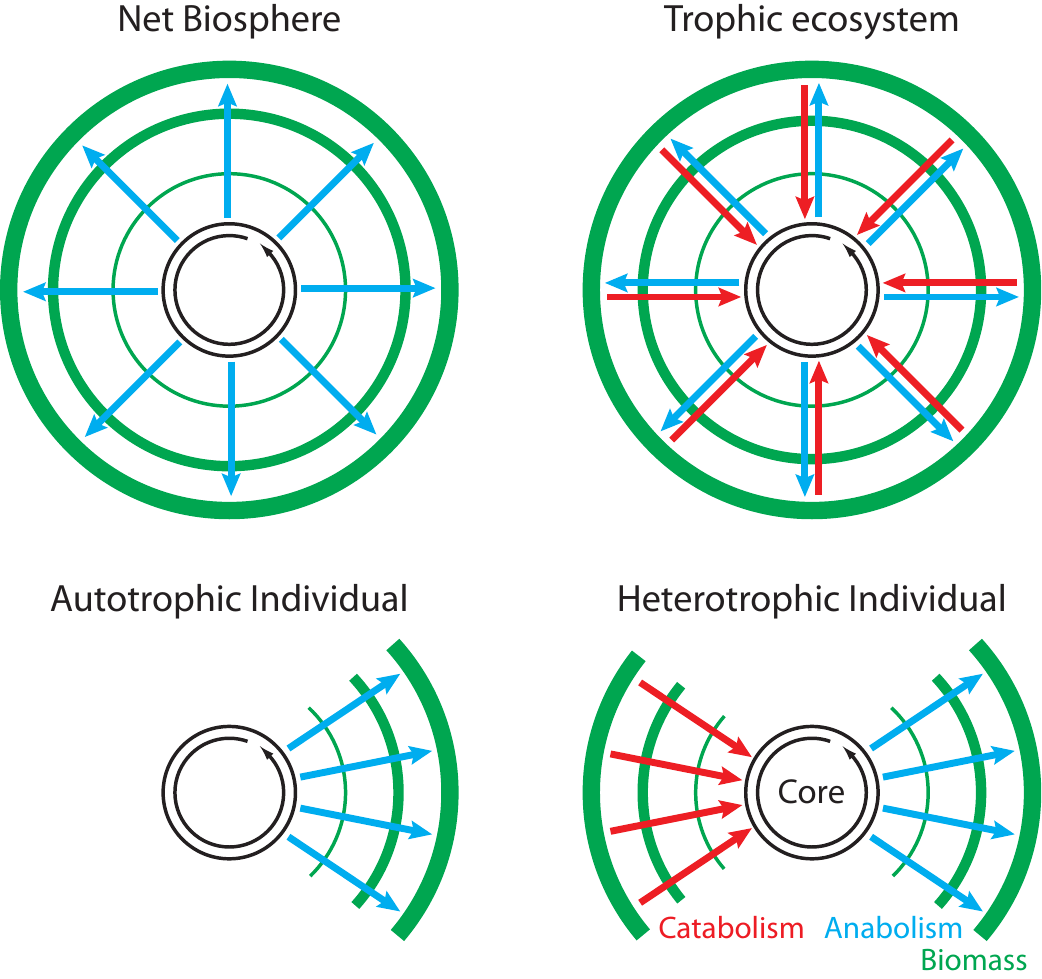}
  \caption{
    Global structure of metabolism.  Anabolic pathways (blue) build
    biomass and catabolic pathways (red) break it down.  Catabolic
      pathways may be direct or approximate reverses of anabolism; see
      main text for further discussion on this point.  Carbon enters
    through the core (black), which is the starting point of anabolism
    and also the endpoint of respiration. Because carbon enters
      the biosphere as ${\mbox{CO}}_2$ (see
      Fig.~\ref{fig:Biosphere_global}), the biosphere today is
      autotrophic as a whole and anabolism is functionally prior to
      catabolism. As a result, both single organisms and assemblies
    of organisms can function using metabolic modes consisting only of
    anabolic pathways that fan outward from the core (blue and green).
    By partitioning pathway directions between anabolic and catabolic
    (joined at the core), organisms can use metabolic modes consisting
    of the familiar ``bowtie'' architecture of derived metabolism (red
    with blue).  Their assembly into trophic ecosystems (blue and red
    radial graph) then both builds and degrades organic compounds
    actively, cycling carbon between environmental ${\mbox{CO}}_2$ and
    biomass (green).  In these graphs, concentric (green) shells
    reflect sequential steps in biosynthesis leading to a hierarchy of
    increasing molecular complexity.
    \label{fig:Metabolism_Overview} 
  }
\end{figure}
\end{center}

Catabolic pathways in a cell may be fed through physiological or
trophic links to other cells or organisms, or they may break down food
produced previously by the same cell and then stored.
Fig.~\ref{fig:Metabolism_Overview} illustrates schematically the
relation of the three classes.  Both catabolic and anabolic pathways
may be large and somewhat diversified; the core itself constitutes no
more than a few hundred small
metabolites~\cite{Srinivasan:aquifex_chart:09,Srinivasan:aquifex_analysis:09},
most of which have functions that are universal throughout the
biosphere.

Whole-organism metabolisms are conventionally divided into two classes
-- autotrophic and heterotrophic -- according to the ways they combine
anabolic and catabolic pathways~\cite{Lengeler:BP:99}.  Autotrophs
synthesize all required metabolites from inorganic precursors, and can
function without catabolism, using only the core and anabolic pathways
radiating from it. Establishing the metabolic self-sufficiency of a
putatively autotrophic organism can prove challenging,
however~\cite{Srinivasan:autotroph:12}.  Heterotrophs, in contrast,
are organisms that must obtain organic inputs from their environments
because they lack essential biosynthetic pathways.  Autotrophy and
heterotrophy are best understood as modes of metabolism, between which
some individual species may switch depending on circumstances, and
which may even be mixed at the level of sub-networks within a given
organism.  Many organisms are obligate autotrophs or heterotrophs, but
others are facultative autotrophs that can switch between
fully-autotrophic and heterotrophic metabolic states, while still
others are mixotrophs that concurrently use both ${\mbox{CO}}_2$ and
organic carbon inputs to synthesize different parts of their
biomass~\cite{Lengeler:BP:99}.  The important distinction for what
follows is that autotrophs and heterotrophs play fundamentally
different ecological roles.

Autotrophic metabolism forms the lowest trophic level in the
biosphere, fixing ${\mbox{CO}}_2$ into organic matter, while
heterotrophic metabolism forms all subsequent levels, determining the
structure of flows of organic compounds in trophic
webs~\cite{Sterner:eco_stoich:02}, and actively cycling carbon from
biomass back to environmental ${\mbox{CO}}_2$.  While all biological
free energy passes at some stage through redox couples, autotrophs
capture a part of this energy by transferring electrons from high
energy reductants to ${\mbox{CO}}_2$~\cite{Smith:universality:04}.
Heterotrophs may exploit incomplete use of this free energy through
internal redox reactions (fermentation), or they may re-oxidize
organic matter back to ${\mbox{CO}}_2$ (respiration).

The role of catabolism in most organisms is closely tied to their
ecological role as heterotrophs.  Heterotrophy provides enormous
opportunity for metabolic
diversification~\cite{Rodrigues:met_nets:09}, in the evolution of
catabolic pathways and the partitioning of essential anabolic
reactions among the constituent species within ecosystems.  However,
the study of metabolism restricted to particular heterotrophic
organisms\footnote{Almost all model organisms have been heterotrophs,
  because these are accessible and are usually connected to humans as
  symbionts, pathogens, or cultivars.  \emph{E.~coli} (in which
  operons were discovered) is a phenotypically and trophically very
  plastic organism as this is required for its complex life cycle.  No
  known multicellular organisms can reduce the triple-bond of
  ${\mbox{N}}_2$, making them reliant either directly on microbial
  nitrogen fixers for reduced nitrogen (${\mbox{NH}}_4^{+}$) or on
  mineralized forms such as ${\mbox{NO}}_3^{-}$ derived ultimately
  from microbial nitrogen metablism (or outputs of human technological
  processes).  The only known autotrophic organisms are bacteria
    and archaea, and none of these is developed nearly to the level
    that standard heterotrophic model systems (such as \emph{E.~coli})
    are.} can obscure much of its universality: heterotrophs may
differ widely, but the aggregate anabolic networks that sustain them
at the level of ecosystems are largely invariant.  Autotrophs show
that much of this diversity is not essential to life, allowing us to
conceptually separate the requirements for biosynthesis from
complexities that originate in processes of individual specialization
and ecological assembly~\cite{Smith:auto_hetero:10}.

The ``bowtie'' motif~\cite{Csete:bowties:04,Zhao:bowties:07} -- a
  paradigm derived from the study of heterotrophs\footnote{The
    paradigm of the metabolic bowtie is also in part a borrowing from
    a conventional paradigm in engineering~\cite{Csete:bowties:04},
    motivated by applications to human physiology and medicine (John
    Doyle, pers.~comm.).} -- can be misleading, as it combines
  universal metabolic pathway dependencies with widely variable
  physiological or ecological specializations.  The core and
anabolism are essential (and we argue more ancestral), and the
reduction in cross-linking with distance from the core may be seen to
reflect an entirely outgoing radial ``fan'' of anabolism.  Biomass is
organized in a sequence of concentric shells spanned by the radial
pathways, which count the number and complexity of biosynthetic
steps~\cite{Hartman75,Srinivasan:aquifex_analysis:09}.  Organisms, in
particular autotrophs, exist which can function without catabolism,
but only the most derived parasites lack anabolism.  For example,
members of the genus \emph{Mycoplasma} can function with remarkably
small genomes, having given up nearly all genes associated with the
\emph{de novo} synthesis of amino acids, cofactors, nucleotides and
lipids~\cite{Fraser95,Himmelreich96,Razin98}, because they live as
intracellular parasites in hosts that synthesize these.

Most catabolic pathways are also, in varying degrees, reversals of
widespread anabolic pathways. In some cases the reversal is exact,
often for short pathways, as in the case of glycine metabolism that we
discuss in detail in the next section. In other cases, such as
gluconeogenesis and glycolysis~\cite{Say:gluco_glyco:10} or fatty acid
metabolism, catabolic pathways resemble their anabolic counterparts
closely but differ in a few intermediates, cofactors, or enzymes,
usually for thermodynamic reasons~\cite{Metzler:BC:03}.  Finally, in
some cases catabolism reflects genuine innovations, as in the
metabolism of the branched-chain amino acids~\cite{Massey76} or of
nucleotides~\cite{Vogels76}, or some salvage pathways.  We find it
significant that even in cases where reversals are only approximate
due to variation in some of the substrates or catalysts, the overall
\emph{sequences} of reactions at the substrate-level are often nearly
completely preserved.  In such cases, substitutions, which may appear
to be large differences from the perspective of enzyme homology,
clearly are often local alterations in energy flow usually involving
interchanged reaction orders. An example of this is the variable order
of thioesterification to form succinyl-CoA in reductive, fermentative,
and oxidative TCA
cycles~\cite{Baughn:Mtb_TCA:09,Watanabe:tuberculosis:11,Zhang:cyano_TCA:11}.
In the last two cases, organisms may use succinate as an
  intermediate in the formation of succinyl-CoA from
  $\alpha$-ketoglutarate, rather than directly performing an oxidative
  decarboxylation, which would constitute strict reversal of the
reductive TCA reaction. Succinate acts as an electron sink, pulling
the reaction forward, but an additional ATP hydrolysis is then needed
to form succinyl-CoA from it.

The preserved reaction sequences may be ``channels" within organic
chemistry with optimal path length or
connectivity~\cite{Melendez94,Melendez96,Noor:min_walk:10} that were
easier to bring or maintain under catalytic control, or for later
innovations they may reflect lock-in by requirements of secondary
metabolism.  Finally, many reversals from anabolism to catabolism can
be explained as consequences of ecological change, with finer
distinctions arising as adaptations to specific ecological or
geochemical environments.

The conceptual difference and asymmetry between autotrophy and
heterotrophy becomes clearer when we examine the metabolic structure
of ecosystems at increasing scales of aggregation.  Entire ecosystems,
to the extent that they are approximately closed, function chemically
as autotrophs.  The biosphere as a whole (see
Fig.~\ref{fig:Biosphere_global}) is not only approximately, but fully
autotrophic, as today it does not depend significantly on
extraterrestrially, atmospherically, or geologically produced
organics.  This observation still admits two possibilities for the
emergence of aggregate metabolism: Either the biosphere has been
autotrophic since its inception, or it was originally heterotrophic
and later switched to using ${\mbox{CO}}_2$ as its sole carbon
source. We have recently shown that assuming autotrophy at least as
far back as the era of a common metabolic ancestor leads to a highly
parsimonious reconstruction for the evolution of carbon-fixation
pathways~\cite{Braakman:carbon_fixation:12}.  The congruence of our
tree of carbon-fixation phenotypes with standard
phylogenies~\cite{Puigbo:trees:09}, which place modern autotrophs as
the conservative descendents of deep
branches~\cite{Reysenbach:vents:02}, together with numerous arguments
drawing evidence from biochemistry and geochemistry that thermophilic
autotrophs are the most-plausible models for deep-ancestral bacteria
and archaea~\cite{Russell:AcetylCoA:04,Martin:OMP:07,Martin:vents:08,%
  Hugler:fix_paths:11,Fuchs11}, permits quite specific and consistent
biochemical proposals for an autotrophic deep-ancestral stage of life.
To our knowledge, there is no equivalent body of evidence leading to
specific and consistent predictions of heterotrophic forms at the
earliest evolutionary times.

For all of these reasons we will interpret the core and anabolic
pathways as the base layer and skeleton of most-fundamental
constraints on metabolism, and will consider the problem of
emergence and early evolution of fully autotrophic systems. 
Reconstructing the emergence of autotrophic metabolism provides
important context to the emergence of life, to which we return in
Sec.~\ref{sec:origins}.  We restrict the discussion here to the
structure and evolution of metabolism, and to conclusions that can
be drawn from biochemistry, phylogenetics, and geochemical and
ecological context.  These conclusions do not depend on speculations
about what chemical stages may have preceded the emergence of
anabolism.  

As long as we do not conflate the chemical condition of autotrophy
(complete anabolism) with assumptions about individuality (whether
complete anabolisms are contained within the regulatory control of
individual organisms)~\cite{Smith:auto_hetero:10}, and as long as we
recognize the ecosystem as potentially the correct level of
aggregation to define autotrophy, we need not assume that the first
fully functioning autotrophic metabolism consisted of individual
cells.  Our interpretation extends equally to populations of organisms
that were physiologically as well as genetically incomplete and
functioned cooperatively~\cite{Woese:univ_anc:98,Woese:HGTtree:00,%
  Woese:HGTcells:02,Goldenfeld:evol_collect:11}. Once organism-level
and species-level organization has been put aside as a separate
question, the chemical distinction between heterotrophy and autotrophy
is one between metabolic partial-systems with unknown and highly
variable boundary conditions, versus whole-systems required to subsist
on ${\mbox{CO}}_2$ and reductant. If we wish to understand the
structure of the biosphere and to interpret the sequence of
innovations in core carbon fixation, the added constraint of
autotrophy provides a framework to do this.

\subsection{Network topology, self-amplification, and levels of structure}
\label{sec:network_autocat}

\begin{figure}
  \begin{center} 
   \includegraphics[scale=0.65]{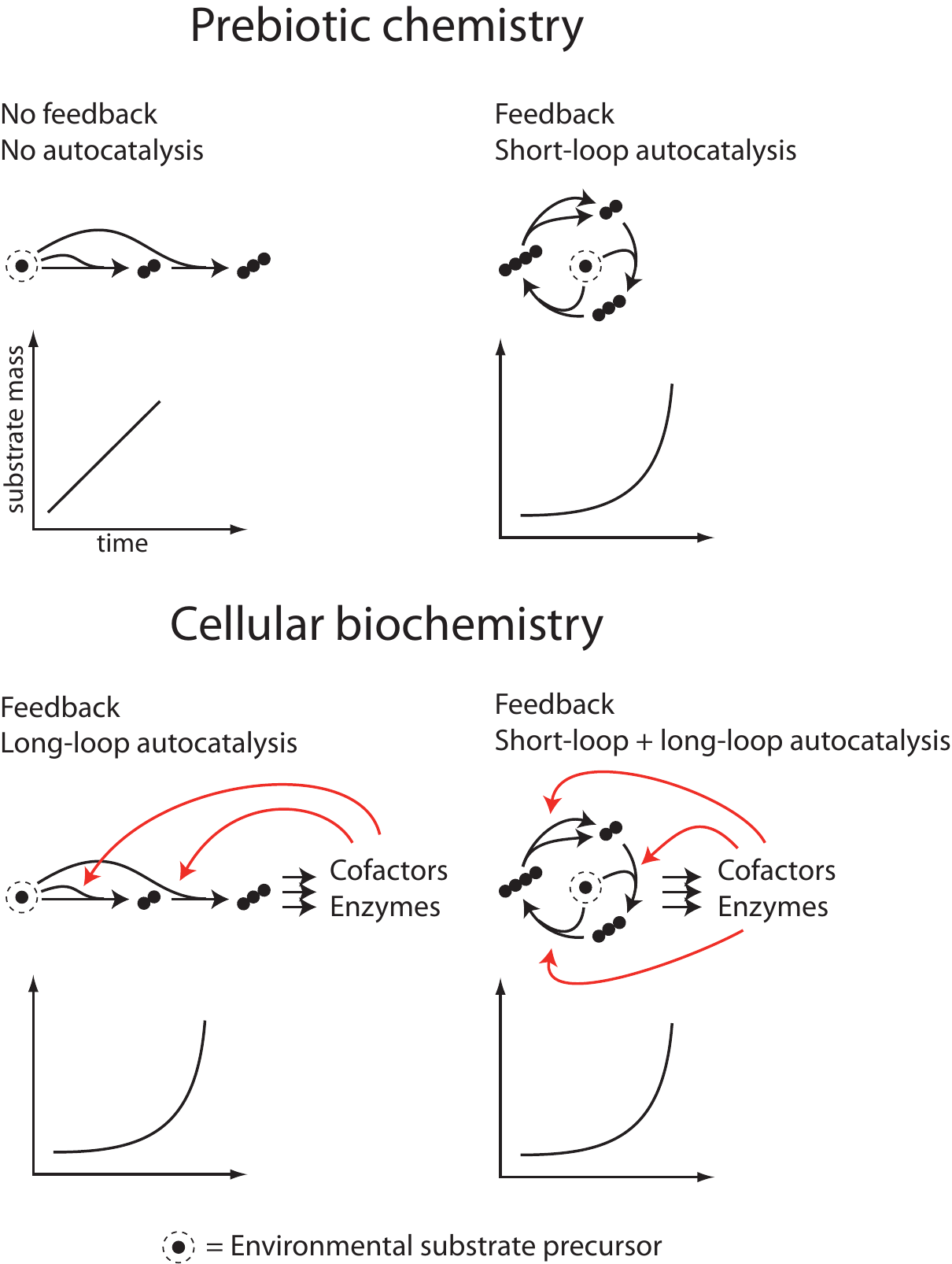}
  \caption{
  Upper limit growth rate curves for chemical reaction networks of
  different classes. To highlight the role of network topologies, the
  chemistry is simplified with only C-C bond forming and cleaving
  reactions shown. Each of the growth curves qualitatively
  represents the upper limits for mass accumulation within the
  participating substrates of the pathways. When fully integrated
  within modern cellular biochemistry, both linear and cyclic pathways
  are network autocatalytic and capable of exponential growth. This
  form of network autocatalysis, however, derives from feedback
  provided by cofactors and enzymes, both of which have elaborate
  synthesis pathways, and is thus classified as ``long-loop''
  autocatalysis. In an abiotic world in which reaction-level catalysis
  is limited to external sources, only cyclic pathways with feedback
  topologies at the substrate level -- correspondingly classified as
  ``short-loop'' autocatalysis -- are capable of exponential growth,
  while linear pathways are limited to linear growth. We contend that
  an early presence of short-loop autocatalysis is important because
  it provides a mechanism to concentrate mass flux within abiotic
  chemical networks, preventing excessive dilution with increasing
  size and complexity of organic molecules, and in turn giving easier
  and more robust access to subsequent long-loop feedback closures.
    \label{fig:network_autocat_demo} 
  }
  \end{center}
\end{figure}

Understanding either the emergence of metabolism, or the robust
persistence of the biosphere, requires understanding life's capacity
for exponential growth.  Exponential growth results from proportional
self-amplification of metabolic and other networks that have an
``autocatalytic''
topology~\cite{Eigen:hypercycle:77,Eigen:hypercycle:78,%
  Kauffman:OriginsOrder:93,Zachar10,Hordijk:autocat:12,Andersen:NP_autocat:12}
(see Fig.~\ref{fig:network_autocat_demo}).  Network autocatalysis is a
term used to describe a topological (stoichiometric) property of the
substrate network of chemical reactions.  In a \emph{catalytic}
network, one or more of the network intermediates is needed as a
substrate to enable the pathway to connect to its inputs or to convert
them to outputs, but the catalytic species is regenerated by the stage
at which the pathway completes.  Network-catalytic pathways must
therefore incorporate feedback and comprise one or more loops with
regard to the internally produced molecules.  An \emph{autocatalytic}
network is a catalytic network augmented by further reactions that
convert outputs to additional copies of the network catalyst,
rendering the pathway self-amplifying.

\emph{Molecular autocatalysis} -- the property that intermediates in a
pathway serve as conventional molecular catalysts for other reactions
in the pathway -- may be understood as a restricted form of network
autocatalysis in which the reaction to which some species is an
essential input is the same reaction that regenerates that species.
Some chemists prefer to use the term ``network autoamplification'' for
the general case, restricting ``autocatalysis'' to apply only when
species are traditionally-defined molecular catalysts.  We will use
``autocatalysis'' for the general case, to reflect the property of
stoichiometry that a pathway regenerates essential inputs.  For us the
distinction between autocatalysis at the single molecule versus more
general network level mainly effects the kinetics and regulation of
pathways.

Network autocatalysis is necessary to maintain dynamical ordered
states, by re-concentrating inputs into a finite number of
intermediates, against the disordering effects of thermodynamic decay
and continual external perturbation.  Therefore all observed
persistent material flows in the biosphere can only be products of
autocatalytic networks, though they may require hard-to-recognize
feedbacks ranging from the level of cell metabolism to trophic ecology
for full regeneration.  This \emph{ex post} observation does not,
however, explain why self-amplification was possible in abiotic
chemistry to give rise to a biosphere.  In addition to topologies
enabling feedback, the latter would have required that intermediates
in the network be produced at rates higher than those at which they
were removed.

The significant observations about autocatalysis in the extant
biosphere, which may also contain information about its emergence,
concern the complexity, number, and particular form of levels in which
autocatalytic feedback can be found.  Where the hierarchical modules
of metabolic structure or function follow the boundaries required for
feedback closure of different autocatalytic sub-networks, it may be
possible to order the appearance of those sub-networks in time. It may
also be possible to infer the geochemical supports they required for
stability and self-amplification, before those supports were attained
through integration into cellular biochemistry.

We wish, in these characterizations, to recognize what we might call
``conditional'' as well as strict autocatalysis.  In extant organisms,
where (essentially) all reactions are catalyzed by macromolecules, and
most cofactors (reductants, nucleoside-triphosphates, coenzymes) are
recharged by cellular processes, strict autocatalysis of any network
is only satisfied in the context of the full complement of integrated
cellular processes.  If, however, inputs provided by cofactors,
macromolecules and energy systems in modern cells could have been
provided externally in earlier stages of life, for instance by
minerals or geochemical processes, then identifying networks in extant
biochemistry that, although simple, \emph{would be} autocatalytic if
given these supports, may give information about intermediate stages
of emergence (see Fig.~\ref{fig:network_autocat_demo}).  The strong
modularity of extant metabolism and its congruence with such
conditionally autocatalytic topologies suggests that a separation into
layers corresponding to stages of emergence may be sensible. 

We will argue that the two most important and functionally distinct
early layers are the small-molecule substrate of core metabolism
itself, and the organic cofactors synthesized downstream from this
core, which feed back through network and molecular catalysis to form
a control layer over the core network (as well as, later, secondary
networks).  The picture of extant metabolism as the outcome of layers
of emergence has been advanced in many
forms~\cite{Szent72,Hartman75,Morowitz:BCL:92}, and the central
importance of feedback through catalysis has also been
emphasized~\cite{Copley:PMRNA:06}.  Here we consider the very specific
relations between reductive core pathways and cofactors, as evidence
that intermediary metabolism is the result of kinetic stabilization
and selection of the core that arose previously.  In addition to
reconstructing historical stages, the mechanisms leading to
autocatalysis in different sub-systems may suggest important
geochemical contexts or sources of robustness still exploited in
modern metabolism.

\subsection{Network-autocatalysis in carbon-fixation pathways}

At the chemically simplest level of description -- that of the
small-molecule metabolic substrates and their reaction-network
topologies -- carbon fixation pathways form two classes.  Five of the
the six known pathways are autocatalytic loops, while one is a linear
reaction sequence. (All uses of autocatalysis in this section refer to
\emph{conditional} autocatalysis as explained above.)  The loop
pathways condense ${\mbox{CO}}_2$ or bicarbonate onto their substrate
molecules, lengthening them.  Each condensation is accompanied
or followed by a reduction, making the average oxidation state of
carbon in the pathway substrate lower than that of the input
${\mbox{CO}}_2$, and resulting in a negative net free energy of
formation in a reducing environment~\cite{Smith:universality:04}.
(Reducing power may originate in the geochemical environment, but in
modern cells electrons are transferred endergonically to more powerful
reductants such as NADH, NADPH, ${\mbox{FADH}}_2$, or reduced
ferredoxin.)  Each fixation loop contains one reaction where the
maximal-length substrate is cleaved to produce two intermediates
earlier in the same pathway, resulting in self-amplification of the
pathway flux.  As long as pathway intermediates are replenished faster
than they are drained by parasitic or anabolic side reactions, the
loop current remains above the autocatalytic threshold.  However, the
threshold is fragile, as pathway kinetics provide no inherent barrier
against flux's falling below threshold and subsequently collapsing.
The autocatalytic threshold and dynamics of growth, saturation, or
collapse are considered in Sec.~\ref{sec:evol_history}.

At the level of network topology, the linear \emph{Wood-Ljungdahl}
(WL) fixation
pathway~\cite{Utter:WL:51,Ljungdahl:WL1:65,Ljungdahl:WL2:65} is
strikingly unlike the five loop pathways.  Instead of covalently
binding ${\mbox{CO}}_2$ onto pathway substrates, which then serve as
platforms for reduction, the WL reactions directly reduce one-carbon
(${\mbox{C}}_1$) groups, and then distribute the partly- or
fully-reduced intermediates to other anabolic pathways where they are
incorporated into metabolites.  The linear sequence of reductions has
no feedback, and the ${\mbox{C}}_1$ groups at intermediate oxidation
states do not increase in complexity.  Instead, these reductions
(leading to intermediate ${\mbox{C}}_1$ states that would be unstable
in solution) are carried out on evolutionarily refined folate
cofactors~\cite{Maden:folates_pterins:00}.  The topology of the WL
pathway becomes self-amplifying only if the larger and more complex
biosynthetic network for these cofactors is considered together with
that of the ${\mbox{C}}_1$ substrate, requiring that a longer
feedback loop be maintained than the mere substrate loop in the
other fixation pathways.  In the network context of the WL fixation
mechanism, the folate cofactors have an intermediate role between
network catalysts and molecular catalysts, as they are passive
carriers, but form stable molecular intermediates rather than mere
complexes as are formed by enzymes with their substrates. We will
characterize this distinction between the loop-fixation pathways and
WL as a distinction between \emph{short-loop} and \emph{long-loop}
autocatalysis (see Fig.~\ref{fig:network_autocat_demo}).

The network catalysts that could be said to ``select'' the short-loop
pathways are the reaction intermediates themselves.  The key
metabolites that have the corresponding selection role for WL are the
folate cofactors produced in a secondary biosynthetic network.
Short-loop and long-loop pathways are therefore distinguished both by
the number of reactions that must be maintained and regulated, and by
the fact that WL spans substrates and the biosynthesis of cofactors,
which we will argue in Sec.~\ref{sec:cofactors} are naturally
interpreted as qualitatively distinct layers within biochemistry.

The appearance of different features suggesting simplicity or
primordial robustness, in different fixation pathways, together with
aspects of their phylogenetic distribution, have led to diverse
proposals about the order of their
emergence~\cite{Berg:carbon_fixn:10,Pereto:orig_metab:12}.  WL is the
only carbon-fixation pathway found in both bacteria and archaea, and
its reactions have been shown to have abiotic mineral
analogues~\cite{Russell:AcetylCoA:04,Martin:OMP:07,Berg:carbon_fixn:10},
suggesting a prebiotic origin.  Yet WL is not self-amplifying and so
lacks the capacity for chemical ``competitive exclusion'' (equivalent
to the capacity for exponential growth).  The cofactors that make it
self-amplifying are complex, and the simple pathway structure of
${\mbox{C}}_1$ reduction does not suggest what would have supported
their formation.

In contrast, autocatalysis within the small-molecule substrate
networks of the loop pathways suggests the inherent capacity for
self-amplification, exponential growth, and chemical competitive
exclusion.  This is an appealing
explanation~\cite{Smith:universality:04} for the role, particularly of
the intermediates in the reductive citric acid
cycle~\cite{Buchanan:rTCA:90,Morowitz:BCL:92} (discussed in
Sec.~\ref{sec:core_carbon}) as precursors of biomass.  Arcs within
this pathway have also been reproduced experimentally in mineral
environments~\cite{Cody:pyruvate:00}, though a self-amplifying system
has not yet been demonstrated.  However, self-amplification requires
complete loops, and even the most compelling candidate for a
primordial form (reductive citric acid cycling) is found only in a
subset of bacterial clades.

We argue in the next section that a joint fixation pathway
incorporating both WL and reductive citric acid cycling resolves many
of these ambiguities in a way that no modern fixation pathway can.
Proposals have previously been made in Ref's.~\cite{Hartman75}
and~\cite{Martin:OMP:07} for WL fixation followed by the use of
citric-acid cycle pathways. However, in Ref~\cite{Hartman75} the
TCA cycle is suggested to run oxidatively, while the primordial
networks proposed in Ref.~\cite{Martin:OMP:07} are forms of
acetogenesis. Neither therefore emphasizes self-amplification and
short-loop autocatalysis as essential early requirements. As a
phylogenetic root, a fully connected network combining WL and the rTCA
cycle defines a template from which the fixation pathways in all
modern clades could have diverged. As a candidate for a primordial
metabolic network, in turn, it provides both chemical selection of
biomass precursors by short-loop autocatalysis, and a form of
protection against the fragility of the autocatalytic threshold.  We
will first describe the biochemistry and phylogenetics of
carbon-fixation pathways in the current biosphere, and then show how
their patterns of modularity and chemical redundancy provide a
framework for historical reconstruction.

\begin{widetext}

\begin{center}
\begin{figure}[ht]
  \begin{center} 
  \includegraphics[scale=0.65]{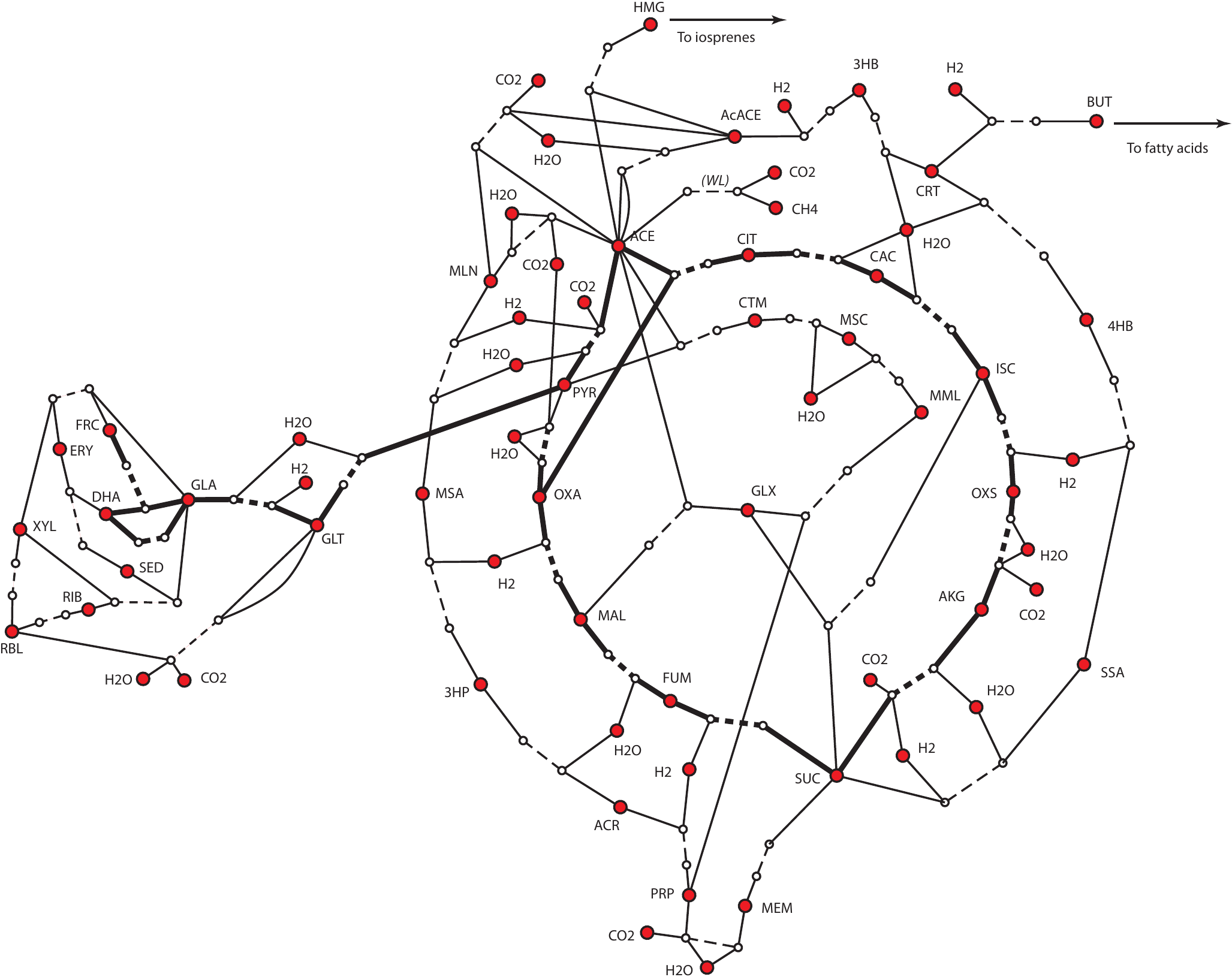}
  \caption{
  The projection of the complete network for core carbon anabolism
  onto its CHO components.  Phosphorylated intermediates and
  thioesters with Coenzyme-A are not shown explicitly.  The bipartite
  graph notation used to show reaction stoichiometry is explained in
  App.~\ref{sec:graph_explanation}.  Arcs of the reductive citric acid
  cycle and gluconeogenesis are bold, showing that these pathways pass
  through the universal biosynthetic precursors.  The Wood-Ljungdahl
  (labeled \emph{WL}) pathway, without its cofactors and reductants
  shown, is represented by the last reaction of the acetyl-CoA
  synthase, which is the inverse of a disproportionation.
  Abbreviations: acetate (ACE); pyruvate (PYR); oxaloacetate (OXA);
  malate (MAL); fumarate (FUM); succinate (SUC);
  $\alpha$-ketoglutarate (AKG); oxalosuccinate (OXS); isocitrate
  (ISC); cis-aconitate (CAC); citrate (CIT); malonate (MLN); malonate
  semialdehyde (MSA); 3-hydroxypropionate (3HP); acrolyate (ACR);
  propionate (PRP); methylmalonate (MEM); succinate semialdehyde
  (SSA); 4-hydroxybutyrate (4HB); crotonate (CRT); 3-hydroxybutyrate
  (3HB); acetoacetate (AcACE); butyrate (BUT); hydroxymethyl-glutarate
  (HMG); glyoxylate (GLX); methyl-malate (MML); mesaconate (MSC);
  citramalate (CTM); glycerate (GLT); glyceraldehyde (GLA);
  dihydroxyacetone (DHA); fructose (FRC); erythrose (ERY);
  sedoheptulose (SED); xylulose (XYL); ribulose (RBL); ribose (RIB).
    \label{fig:carbon_anabolism_CHO} 
  }
  \end{center}
\end{figure}
\end{center}

\end{widetext}

\section{Core carbon metabolism}
\label{sec:core_carbon}

Currently six carbon fixation pathways are
known~\cite{Hugler:fix_paths:11,Fuchs11}.  While they are distinct as
complete pathways, they have significant overlaps at the level of
individual reactions, and even greater redundancy in local-group
chemistry.  They are also, as shown in
Fig.~\ref{fig:carbon_anabolism_CHO} (above), tightly integrated with
the main pathways of core carbon metabolism, including lipid
synthesis, gluconeogenesis, and pentose-phosphate synthesis.

An extensive analysis of their chemistry under physiologically
relevant conditions has shown that individual fixation pathways
contain two groups of thermodynamic bottlenecks: carboxylation
reactions, and carboxyl reduction reactions~\cite{BarEven12}. In
isolation these reactions generally require ATP hydrolysis to proceed,
and the way pathways deal with (or avoid) these costs has been
concluded to form an important constraint on their internal
structure~\cite{BarEven12}.  We will further show how the elaborate
and complex catalytic mechanisms associated with these reactions form
essential \emph{evolutionary} constraints on metabolism.

We will first describe the biochemical and phylogenetic details of the
individual pathways, and then diagram their patterns of redundancy,
first at the level of modular reaction sequences, and then in
local-group chemistry.  Finally we will use this decomposition
together with evidence from gene distributions to propose their
historical relation and identify constraints that could have spanned
the Darwinian and pre-cellular eras.

\subsection{Carbon fixation pathways}

\subsubsection{Overview of pathway chemistries, phylogeny and
  environmental context}
\label{sec:pathway_overview}

{\bf Wood-Ljungdahl:} The Wood-Ljungdahl (WL)
pathway~\cite{Utter:WL:51,Ljungdahl:WL1:65,Ljungdahl:WL2:65,%
  Berg:carbon_fixn:10} consists of a sequence of five reactions that
directly reduce one ${\mbox{CO}}_2$ to a methyl group, a parallel
reaction reducing ${\mbox{CO}}_2$ to CO, and a final reaction
combining the methyl and CO groups with each other and with a molecule
of Coenzyme-A (CoA) to form the thioester acetyl-CoA.  The reactions
are shown below in Fig.~\ref{fig:C1_metabolism}, and discussed in
detail in Sec.~\ref{sec:cofactors}.  The five steps reducing
${\mbox{CO}}_2$ to $-{\mbox{CH}}_3$ make up the core pathway of folate
(vitamin ${\mbox{B}}_9$) chemistry and its archaeal analog, which we
consider at length in Sec.~\ref{sec:modularity}.  The reduction to CO,
and the synthesis of acetyl-CoA, are performed by the bi-functional
CO-Dehydrogenase/Acetyl-CoA Synthase (CODH/ACS), a highly conserved
enzyme complex with Ni-[${\mbox{Fe}}_4{\mbox{S}}_5$] and
Ni-Ni-[${\mbox{Fe}}_4{\mbox{S}}_4$]
centers~\cite{Dobbek:CODH:01,Darnault03,Seravalli04,Svetlitchnyi04}.
Methyl-transfer from pterins to the ACS active site is performed by a
corrinoid iron-sulfur protein (CFeSP) in which the cobalt-tetrapyrrole
cofactor cobalamin (vitamine ${\mbox{B}}_{12}$) is part of the active
site~\cite{Banerjee03,Bender11}.


\begin{figure}[ht]
  \begin{center} 
  \includegraphics[scale=0.6]{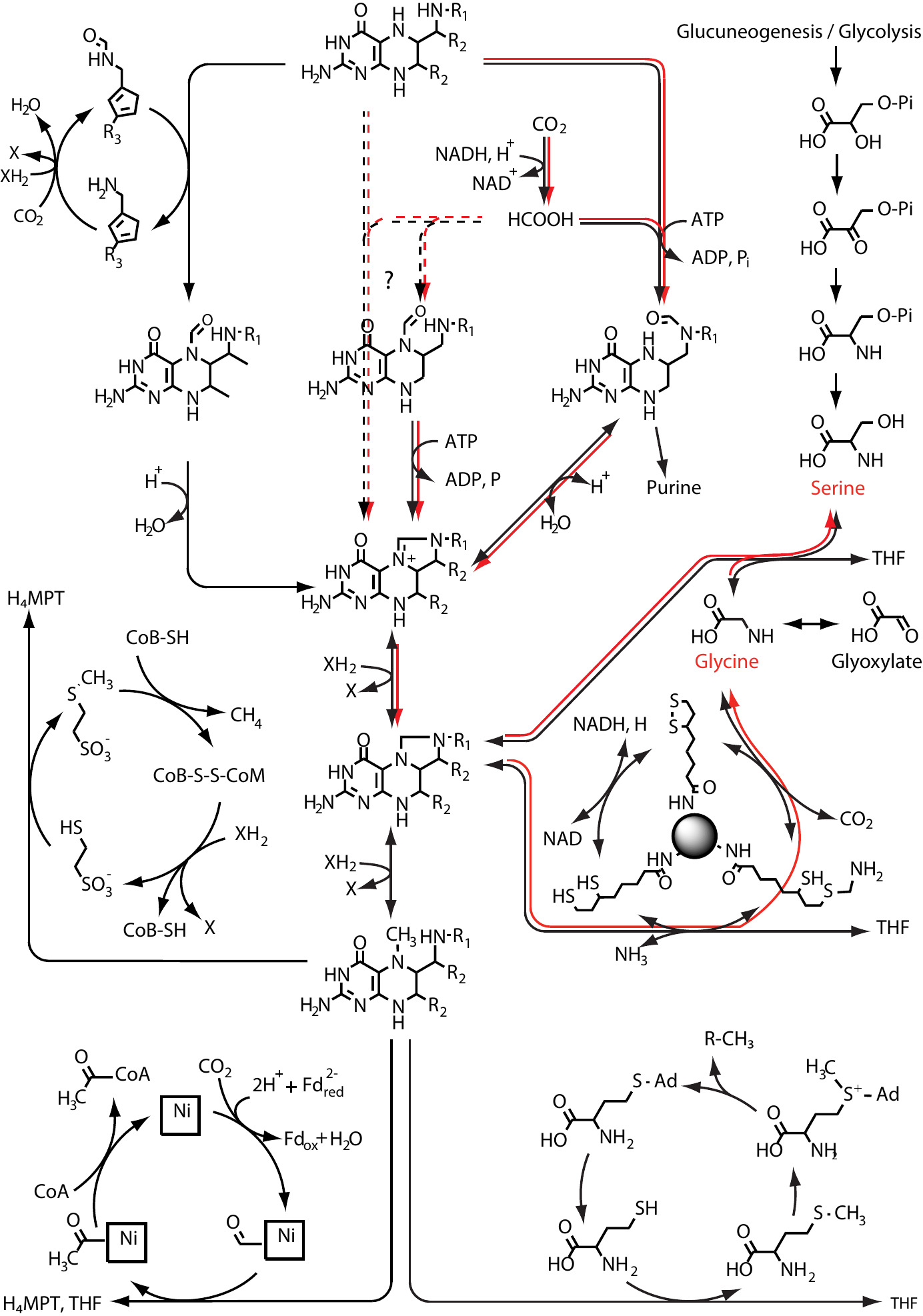}
  \caption{
  The reactions in the Wood-Ljungdahl pathway of direct
  ${\mbox{C}}_1$-reduction.  The main sequence on pterins is shown,
  with five outputs for formyl, methylene, or methyl groups. The
  semi-independent submodule often used to directly synthesize glycine
  and serine from ${\mbox{CO}}_2$, even when acetyl-CoA synthesis is
  absent, is highlighted in red. Alternative pathways to glycine and
  serine, from 3-phosphoglycerate in gluconeogenesis/glycolysis and
  glyoxylate, are shown in the upper right quadrant. Finally, the
  dashed arrows represent a suggested alternative form of formate
  uptake based on binding at ${\mbox{N}}^5$ rather than
  ${\mbox{N}}^{10}$ of folate before cyclization to
  methenyl-THF~\cite{Braakman:carbon_fixation:12}.
    \label{fig:C1_metabolism} 
  }
  \end{center}
\end{figure}


Phylogenetically, WL is a widely distributed pathway, found in a
variety of both bacteria and archaea, including acetogens,
methanogens, sulfate reducers, and possibly anaerobic ammonium
oxidizers~\cite{Hugler:fix_paths:11}.  The full pathway is found only
in strict anaerobes, because the CODH/ACS is one of the most
oxygen-sensitive enzymes
known~\cite{Ragsdale:CODH:83,Zarzycki:3HP_finis:09}.  However, as we
have argued in Ref.~\cite{Braakman:carbon_fixation:12}, the
folate-mediated reactions form a partly-independent sub-module. This
module combines with the equally-distinctive CODH/ACS enzyme to
form the complete WL pathway, but can serve independently as partial
carbon-fixation pathways even in the absence of the final step to
acetyl-CoA (see Fig.~\ref{fig:C1_metabolism}). In this role it is
found almost universally among deep bacterial clades. In addition to
its being highly oxygen sensitive, recent results (S.~Ragsdale,
pers.~comm.) suggest that the CODH/ACS is also sensitive to sulfides
and perhaps other oxidants, a point to which we will return in
Sec.~\ref{sec:phylomet_tree} when discussing evolutionary divergences
between the complete and incomplete forms of WL.

All carbon fixation pathways in extant organisms employ some essential
and apparently unique enzymes and most also rely in essential ways on
certain cofactors. For example, the 3-hydroxypropionate pathway relies
on biotin for reactions shared with (or homologous to) those in fatty
acid synthesis.  The reductive citric-acid cycle relies on reduced
ferredoxin~\cite{Kim:ferredoxins:12}, a simple iron-sulfur enzyme, and on thiamine in its reductive
carbonyl-insertion reaction~\cite{Chabriere:oxidoreductases:99}, and
also on biotin for its $\beta$-carboxylation
steps~\cite{Lombard:biotin:11,Aoshima:AKG_carbox:04} (all of these
examples will also be discussed along with the pathways in which they
are used below).  Among the uses of cofactors in carbon-fixation
pathways, however, the function provided by pterin cofactors in WL is
distinct and arguably the most complex. (\emph{Pterin} is a name
referring to the class of cofactors including folates and the
methanopterins, which are both derived from a neopterin precursor.)
Whereas most cofactors act as transfer agents cycling between two or
three states, pterins undergo elaborate multi-step cycles, mediating
capture of formate, reduction of carbon bound to one or two nitrogen
atoms, and transfer of formyl, methylene, or methyl groups. This
diversity of roles has led to the folate pathway's being termed the
``central superhighway'' of ${\mbox{C}}_1$
chemistry~\cite{Maden:folates_pterins:00}. The distinctive use of
cofactors within WL continues with the dependence of the acetyl-CoA
synthesis on cobalamin, a highly reduced tetrapyrrole capable of
two-electron transfer~\cite{Banerjee03}.  In this sense the simple
network topology of direct ${\mbox{C}}_1$ reduction seems to require a
more elaborate dependence on cofactors than is seen in other pathways.

{\bf Reductive citric-acid cycle:} The reductive citric-acid
(reductive Tricarboxylic Acid, or rTCA)
cycle~\cite{Evans:rTCA_disc:66,Buchanan:rTCA:90} is the reverse of the
oxidative Krebs cycle. It is a sequence of eleven intermediates and
eleven reactions, highlighted in Fig.~\ref{fig:carbon_anabolism_CHO},
which reduce two molecules of ${\mbox{CO}}_2$, and combine these
through a substrate-level phosphorylation with CoA, to form one
molecule of acetyl-CoA.  In the cycle, one molecule of oxaloacetate
grows by condensation with two ${\mbox{CO}}_2$ and is reduced and
activated with CoA.  The result, citryl-CoA, undergoes a retro-aldol
cleavage to regenerate oxaloacetate and acetyl-CoA. Here we separate
the formation of citryl-CoA from its subsequent retro-aldol cleavage,
as this is argued to be the original reaction sequence, and the one
displaying the closest homology in the substrate-level phosphorylation
with that of
succinyl-CoA~\cite{Aoshima:citrate_synthase:04,Aoshima:citrate_lyase:04}.
A second arc of reactions sometimes termed
\emph{anaplerotic}~\cite{Lengeler:BP:99} then condenses two further
${\mbox{CO}}_2$ with acetyl-CoA to produce a second molecule of
oxaloacetate, completing the network-autocatalytic topology and making
the cycle self-amplifying.  The distinctive reaction in the rTCA
pathway is a carbonyl insertion at a thioester (acetyl-CoA or
succinyl-CoA), performed by a family of conserved ferredoxin-dependent
oxidoreductases which are triple-${\mbox{Fe}}_4{\mbox{S}}_4$-cluster
proteins~\cite{Chabriere:oxidoreductases:99}. The cycle is found in
many anaerobic and microaearophylic bacterial lineages, including
aquificales, chlorobi, and $\delta$- and $\epsilon$-proteobacteria.

Enzymes from reductive TCA reactions are very widely distributed among
bacteria, where in addition to complete cycles they support
fermentative pathways that break cycling and use intermediates such as
succinate as terminal electron
acceptors~\cite{Watanabe:tuberculosis:11}.  The distribution of full
rTCA cycling correlates with clades whose origins are placed in a
pre-oxygenic earth, while fermentative TCA arcs and the oxidative
Krebs cycle are found in clades that by phylogenetic position (such as
$\gamma$-, $\beta$- and $\alpha$-proteobacteria) and biochemical
properties (membrane quinones and intracellular redox couples
involving them~\cite{Schoepp:menaquinones:09}) arose during or after
the rise of oxygen.  The co-presence of enzymes descended from both
the reductive and oxidative
cycles~\cite{Tian:MTB_TCA:05,Baughn:Mtb_TCA:09,Zhang:cyano_TCA:11} in
members of clades that at a coarser level are known to straddle the
rise of oxygen, such as actinobacteria and
proteobacteria~\cite{Schoepp:menaquinones:09}, may provide detailed
mechanistic evidence about the reversal of core metabolism from
ancestral reductive modes to later derived oxidative modes.  

Functionally, the transition to an oxidizing earth is complex in two ways.  First, oxygenation is unlikely to have occurred homogeneously within the oceans~\cite{Fenchel:OEL:02,Kasting:anc_oxygen:06,Shields:oxygen:11}. For bacteria not harbored in sheltered environments, this likely created complex needs to responsive phenotypes.  Second, the co-presence of strong oxidants with geochemical reductants could in some cases provide an energy source during oxygenation, but as we see on the current Earth as well as in the history of banded iron, oxidants eventually scavenge reducing equivalents from most environments, likely leaving fermentative pathways as a fallback energy source for many organisms. It is plausible that both of these geochemical responses served as pre-adaptations enabling the complex host-associated lifecycles of bacteria and archaea descended within clades that straddled this transition.

Earlier work~\cite{Melendez96} that also recognized the central
position of TCA reactions has attempted to argue for an ancestral
\emph{oxidative} TCA cycle.  However, the optimal-path part of this
argument, and similar path-length optimality arguments in
Ref's.~\cite{Melendez94,Noor:min_walk:10,Riehl:coli_paths:10}, apply
broadly to intermediary metabolism as a consequence of its limited
reaction types or cross-linking as explained in
Sec.~\ref{sec:ana_cata}, without implying directionality. Furthermore,
the functional criterion of acetate oxidation to produce reducing
equivalents to drive ATP production through oxidative phosphorylation
relies on the assumption that the organic interconversion of acetate
to ${\mbox{CO}}_2$ was the main redox couple of early
metabolism. Prior to the rise of oxygen, however, we find many
inorganic electron acceptors such as elemental sulfur in redox couples
with inorganic electron donors such as ${\mbox{H}}_2$ and not with
organics of stoichiometry
${\mbox{CH}}_2\mbox{O}$~\cite{Guiral:aquif_complx:05,%
Guiral:Aquifex_H_metab:05,Guiral:aquif_complx:09}. For these reasons,
combined with the strict absence of oxidative TCA as a plausible
genetically reconstructed form in deep-branching clades, we find
arguments for an ancestral rTCA
cycle~\cite{Wachtershauser:rTCA:90,Smith:universality:04,Morowitz:EFOoL:07},
replaced possibly via fermentative intermediates, by a later oxidative
Krebs cycle, more convincing.

\begin{figure}[ht]
  \begin{center} 
  \includegraphics[scale=0.65]{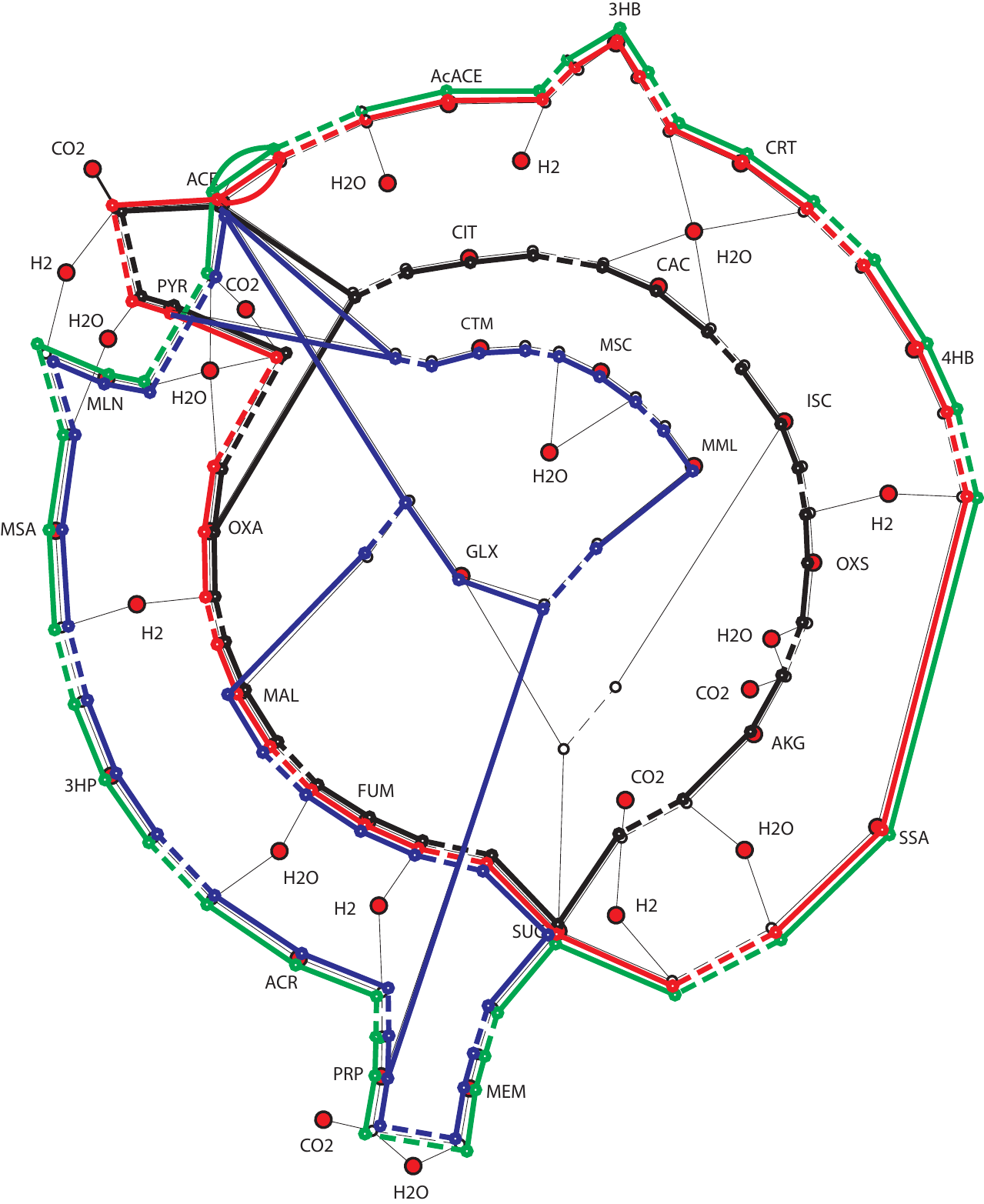}
  \caption{
  The four loop carbon-fixation pathways that pass through some or all
  of the universal biosynthetic precursors, from the graph of
  Fig.~\ref{fig:carbon_anabolism_CHO}.  rTCA is black, DC/4HB is red,
  3HP-bicycle is blue, and 3HP/4HB is green.  The one aldol reaction
  from the glyoxylate shunt that is not part of the 3HP-bicycle is
  shown in fine lines.  The module-boundary nature of acetate (ACE)
  and succinate (SUC) is shown by the intersection of multiple paths
  in these compounds.  Radially aligned reactions are homologous in
  local-group chemistry; deviations from strict homology in different
  pathways appear as excursions from concentric circles. 
    \label{fig:pathways_compare_ring_4col} 
  }
  \end{center}
\end{figure}

{\bf Dicarboxylate/4-hydroxybutyrate cycle:} The
dicarboxyate/4-hydroxybutyrate (DC/4HB)
cycle~\cite{Huber:4HB_discovery:08,Berg:carbon_fixn:10}, illustrated
in Fig.~\ref {fig:pathways_compare_ring_4col} is, like rTCA, a
single-loop network-autocatalytic cycle, but has a simpler form of
autocatalysis in which acetyl-CoA rather than oxaloacetate is the
network catalyst.  Only two ${\mbox{CO}}_2$ molecules are attached in
the course of the cycle to form acetoacetyl-CoA, which is then
thioesterified at the second acetyl moiety and cleaved to directly
regenerate two molecules of acetyl-CoA.  An extra copy of the network
catalyst is thus directly regenerated (with suitable CoA activation)
without the need for anaplerotic reactions.  The cycle has so far been
found only in anaerobic crenarchaeota, but within this group it is
believed to be widely distributed
phylogenetically~\cite{Huber:4HB_discovery:08,Berg:carbon_fixn:10}.
The first five reactions in the cycle (from acetyl-CoA to
succinyl-CoA) are identical to those of rTCA.  The second arc of the
cycle begins with reactions found also in 4-hydroxybutyrate and
$\gamma$-aminobutyrate fermenters in the Clostridia (a subgroup of
Firmicutes within the bacteria), and terminates in the reverse of
reactions in the isoprene biosynthesis pathway.  The DC/4HB pathway
thus uses the same ferredoxin-dependent carbonyl-insertion reaction
used in rTCA (though only at acetyl-CoA), along with distinctive
reactions associated with 4-hydroxybutyrate fermentation. In
particular, the dehydration/isomerization sequence from
4-hydroxybutyryl-CoA to crotonyl-CoA is performed by a
flavin-dependent protein containing an [Fe$_4$-S$_4$] cluster, and
involves a ketyl-radical intermediate~\cite{Muh96,Martins04}.

\begin{figure}[ht]
  \begin{center} 
  \includegraphics[scale=0.6]{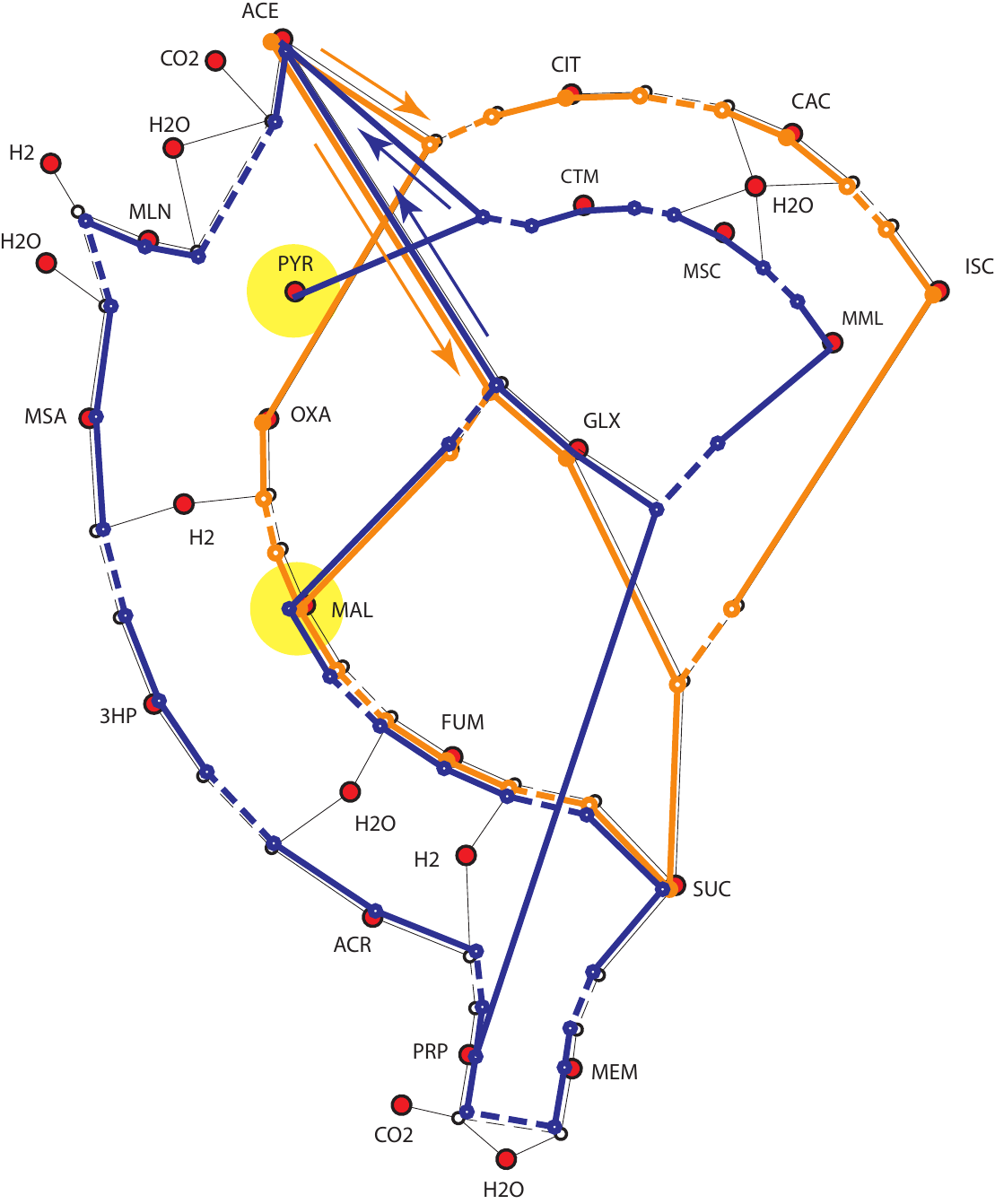}
  \caption{
  The 3-hydroxypropionate bicycle (blue) and the glyoxylate shunt
  (orange) compared.  Directions of flow are indicated by arrows on
  the links to acetate (ACE).  The common core that enables flux
  recycling in both pathways is the aldol reaction between glyoxylate
  (GLX), acetate, and malate (MAL).  The four other aldol reactions
  (labeled by their cleavage direction) are from isocitrate (ISC),
  methyl-malate (MML), citrate (CIT), and citramalate (CTM).  Malate
  is a recycled network catalyst in both pathways.  Carbon is fixed in
  the 3HP-bicycle as pyruvate (PYR), so the cycle only becomes
  autocatalytic if pyruvate can be converted to malate through
  anaplerotic (rTCA) reactions.  
    \label{fig:bicycle_shunt_compare_pared} 
  }
  \end{center}
\end{figure}

{\bf 3-hydroxypropionate bicycle:} The 3-hydroxypropionate (3HP)
bicycle~\cite{Zarzycki:3HP_finis:09}, highlighted in Fig.~\ref
{fig:pathways_compare_ring_4col} and in
Fig.~\ref{fig:bicycle_shunt_compare_pared} (below), has the most
complex network topology of the fixation pathways, using two linked
cycles to regenerate its network catalysts and to fix carbon. The
network catalysts in both loops are acetyl-CoA and the outlet for
fixed carbon is pyruvate. The reactions in the cycle begin with the
biotin-dependent carboxylation of Acetyl-CoA to form Malonyl-CoA, from
the fatty-acid synthesis pathway, followed by a distinctive
thioesterification~\cite{Alber08} and a second, homologous
carboxylation of propionyl-CoA (to methylmalyl-CoA) followed by
isomerization to form succinyl-CoA.  The first cycle then proceeds as
the \emph{oxidative} TCA arc, followed by retro-aldol reactions also
found in the glyoxylate shunt (described below).  A second cycle is
initiated by an aldol condensation of propionyl-CoA with glyoxylate
from the first cycle to yield $\beta$-methylmalyl-CoA, which follows a
sequence of reduction and isomerization through an enoyl intermediate
(mesaconate) similar to the second cycle of rTCA or the 4HB pathway.
This complex pathway was discovered in the Chloroflexi and is believed
to represent an adaptation to alkaline environments in which the
${\mbox{CO}}_2$/${\mbox{HCO}}_3^{-}$ (bicarbonate) equilibrium
strongly favors bicarbonate. All carbon fixations proceed through
activated biotin, thus avoiding the carbonyl insertion of the rTCA and
DC/4HB pathways.  While topologically complex, the bicycle makes
extensive use of relatively simple aldol chemistry, which we will
argue in Sec.~\ref{sec:aldol} made its evolutionary innovation
less improbable than the topology alone might suggest.

{\bf 3-hydroxypropionate/4-hydroxybutyrate cycle:} The
3-hydroxypropionate/4-hydroxybutyrate (3HP/4HB)
cycle~\cite{Teufel:3HP_4HB:09}, shown in
Fig.~\ref{fig:pathways_compare_ring_4col}, is a single-loop pathway in
which the first arc is the 3HP pathway, and the second arc is the 4HB
pathway.  Like DC/4HB, 3HP/4HB uses acetyl-CoA as network catalyst and
fixes two ${\mbox{CO}}_2$ to form acetoacetyl-CoA.  The pathway is
found in the Sulfolobales (crenarchaeota), where it combines the
crenarchaeal 4HB pattern of autotrophic carbon fixation with the
bicarbonate adaptation of the 3HP pathway. Like the 3HP bicycle, the
3HP/4HB pathway is thought to be an adaptation to alkalinity, but
because the 4HB arc does not fix additional carbon, this adaptation
resulted in a simpler pathway structure than the bicycle.

{\bf Calvin-Benson-Bassham cycle:} The Calvin-Benson-Bassham (CBB)
cycle~\cite{Bassham:Calvin_cycle:54,Tabita:Calvin_cycle:04} is
responsible for most of known carbon fixation in the biosphere.  In
the same way as WL adds only the distinctive CODH/ACS reaction to an
otherwise very-widely-distributed folate
pathway~\cite{Braakman:carbon_fixation:12}, the CBB cycle adds a
single reaction to the otherwise-universal network of aldol reactions
among sugar-phosphates that make up the gluconeogenic pathway to
fructose 1,6-bisphosphate and the reductive pentose phosphate pathway
to ribose and ribulose 1,5-bisphosphate.\footnote{The universality of
  this network requires some qualification.  We show a canonical
  version of the network in Fig,~\ref{fig:carbon_anabolism_CHO}, and some variant on
  this network is present in every organism that synthesizes ribose.
  However, the ${\left( {\mbox{CH}}_2 \mbox{O} \right)}^n$
  stoichiometry of sugars, together with the wide diversity of
  possible aldol reactions among sugar-phosphates, make sugar
  re-arrangement a problem in the number theory of the small integers,
  with solutions that may depend sensitively on allowed inputs and
  outputs.  Other pathways within the collection of attested
  pentose-phosphate networks are shown in
  Ref.~\cite{Andersen:NP_autocat:12}.}  The distinctive CBB reaction
that extends reductive pentose-phosphate synthesis to a carbon
fixation cycle is a carboxylation performed by the Ribulose
1,5-bisphosphate Carboxylase/Oxygenase (RubisCO), together with
cleavage of the original ribulose moiety to produce two molecules of
3-phosphoglycerate.  The Calvin cycle resembles the 4HB pathways in
regenerating two copies of the network catalyst directly, not
requiring separate anaplerotic reactions for autocalysis.  In addition
to carboxylation, RubisCO can react with oxygen in a process known as
\emph{photorespiration}~\cite{Eisenhut:photorespiration:06,%
  Eisenhut:photorespiration:08,Foyer:photorespiration:09} to produce
2-phosphoglycolate (2PG), a precursor to glyoxylate that is
independent of rTCA-cycle reactions. The CBB cycle is widely
distributed among cyanobacteria,in chloroplasts in plants, and in some
secondary endosymbionts.

{\bf The glyoxylate shunt:} Although it is not an autotrophic
carbon-fixation pathway, the glyoxylate shunt (or glyoxylate bypass)
is of interest because it shares intermediates and reactions with many
of the above fixation pathways, and because it resembles a fixation
pathway in certain topological features.  The pathway is shown
in Fig.~\ref{fig:bicycle_shunt_compare_pared}.  All aldol reactions
that can be performed starting from rTCA intermediates appear in this
pathway, either as cleavages or as condensations.  In addition to
condensation of acetate and oxaloacetate to form citrate, these
include cleavage of isocitrate to form glyoxylate and succinate, and
condensation of glyoxylate and acetate to form malate.  The shunt is a
weakly oxidative pathway (generating one ${\mbox{H}}_2$-equivalent
from oxidizing succinate to fumarate), and is otherwise a network of
internal redox reactions.  It is therefore a very widely-used
facultative pathway under conditions where carbon for biosynthesis,
more than reductant, is limiting.

Two of the arcs of the shunt overlap with arcs in the oxidative Krebs
cycle, but the entire pathway is a bicycle much like the 3HP-bicycle,
sharing many of the same intermediates, but running in the opposite
direction.  Oxidative pathways such as the Krebs cycle are ordinarily
catabolic, and hence not self-maintaining.  The glyoxylate shunt may
be regarded as a network-autocatalytic pathway for intake of acetate,
using malate as the network catalyst and regenerating a second
molecule of malate from two acetate molecules. This may be part of the
reason that the shunt is up-regulated in the Deinococcus-Thermus
family of bacteria in response to radiation exposure~\cite{Liu03},
providing additional robustness from network topology under conditions
when metabolic control is compromised.

\subsubsection{Thermodynamic constraints on pathway structure}

The central energetic costs of carbon-fixation pathways are associated
with carboxylation reactions in which ${\mbox{CO}}_2$ molecules are
added to the growing substrate, and the subsequent reactions in which
the carboxyl group is reduced to a carbonyl~\cite{BarEven12}. In
isolation these reactions require ATP hydrolysis, but these costs can
be avoided in several ways. In some cases a thioester intermediate is
used to effectively couple together a carboxyl reduction and a
subsequent carboxylation, allowing the two reactions to be driven by a
single ATP hydrolysis. An unfavorable (endergonic) reaction can also
be coupled to a highly favorable (exergonic) reaction, allowing the
reactions to proceed without ATP hydrolysis.

Individual pathways employ such couplings to varying degrees,
resulting in a range of ATP costs associated with carbon fixation. At
the low end, WL eliminates nearly all use of ATP through its unique
pathway chemistry, requiring only a single ATP in the synthesis of
pyruvate from ${\mbox{CO}}_2$. This ATP is associated with the
attachment and activation of formate on folates. Reducing
${\mbox{CO}}_2$ to free formate prior to attachment, and further
reducing the activated formate on folates prior to incorporation into
growing substrates saves one ATP associated with carboxylation.
Additional ATP costs are saved by coupling the endergonic reduction of
${\mbox{CO}}_2$ to CO to the subsequent exergonic synthesis of
acetyl-CoA. Finally, the activated thioester bond of acetyl-CoA allows
the subsequent carboxylation to pyruvate to also proceed without
additional ATP. In methanogens even the ATP cost of attaching formate
to folates has been eliminated by modifying the structure of THF to
that of ${\mbox{H}}_4 \mbox{MPT}$~\cite{Braakman:carbon_fixation:12},
enabling a membrane-bound iron-sulfur system to serve as energy
source.  Similarly, rTCA has high energetic efficiency as a result of
extensive reaction coupling, requiring only 2 ATP to synthesize
pyruvate from
${\mbox{CO}}_2$~\cite{Berg:carbon_fixn:10,BarEven12}. Two ATP are
saved by coupling carboxyl reductions to subsequent carboxylations
using thioester intermediates, and an additional ATP is saved by
coupling the carboxylation of $\alpha$-ketoglutarate to the subsequent
carbonyl reduction leading to isocitrate.

At the high end of energetic cost of carbon-fixation are pathways that
couple unfavorable reactions less effectively, or not at all, or even
hydrolyze ATP for reactions other than carboxylation or carboxyl
reduction. Both the DC/4HB pathway and the 3HP bicycle decouple one or
more of the thioester-mediated carboxyl reduction + carboxylation
sequences of the kind used in rTCA, and neither couples endergonic
carboxylations to exergonic reductions.  As a result DC/4HB requires 5
ATP and the 3HP bicycle 7 ATP to synthesize pyruvate from
${\mbox{CO}}_2$~\cite{Berg:carbon_fixn:10,BarEven12}. The 3HP/4HB
pathway has the highest cost of any fixation pathway, with 9 ATP
required to synthesize pyruvate from ${\mbox{CO}}_2$. This is partly
because it also decouples thioester-mediated carboxyl reduction +
carboxylation sequences, and partly because pyruvate is synthesized by
diverting and ultimately \emph{decarboxylating}
succinyl-CoA~\cite{Fuchs11,Berg:carbon_fixn:10,BarEven12}. Finally,
CBB is also at the high end in terms of cost, requiring 7 ATP to
synthesize pyruvate from ${\mbox{CO}}_2$. Although this pathway avoids
the cost of carboxylation reactions by coupling them to exergonic
cleavage reactions, CBB is the only fixation pathway that invests ATP
hydrolysis in chemistry other than carboxylations or carboxyl
reductions, thereby increasing its relative cost~\cite{BarEven12}.

\subsubsection{Centrality and universality of the reactions in the
  citric-acid cycle, and the pillars of anabolism}
\label{sec:graph_pillars}

The apparent diversity of six known fixation pathways is unified by
the role of the citric-acid cycle reactions, and secondarily by that
of gluconeogenesis and the pentose-phosphate pathways.
Fig.~\ref{fig:carbon_anabolism_CHO} showed the C,H,O
stoichiometry for a network of reactions that includes all six known
pathways. Here, by \emph{stoichiometry} we refer to the mole-ratios of
reactants and products for each reaction, with molecules represented
by their CHO constituents, and attached phosphate or thioester groups
omitted.  Where phosphorylation or thioesterification mediates a net
dehydration, we have represented the dehydration directly in the
figure. The network contains only 35 organic intermediates, because
many intermediates and reactions appear in multiple pathways.
Hydroxymethyl-glutarate and butyrate are also shown, to indicate
points of departure to isoprene and fatty acid synthesis,
respectively.

In Fig.~\ref{fig:carbon_anabolism_CHO} the TCA cycle and the
gluconeogenic pathway are highlighted.  Beyond being mere points of
departure for alternative fixation pathways and for diversifications
in intermediary metabolism, they are invariants under diversification
because they determine carbon flow among the universal precursors of
biosynthesis.

\begin{figure}[ht]
  \begin{center} 
  \includegraphics[scale=0.375]{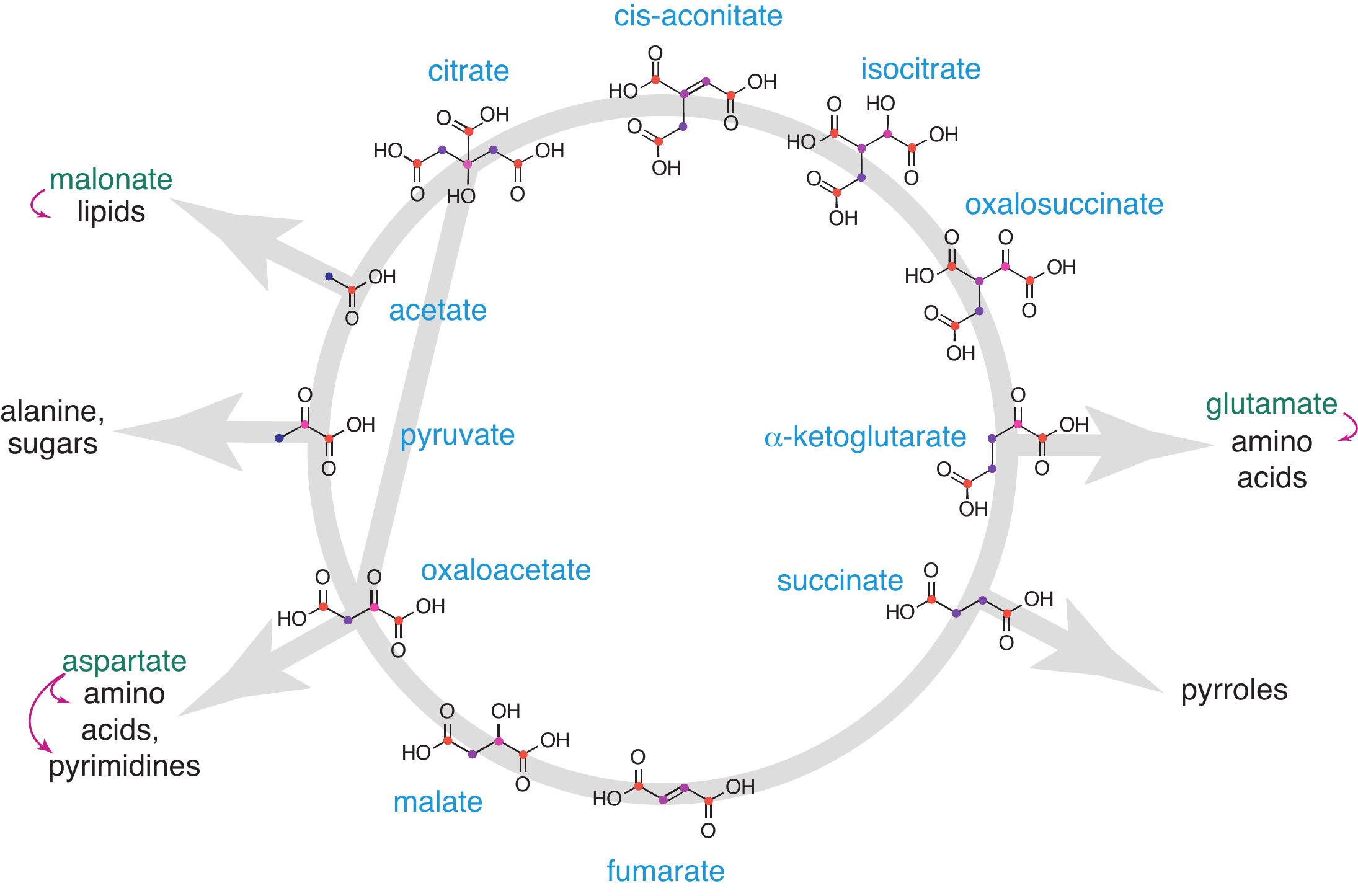}
  \caption{
  The pillars of anabolism, showing lipids, sugars, amino acids,
  pyrimidines and purines, and tetrapyrroles from either succinate or
  AKG.  Molecules with homologous local chemistry are at opposite
  positions on the circle.  Oxidation states of internal carbon atoms
  are indicated by color (red = oxidized, blue = reduced). 
    \label{fig:pillars_anabolism} 
  }
  \end{center}
\end{figure}

Almost all anabolic pathways in extant organisms originate in one of
five intermediates in the TCA cycle -- acetate (as acetyl-CoA),
pyruvate, oxaloacetate, succinate (or succinyl-CoA) or
$\alpha$-ketoglutarate -- which have been dubbed the ``pillars of
anabolism''~\cite{Srinivasan:aquifex_analysis:09}. Succinyl-CoA can
serve as the precursor to pyrroles (metal-coordinating groups in many
cofactors) -- mainly in $\alpha$-proteobacteria and mitochondria --
but these are more commonly made from $\alpha$-ketoglutarate
\emph{via} glutamate in what is known as the C5
pathway~\cite{Wettstein95}.  A phylogenetic analysis of these pathways
confirms that the C5 pathway is the most plausible ancestral route to
pyrrole synthesis~\cite{Braakman:aquifex_model:12}.  (An even
stronger claim has been made, from a similar combination of
biochemical and phylogenetic inputs, that it is a likely relic of the
RNA world~\cite{Benner:palimpsest:89}.)  Thus as few as four TCA
intermediates provide the organic inputs to all anabolic pathways.
Fig.~\ref{fig:pillars_anabolism} shows the major molecule classes
associated with each intermediate.  The only exceptions to this
universality, which form a biosynthetic sequence, are glycine, serine,
and a few compounds synthesized from them; this sequence can be
initiated directly from ${\mbox{CO}}_2$ outside of the pillars (see
Fig.~\ref{fig:C1_metabolism}). This observation becomes key in
reconstructing the evolutionary history of carbon-fixation (see
Sec.~\ref{sec:coarse_graining}).  The gluconeogenic pathway then forms
a similarly unique bridge between the TCA intermediate pyruvate (in
the activated form phosphoenolpyruvate) and the network of
sugar-phosphate reactions known as the \emph{pentose-phosphate
pathway}.

Carbon-fixation pathways must reach all four (or five) of the
universal anabolic starting compounds.  They may do this either by
producing them as pathway intermediates, or by means of secondary
reactions converting pathway intermediates into the essential
precursors. The degree to which a pathway passes through all essential
biosynthetic precursors may suggest its antiquity.  In
metabolism-first theories of the origin of
life~\cite{Morowitz:BCL:92}, the limited set of compounds selected and
made available in high concentration by proto-metabolism determined
the opportunities for further biosynthesis, thus establishing
themselves as the precursors of anabolism.

Among the five network-autocatalytic fixation pathways, the CBB
pathway is unique in not passing through \emph{any} universal anabolic
precursors. When used as a fixation pathway, CBB reactions must thus
be connected to the rest of anabolism through several reactions in the
glycolytic pathways connecting 3-phosphoglycerate (3PG) to pyruvate.
Pyruvate is then connected to the remaining precursors through partial
TCA sequences.  The glycolytic pathway is the primary connection of
CBB to the anabolic precursors, but 2-phosphoglycolate (2PG) produced
during photorespiration may also be converted to glyoxylate and
subsequently to glycine and serine (see Fig.~\ref{fig:C1_metabolism}).
Glycine synthesis from photorespiration is not itself a carbon-fixing
process, but rather a salvage pathway to compensate for poor
discrimination of the enzyme RuBisCO.  2PG is produced from
ribulose-1,5-bisphosphate when ${\mbox{O}}_2$ replaces ${\mbox{CO}}_2$
in the RuBisCo uptake reaction.  This toxin inhibits RuBisCO, and
would require excretion if it could not be recycled, leading to a net
loss of carbon from the pentose-phosphate network.  However, most
RuBisCO uptake events do fix ${\mbox{CO}}_2$, and the carbon
circulating in the pentose-phosphate pathway in CBB organisms is the
product of these successful fixation cycles. Thus, photorespiration
with glycine salvage amounts to a variant elaboration of the fixation
pathway to include null cycles, and the connection of this more
complex process to the precursor set.

Among the remaining loop-fixation pathways, only rTCA passes through
all five anabolic pillars. Through its partial overlap with rTCA,
DC/4HB passes through four, excluding $\alpha$-ketoglurate.  The
3HP-bicycle further bypasses oxaloacetate, while the 3HP/4HB loop and
WL include only acetyl-CoA.  All of the latter pathways require
anaplerotic reactions in the form of incomplete (either oxidative or
reductive) TCA arcs; when these combine (in various ways) with WL
carbon fixation, they are known collectively as the \emph{reductive
  acetyl-CoA pathways}.

The most parsimonious explanation for the universality of the TCA arcs
as anaplerotic reactions is lock-in by downstream anabolic pathways,
to which metabolism was committed by the time carbon-fixation
strategies diverged.  This can also be understood as a direct
extension of the metabolism-first assumption that anabolic pathways
themselves formed around proto-metabolic selection of the rTCA
intermediates.\footnote{Harold Morowitz summarizes this assumption
  with the phrase \emph{metabolism recapitulates
    biogenesis}~\cite{Morowitz:BCL:92}.}  (A similar but later form of
commitment has been argued to convert basal gene regulatory networks
in metazoan development into \emph{kernels}, which admit no variation
and act as constraints on subsequent evolutionary
dynamics~\cite{Davidson:kernels:06,Erwin:kernels:09}.)  If lock-in
provides the correct interpretation of TCA universality, then much of
the burden of accounting for the inventory of small metabolites is
shifted away from Darwinian selection for function in a post-RNA
world, and onto constraints of biosynthetic simplicity and network
context.  We show below that phylogenetic reconstruction is consistent
with a selective role for rTCA cycling in the root metabolism of
cellular life, though only as part of a larger network than the modern
rTCA cycle.

\subsection{Modularity in the internal structure and mutual
  relationships of the known fixation pathways}
\label{sec:modularity}

Fig.~\ref{fig:carbon_anabolism_CHO} shows that the number of molecules
and reactions required to include all carbon fixation pathways is much
smaller than might have been expected from their nominal diversity,
because many reactions are used in multiple pathways, and all pathways
remain close to the universal precursors.  We have already noted in
the previous section that this re-use goes beyond the requirements of
autocatalysis, to the anaplerotic role of rTCA arcs adapting variant
fixation pathways to an invariant set of biosynthetic precursors.

The aggregate network also shows many kinds of structure: clusters,
concentric rings, and ladders reflecting parallel sequences of the
same inputs and outputs in different pathways.  We will show in this
section that these result from re-use of local-group chemistry in
transformations of distinct molecules.

At the end of the section we will describe a third form of re-use not
represented in the aggregate graph.  The folate-mediated direct
${\mbox{C}}_1$ reduction sequence of Wood-Ljungdahl, responsible for production of the methyl group used in the synthesis of acetyl-CoA in Fig.~\ref{fig:carbon_anabolism_CHO}, is also found as a free-standing fixation pathway across the bacterial tree, often as one component in
a disconnected autotrophic network using one of the loop fixation
pathways as its other component.

Because of such extensive redundancy, little innovation is required to
explain the extant diversity of carbon fixation.  All known carbon
fixation strategies can be described as assemblies of a small number
of strongly-defined \emph{modules}, which govern not only the function
of pathways, but also their evolution.

\subsubsection{Modularity in carbon fixation loops from re-use of
  pathway segments}

Fig.~\ref{fig:pathways_compare_ring_4col} shows the sub-network from
Fig.~\ref{fig:carbon_anabolism_CHO} containing the four
loop-autotrophic carbon fixation pathways that pass through some or
all universal precursors, together with reactions in the glyoxylate
shunt.  The four loop pathways are shown in four colors, with the
organic pathway-intermediates (but not environmental precursors or
reductants) highlighted.

The figure shows that these pathways re-use intermediates by combining
entire \emph{pathway segments}.  The combinatorial assembly of these
segments is possible because they all pass through acetate (as
acetyl-CoA), succinate (usually as succinyl-CoA), and all except the
second loop of the 3HP bicycle pass through both.  Thus the conserved
reactions among the autocatalytic loop carbon-fixation pathways are
shared within strictly preserved sequences, which have key molecules
as the boundaries at which segments may be combined.

\subsubsection{Homologous local-group chemistry across pathway
  segments} 

\begin{widetext}

\begin{figure}[ht]
  \begin{center} 
  \includegraphics[scale=0.65]{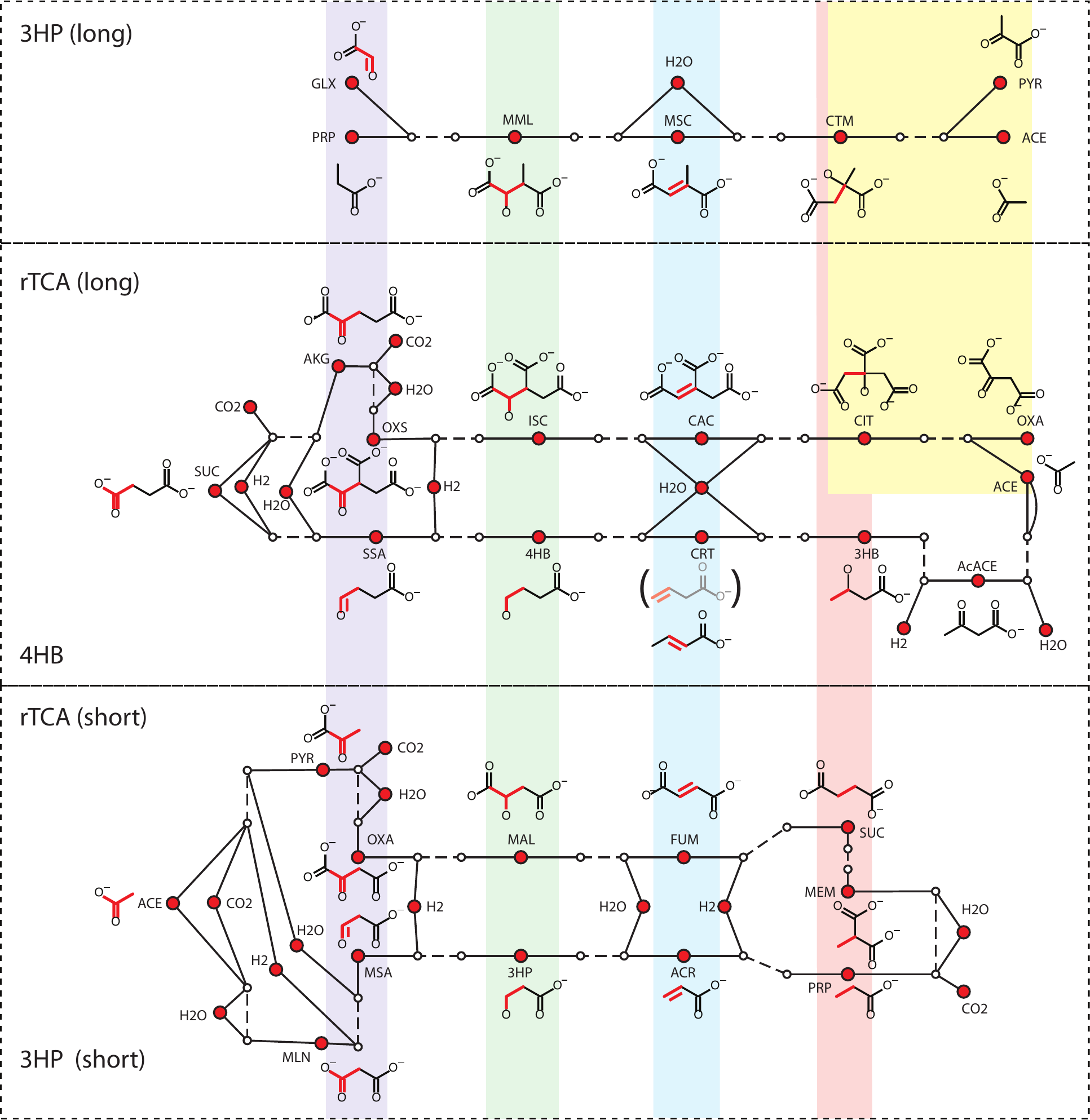}
  \caption{
  Comparison of redundant reactions in the loop carbon fixation
  pathways.  Pathways are divided into ``long-molecule'' (upper-ranks)
  and ``short-molecule'' (lower-ranks) segments; long-molecule
  segments occupy roughly the upper-right half-plane in
  Fig.~\ref{fig:pathways_compare_ring_4col}, and abbreviations are as
  in Fig.~\ref{fig:carbon_anabolism_CHO}.  Molecule forms are shown
  next to the corresponding tags.  Bonds drawn in red are the active
  acetyl or semialdehyde moieties in the respective segments.
  Vertical colored bars align homologous carbon states.  The yellow
  block shows retro-aldol cleavages of citrate or citramalate.  Two
  molecules are shown beneath the tag CRT (crotonate): the greyed-out
  molecule in parentheses would be the homologue to the other
  aconitase-type reactions; actual crotonate (full saturation)
  displaces the double bond by one carbon, requiring the abstraction
  of the $\alpha$-proton in 4-hydroxybutyrate via the ketyl-radical
  mechanism that is distinctive of this pathway.
    \label{fig:pathways_compare_1_annot} 
  }
  \end{center}
\end{figure}

\end{widetext}

In addition to the re-use of complete reactions in pathway segments,
variant carbon-fixation pathways have extensively re-used
transformations at the level of local functional groups.  The network
of Fig.~\ref{fig:pathways_compare_ring_4col} is arranged in concentric
rings, in which the arcs of the rTCA cycle align with the 3HP or 4HB
pathways, or with the mesaconate arc of the 3HP bicycle.  The
``ladder'' structure of inputs and outputs of reductant
(${\mbox{H}}_2$) or water between these rings shows the similar
stoichiometric progression in these alternative pathways.
Fig.~\ref{fig:pathways_compare_1_annot} decomposes the aggregate
network into two pairs of short-molecule and long-molecule arcs, and
the mesaconate arc, and shows the pathway intermediates in each arc.
The figure makes clear that, both \emph{within} the arcs of the loop
pathways, and \emph{between} alternate pathways, the type, sequence,
and position of reactions is highly conserved.  In particular, the
reduction sequence from $\alpha$-ketones or semialdehydes, to
alcohols, to isomerization through enoyl intermediates, is applied to
the same bonds on the same carbon atoms from input acetyl moieties in
rTCA, 3HP, and 4HB pathways, and to analogous functional groups in the
bicycle.  Finally, in the cleavage of both citryl-CoA and
citramalyl-CoA, the bond that has been isomerized through the enoyl
intermediate is the one cleaved to regenerate the network catalyst.

Even the distinctive step to crotonyl-CoA in the 4HB pathway creates
an aconate-type intermediate, and the enzyme responsible has high
homology to the acrolyl-CoA
synthetase~\cite{Hetzel:AcrolylCoA:03,Herrmann:flavoproteins:08},
whose output (acrolyl-CoA) follows the standard pattern.  Only the
position of the double bond breaks the strict pattern in crotonyl-CoA,
and the abstraction of the un-activated proton required to produce
this bond requires the unique ketyl-radical
intermediate~\cite{Buckel96}.  From crotonyl-CoA, the sequence to
3-hydroxybutyrate is then followed by a surprising \emph{oxidation}
and re-hydration, resulting in a 5-step, redox-neutral, sequence.  The
net effect of this sequence is to shift the carbonyl group (of
succinate semialdehyde, SSA) by one carbon (in acetoacetate, AcACE).
Because the 4HB pathway takes in no new ${\mbox{CO}}_2$ molecules,
this isomerization enables regeneration of the network catalysts in
the same way the reduction/aldol-cleavage sequence enables
regeneration for rTCA or the 3HP bicycle.

Duplication of reaction sequences in diverse fixation pathways appears
to have resulted from retention of gene sets as organism clades
diverged.  Duplication of local-group chemistry in diverse reactions
similarly appears to have resulted (at least in most cases) from
retention of reaction mechanisms as enzyme families diverged.  All
enoyl intermediates are produced by a widely diversified family of
aconitases~\cite{Gruer:aconitases:97}, while biotin-dependent
carboxylations are performed by homologous enzymes acting on pyruvate
and $\alpha$-ketoglutarate~\cite{Aoshima:AKG_carbox:04}, and
substrate-level phosphorylation and thioesterification are similarly
performed by homologous enzymes on citrate and succinate in
rTCA~\cite{Aoshima:citrate_synthase:04,Aoshima:citrate_lyase:04}.
Similar to the synthesis of citryl-CoA we separate here the
carboxylation of $\alpha$-ketoglutarate from the subsequent reduction
of oxalosuccinate to isocitrate -- performed by a single enzyme in
most organisms -- because it is argued to be the ancestral
form~\cite{Aoshima:oxalosucc:06,Aoshima:isocit_dehydro:08}. However,
the thioesterification of propionate in the 3HP pathway is performed by
distinct enzymes in bacteria and archaea, an observation that has been
interpreted to suggest convergent
evolution~\cite{Teufel:3HP_4HB:09,Zarzycki:3HP_finis:09}. The
widespread use of a few reaction types by a small number of enzyme
classes/homologues may reflect their early establishment by
promiscuous
catalysts~\cite{Copley:promiscuity:03,Pereto:orig_metab:12}, followed
by evolution toward increasing specificity as intermediary metabolic
networks expanded and metabolites capable of participating in carbon
fixation diversified.

A functional identification of modules that seeks to minimize
influence from historical effects (such as lock-in) has been carried
out by Noor \emph{et al.}~\cite{Noor:min_walk:10}, and identifies
similar module boundaries.  Using as data the first three numbers of
the EC classification of enzymes -- which distinguish reaction types
but coarse-grain over both substrate specificity and enzyme homology
-- they show that many pathways in core metabolism are the shortest
routes possible between inputs and products. This work builds on
earlier studies showing that under very simple rules, the pentose
phosphate pathway uses the minimal number of steps to connect inputs
to outputs~\cite{Melendez85,Melendez94}.  In both of those studies,
the authors suggested that Darwinian selective pressure may have led
to a such minimal pathways as optimal network connections between
\emph{given} pairs of metabolites, with the implication that the
selection of metabolites was based on some aspects of phenotypic
function aside from their network positions.

If we do not presume that phenotypic selection preceded metabolism,
however, the problem of pathway optimization ceases to be one with
fixed endpoints, and causation may even run from pathways to the
metabolite inventory. In this view minimal pathways may have been
selected because their kinetics and regulation were easier to control.
Starting points of downstream intermediary metabolism could then have
been selected from the intermediates made available by fixation
pathways.  We argue in favor of a selection hierarchy of this form in
Sec.~\ref{sec:character_root}: Shorter fixation pathways capable of
attaining autocatalytic feedback offer fewer opportunities for
dilution by parasitic side-reactions~\cite{Huber:Ace_CoA:00}, and
(reaction chemistry otherwise being equal) require less regulatory
control.  They may thus be the only sustainable forms.

Where the pathways analyzed by Noor et al.~overlap with those
we have shown, many of their minimal sequences overlap with the
modules in Fig.~\ref{fig:pathways_compare_ring_4col}, as well as with
others in gluconeogenesis which we do not consider here. Thus,
from the perspective of an emerging metabolism, it may be that
historical retention of a small number of reaction types reflects
facility of the substrate-level chemistry, and that this has placed
time-independent constraints on evolution.

The functional-group homology shown in
Fig.~\ref{fig:pathways_compare_1_annot} allows us to separate
stereotypical sequences of widely diversified reactions from key
reactions that distinguish pathways.  The stereotypical sequences lie
downstream of reactions such as the ferredoxin-dependent carbonyl
insertion (rTCA), or biotin-dependent carboxylation (3HP), which are
associated with highly conserved enzymes or cofactors.  The downstream
reactions are also more ``elementary'', in the sense that they are
common and widely diversified in biochemistry, compared to the
pathway-distinguishing reactions. 

\subsubsection{Association of the initiating reactions with
  transition-metal sulfide mineral stoichiometries and other
  distinctive metal-ligand complexes}
\label{sec:mineral_association}

The observation that alternative fixation pathways are not
distinguished by their internal reaction sequences, but primarily by
their initiating reactions, suggests that these reactions were the
crucial bottlenecks in evolution, and perhaps reflect the limiting
diversity of chemical mechanisms for carbon bond formation. Mechanisms
of organosynthesis in aqueous solution are especially limited by the
instability of radical intermediates, which may be stabilized by
association with metal centers. The distinctive use of metals in the
(often highly conserved) enzymes and cofactors for these initiating
reactions may thus suggest a direct link between prebiotic mineral and
metal-ligand chemistry~\cite{Morowitz:ligands:10}, and constraints
inferable from the long-term structure of later cellular evolution.

Several enzyme iron-sulfur centers have been
recognized~\cite{Russell:FeS:06,Fontecilla:metalloenzymes:09} to use
strained versions of the unit cells found in Fe-S minerals,
particularly Mackinawite and Greigite.  These are particular instances
within a wider use of transition-metal-sulfide chemistry in
core-metabolic enzymes.

Pyruvate:ferredoxin oxidoreductase (PFOR), which catalyzes the
reversible carboxylation of acetyl-CoA to pyruvate, contains three
[${\mbox{Fe}}_4{\mbox{S}}_4$] clusters and a thiamin pyrophosphate
(TPP) cofactor.  The [${\mbox{Fe}}_4{\mbox{S}}_4$] clusters and TPP
combine to form an electron transfer pathway into the active site, and
the TPP also mediates carboxyl transfer in the active site
\cite{Chabriere:oxidoreductases:99}.

The bifunctional carbon monoxide dehydrogenase/acetyl-CoA synthase
(CODH/ACS) enzyme that catalyzes the final acetyl-CoA synthesis
reaction in the WL pathway employs even more elaborate
transition-metal chemistry.  Like PFOR, this enzyme uses
[${\mbox{Fe}}_4{\mbox{S}}_4$] clusters for electron transfer, but its
active sites contain additional, more unusual metal centers.  The CODH
active site contains an asymmetric Ni-[${\mbox{Fe}}_4{\mbox{S}}_5$]
cluster on which ${\mbox{CO}}_2$ is reduced to
CO~\cite{Dobbek:CODH:01}, while the ACS active site contains a
Ni-Ni-[${\mbox{Fe}}_4{\mbox{S}}_4$] cluster on which CO (from CODH)
and a methyl group from folates are joined to form
acetyl-CoA~\cite{Darnault03,Seravalli04,Svetlitchnyi04}. It was
originally thought that a variant form of the ACS active site contains
a Cu-Ni-[${\mbox{Fe}}_4{\mbox{S}}_4$]
cluster~\cite{Doukov:CODHACS:02,Seravalli03}, but it was subsequently
shown that the Cu-containing cluster represents an inactivated form of
ACS~\cite{Seravalli04}.  Similarly, it has also been shown that the
open form of the Ni-Ni ACS may switch to a closed, inactivated, form
by exchanging one of the nickel atoms for a zinc
atom~\cite{Darnault03}.

Finally, methyl-group transfer to the ACS active site mediated by the
corrinoid iron-sulfur protein (CFeSP) containing the cofactor
cobalamin also involves elaborate metal
chemistry~\cite{Banerjee03,Bender11}. In accepting the methyl-group
from folates, the cobalt in cobalamin becomes oxidized from the
${\mbox{Co}}^+$ to the ${\mbox{Co}}^{3+}$ state. Donation of the
methyl group to the ACS active site restores the ${\mbox{Co}}^+$
state, while in turn oxidizing the active Nickel in the ACS active
site from the ${\mbox{Ni}}^0$ to the ${\mbox{Ni}}^{2+}$ state. Release
of acetyl-CoA then reduces the active nickel back to the ${\mbox{Ni}}^0$ state,
allowing the cycle to start over~\cite{Bender11}.

Perhaps not surprisingly, all these examples of metal-cluster enzymes
concern catalysis not just of the formation of C-C bonds, but of the
incorporation of the small gas-phase molecule ${\mbox{CO}}_2$.  In
general, enzymes involved in the processing of small gas-phase
molecules (including ${\mbox{H}}_2$ and ${\mbox{N}}_2$) are among the
most unique enzymes in biology -- all but one of the known
Nickel-containing enzymes belong to this group~\cite{Ragsdale09} --
always containing highly complex metal centers in their active
sites~\cite{Volbeda95,Peters98,Georgiadis92,Kim92,Lancaster11,Spatzal11}.
This indicates both the difficulty of controlling the catalysis of
these reactions, and the importance of understanding their functions
in the context of the emergence of
metabolism~\cite{Fontecilla:metalloenzymes:09}.

\subsubsection{Complex network closures: diversity and opportunity
  created by aldol reactions}
\label{sec:aldol}

The network closures that retain carbon flux and enable autocatalysis
in rTCA, DC/4HB, and 3HP/4HB pathways are all topologically rather
simple, and are quite similar due to the homology among most of the
pathway intermediates.  Their module boundaries also are all defined
by acetate and succinate, and at least in the case of acetate, were
probably facilitated by its multiple pre-existing roles as the
redox-drain of the rTCA cycle~\cite{Morowitz:EFOoL:07} and the
starting point for both isoprenoid and fatty-acid lipid biosynthesis.

In contrast, the topology of the 3HP-bicycle appears complex, and
perhaps an improbable solution to the problem of recycling all carbon
flux through core pathways. This form of complexity arises from the
requirement to complete an autocatalytic network topology while
avoiding reactions based on ${\mbox{CO}}_2$ in favor of those based on
${\mbox{HCO}}_3^{-}$. It is thus different from the topological
complexity within the bowtie, where dense cross-linking in the core
arises at the intersection of many minimal pathways. If we are to
argue that the emergence or evolution of network closures such as that
in the bicycle is facilitated by a form of modularity, it must exist
at the level of reaction mechanisms that render the evolutionary
innovation of such topologties plausible.  For the 3HP bicycle and
the related glyoxylate shunt -- and to a lesser degree also for rTCA
-- the mechanism of interest is the aldol reaction.

The aldol reaction is an internal oxidation-reduction reaction, which
means that it exploits residual free energy from organosynthesis, and
also that it can take place independently of external electron donors
or acceptors.  Many aldol reactions are also kinetically
\emph{facile}, occurring at appreciable rates without the aid of
catalysts.  We therefore expect that among compounds capable of
participating in them, aldol reactions would have been common in the
prebiotic world, providing opportunities for pathway generation.
Since their diversity is difficult to suppress except by special
mechanisms~\cite{Ricardo:borate:04}, we expect that potential aldol
reactions among metabolites would either have become regulated
(perhaps through phosphoryl occupation of hydroxyl groups) or else
incorporated into actively-used biochemical pathways.

Aldol reactions are important generators of diversity in organic
chemistry, notorious for the very-complex network known as the
\emph{formose
  reaction}~\cite{Weber:SugarWorld:00,Weber:SugarWorld:01,%
  Weber:Energetics:02,Weber:Kinetics:04}, initiated from formaldehyde
and glycolaldehyde.  Many aldol reactions are possible for sugars, and
the reductive pentose phosphate pathway is indeed a network of
selected aldol condensations and cleavages among
sugar-phosphates~\cite{Andersen:NP_autocat:12}.

Fewer aldol reactions are possible among intermediates of the rTCA
cycle and their homologues such as methyl-malate or citramalate in
other carbon-fixation pathways, but all possibilities are indeed used
either in intermediary metabolism or in carbon fixation. The complex
topology of the 3HP bicycle therefore also suggests that a diverse
inventory of pathway segments were available at the time of its
emergence. Fig.~\ref{fig:bicycle_shunt_compare_pared} shows the
overlap between the bicycle and the closely-related glyoxylate shunt,
which is thus a possible precursor to the bicycle. In both pathways,
the network topologies that regenerate all carbon flux or achieve
autocatalysis are created by aldol reactions.  The retention of carbon
within the shunt appears to be a reason for its widespread
distribution and frequent use~\cite{Liu03,Fischer03,Beste11}, even
when energetically more-efficient pathways such as the Krebs cycle
exist as alternatives within organisms. The re-use of reaction
mechanisms on different substrates is a distinct form of simplicity
and redundancy that we consider more generally in
Sec.~\ref{sec:innovation}.

\subsubsection{Re-use of the direct C1 reduction pathway and hybrid
  fixation strategies}
\label{sec:hybrid_fixation}

A unique form of re-use is found for the sequence of reactions that
directly reduce one-carbon (${\mbox{C}}_1$) groups on pterin
cofactors.  We have argued
elsewhere~\cite{Braakman:carbon_fixation:12} that even when a
complete, autotrophic WL pathway is not present due to the loss of the
oxygen-sensitive CODH/ACS enzyme, the direct ${\mbox{C}}_1$-reduction
sequence on pterins is often still present and being used as a partial
fixation pathway.  The reaction sequence supplies the diverse
methyl-group chemistry mediated by S-adenosyl-methionine, and the
direct synthesis of glycine and serine from methylene groups,
reductant, and ammonia.  Serine then serves as a precursor to cysteine
and tryptophan.  The pathway may exist in either a complete
(8-reaction) or a previously-unrecognized but potentially widespread
(7-reaction) form that involves uptake on ${\mbox{N}}^5$ rather than
${\mbox{N}}^{10}$ of THF~\cite{Braakman:carbon_fixation:12} (see
Fig.~\ref{fig:C1_metabolism}.)

The widely distributed and diversified form of direct ${\mbox{C}}_1$
reduction functions much as auxiliary catabolic pathways function in
mixotrophs~\cite{Lengeler:BP:99}, operating in parallel to an
independent ``primary'' fixation pathway, with the primary and the
direct-${\mbox{C}}_1$ pathway supplying carbon to different subsets of
core metabolites.  In many cases where the CODH/ACS is lost, this loss
disconnects the primary and direct-${\mbox{C}}_1$ pathway segments,
creating the novel feature of a \emph{disjoint} carbon fixation
pathway. The existence of parallel fixation pathways in autotrophs had
previously been recognized only in one (relatively late-branching)
$\gamma$-proteobacterium, the uncultured endosymbiont of the deep-sea
tube worm \emph{Riftia pachyptila}, which was found to be able to use
both the rTCA and CBB cycles~\cite{Markert07}. In that case, however,
the two pathways are not disjointed, but rather connected through
intermediates in the glycolytic/gluconeogenic pathways. In addition,
the capacity for using either cycle is thought to reflect an ability
to adapt to variation in the availability of environmental energy
sources, with an apparent up-regulation of the more efficient rTCA
cycle under energy-poor conditions~\cite{Markert07}. Our phylogenetic
reconstruction~\cite{Braakman:carbon_fixation:12}, however, indicates
that parallel disjoint pathways were the majority phenotype in the
deep tree of life, in which a reductive C1 sequence to glycine and
serine is preserved in combination with with rTCA in Aquificales and
Nitrospirae, with CBB in Cyanobacteria, with the 3HP bicycle in
Chloroflexi (all bacteria), and with DC/4HB in Desulfurococcales and
Acetolobales, and the 3HP/4HB cycle in Sulfolobales (all archaea).  In
contrast, the full WL pathway is found only in a subset of lineages of
bacteria (especially acetogenic Firmicutes) and archaea (methanogenic
Euryarcheota).

Apparently as a result of the flexibility enabled by parallel carbon
inputs to core metabolites, the direct ${\mbox{C}}_1$ reduction
sequence is more universally distributed than any of the other
loop-networks (whether paired with ${\mbox{C}}_1$ reduction or used as
exclusive fixation pathways), or than the complete WL pathway.  The
status of the pterin-mediated sequence as a module appears more
fundamental than its integration into the full WL pathway, and
comparable to the arcs identified within rTCA, which may function as
parts of fixation pathways or alternatively as anaplerotic extensions
to other pathways.  The two types of pathways also serve similar
functional roles in our phylogenetic reconstruction of a root
carbon-fixation phenotype, as the key components enabling and
selecting the core anabolic precursors.

The reductive synthesis of glycine furnishes a potent reminder of the
importance of taking evolutionary context into account when
interpreting results from studies of metabolism.  The goal of
understanding human physiology and disease states has historically
been a major driver in the study of biochemistry and metabolism.
Although microbial biochemistry is currently better-understood
(because it is simpler) than human biochemistry~\cite{Duarte07}, model
systems and interpretations have continued to emphasize heterotrophy.
An example of the interpretive bias that can result is the common
reference to the reductive citric acid cycle as the ``reverse'' citric
acid cycle, despite its likely being the original form as we (and many
others) have argued.  Similarly, the ``glycine cleavage system''
(GCS) was originally studied in rat and chicken
livers~\cite{Kikuchi73}, before being recognized as phylogenetically
widespread.  The distribution of this system is now known to be nearly
universal across the tree of life (with methanogens being the main
systematic exception, for reasons explained
elsewhere~\cite{Braakman:carbon_fixation:12}), suggesting that it was
present already in the LUCA.  The lipoyl-protein based system has long
been known to be fully reversible~\cite{Kikuchi73,Barker41,Waber79},
and has nearly neutral thermodynamics at physiological
conditions~\cite{BarEven12}. Thus, the LUCA could have used this
system either to synthesize or to cleave glycine.  From this
perspective the former possibility (synthesis) seems a more likely
interpretation, even without additional data.  From a perspective less
strongly focused on heterotrophs, the choice between these
alternatives might have become clear much sooner.

\subsection{A coarse-graining of carbon-fixation pathways}
\label{sec:coarse_graining}

\begin{figure}[ht]
  \begin{center} 
  \includegraphics[scale=0.5]{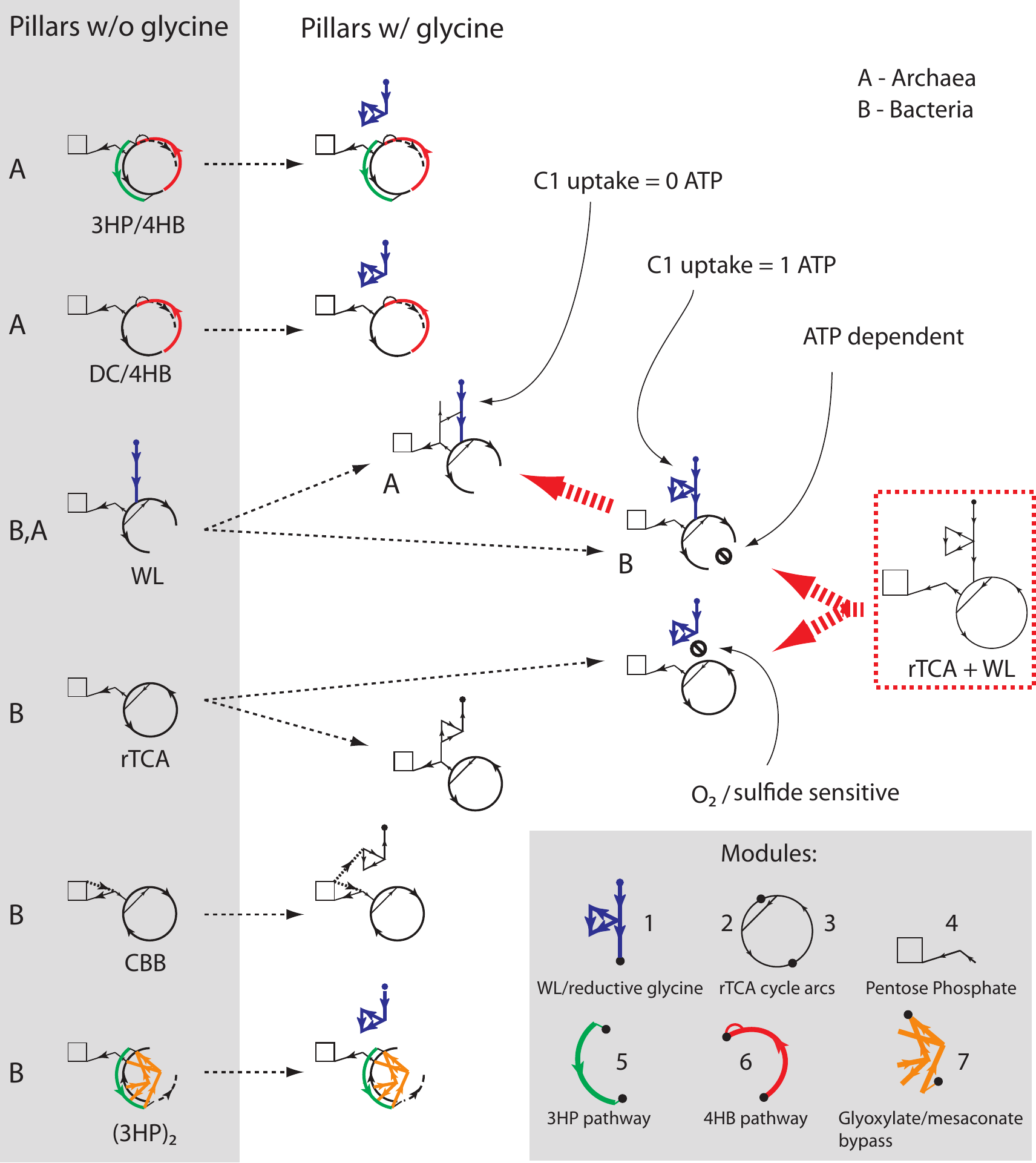}
  \caption{
  Coarse-grained summary of carbon-fixation pathways. The left panel
  shows the six pathways as they are known from extensive laboratory
  characterization. Including glycine along with the anabolic pillars
  as the molecules that must be reached in carbon-fixation then adds
  resolution, allowing finer distinctions among forms and generally
  increasing their similarity. As a result, underlying evolutionary
  templates and patterns begin to emerge. The panel on the bottom
  right shows the modules from which all carbon-fixation pathways are
  constructed, as outlined in the main text.
  \label{fig:coarse_grained_summary} 
  }
  \end{center}
\end{figure}

We can combine all the previous observations on modularity in
carbon-fixation -- the sharing of arcs between loop pathways, the
re-use of TCA and reductive C1 sequence to complete the set of
anabolic pillars -- to perform a ``coarse-graining'' of
carbon-fixation.  Combining the decomposition of
Fig.~\ref{fig:pathways_compare_ring_4col} with the gluconeogenic and
WL pathways in Fig.~\ref{fig:carbon_anabolism_CHO}, we may list the
seven modules from which all known autotrophic carbon-fixation
pathways are assembled: 1) direct one-carbon reduction on folates or
related compounds, with or without the CODH/ACS terminal reaction of
WL; 2) the short-molecule rTCA arc from acetyl-CoA to succinyl-CoA; 3)
the long-molecule rTCA arc from succinyl-CoA to citryl-CoA; 4) the
gluconeogenic/reductive pentose-phosphate pathway, with or without the
RubisCO reaction of CBB; 5) the 3HP arc from acetyl-CoA to
succinyl-CoA; 6) the long-molecule 4HB pathway from succinyl-CoA to
acetoacetyl-CoA; 7) the glyoxylate-shunt/mesaconate pathway to
citramalate, which is the long-molecule loop in the 3HP bicycle.
Fig.~\ref{fig:coarse_grained_summary} shows the summary of these
modules at the pathway level, as well as their different combinations
to form complete autotrophic carbon-fixation pathways.

The importance of including glycine in the set of anabolic pillars
immediately becomes clear in this coarse-grained view. The general
similarity among different carbon-fixation pathways increases
significantly, while finer distinction between forms becomes possible.
In particular, both of the pathways that have been most commonly
discussed in the context of ancestral carbon-fixation and the origin
of life, WL and
rTCA~\cite{Wachtershauser:rTCA:90,Smith:universality:04,%
  Martin:vents:08,Fuchs11}, separate into deep- and late-branching
forms. The increased similarity of the deep-branching forms of these
pathways suggests an underlying template that combines both WL and
rTCA in a fully connected network.  WL and rTCA differ from this
linked network by single reactions associated either with energy (ATP)
economy or oxygen (or perhaps other oxidant) sensitivity.  Combining
information on the synthesis, structural variation, ecology and
phylogenetics of the pterin molecules upon which direct $\mbox{C}_1$
reduction is performed similarly suggests a distinction between the
acetogenic (bacterial) and methanogenic (archaeal) forms of WL
associated with energy economy~\cite{Braakman:carbon_fixation:12}. A
``proto-tree'' of carbon-fixation emerges from the pooling of these
different observations, which in turn makes it possible to reconstruct
a complete phylometabolic tree of carbon-fixation, as discussed in
detail in section~\ref{sec:evol_history} below.

\subsubsection{How the inventory of elementary modules has constrained
  innovation and evolution}

The essential invariance across the biosphere of the seven
sub-networks listed above allows us to represent all carbon-fixation
phenotypes in terms of the presence or absence, connectivity, and
direction of these basic modules.  In this representation, metabolic
innovation at the modular level retains the character of individual
discrete events, even if the pathway segments involved incorporate
multiple genes.  In cases where multiple genes must be acquired to
constitute a module, this innovation may take place at higher levels
of metabolism, after which their incorporation as fixation pathways
appears as a single innovation. For example, most of the reaction
sequences used in the autotrophic 4HP pathway appear in diversified
forms in fermentative secondary metabolism from hydroxybutyrate or
aminobutyrate, which is both outside the rTCA/folate core and
ecologically heterotrophic.  It is plausible (and we think
likely~\cite{Braakman:carbon_fixation:12}) that these pathways were
recruited for autotrophy from an organism similar to \emph{Clostridium
kluyveri}.  (See also discussions in Sec.~\ref{sec:phylomet_tree}).

Because the module boundaries are defined by particular (often
universal) molecular species (\emph{e.g.}, acetyl-CoA, succinyl-CoA,
and ribulose-1,5-bisphosphate) it often remains true that innovation
can be traced to the change in single genes.  This is true for the
loss of the CODH/ACS from acetyl-CoA phenotypes, the innovation of
RubisCO in CBB bacteria, or the loss of substrate-level
phosphorylation to acetyl-CoA or succinyl-CoA in acetogens.  A case
with only slightly greater complexity is the apparently repeated,
convergent evolution of an oxidative pathway to form serine from
3-phosphoglycerate (3PG), which involves three common and widely
diversified reactions: a dehydrogenation, a reductive transamination,
and a dephosphorylation.  The evolution of this bridge pathway
creates a secondary connection between the previously disjoint
carbon-fixation pathways described in Sec.~\ref{sec:hybrid_fixation}.
As a full evolutionary reconstruction (described next) shows, such a
bridge may permit subsequent loss of direct-${\mbox{C}}_1$ reduction
as a fixation route, as in the proteobacteria, or it may release a
constraint, permitting change in pterin cofactor chemistry as in
methanogens.  

At the module level, we may represent changes in carbon
fixation pathways between closely-related phenotypes in terms of
single connections, disconnections, or overall changes of direction
within the subsets of the seven modules which are present.  The change
of direction within modules is usually complete, even if it is partial
or intermediate at the level of whole pathways.  An example is the
switch from autotrophic rTCA to fermentative TCA using a reductive
small-molecule arc and an oxidative large-molecule
arc~\cite{Watanabe:tuberculosis:11}.  Such fermentative pathways may
alternate with fully oxidative TCA (Krebs) cycling, and they often
occur in organisms that carry homologues to genes for both oxidative
and reductive pathway
directions~\cite{Tian:MTB_TCA:05,Baughn:Mtb_TCA:09,Zhang:cyano_TCA:11}.

An important exception to this pattern is the partial reversal of the
formyl-to-methylene sequence on folates, between its carbon-fixation
role and its role in the catabolic cleavage of glycine.  We refer in
Ref.~\cite{Braakman:carbon_fixation:12} to the module formed by
combining the GCS with the methylene-serine transferase as the
\emph{glycine cycle}.  The combination of the complex free energy
landscape provided by the folates~\cite{Maden:folates_pterins:00} with
the reversibility and nearly neutral thermodynamics of the glycine
cycle~\cite{Kikuchi73,BarEven12} permits a high degree of flexibility
within this module.  Carbon can enter either directly through
${\mbox{CO}}_2$, through serine (from 3PG), or through glycine (from
glyoxylate), and from any of these sources it may be redirected to all
of ${\mbox{C}}_1$ chemistry.  The topology of the main reaction
sequence is preserved in all of the above cases of reversal, though new
enzymes or cofactors may be recruited to reverse some reactions. 

A representative example of complete module reversal (and in this
case, complete cycle reversal) enabled by reductant and cofactor
substitution is given by the relation between reductive and oxidative
TCA cycles. The electron donor in rTCA, reduced ferredoxin, is
replaced by lipoic acid as an electron acceptor in the Krebs cycle, in
the TPP-dependent oxidoreductase reaction.  The enzymes catalyzing the
retro-aldol cleavage of rTCA, which have undergone considerable
re-arrangement even within the reductive
world~\cite{Aoshima:citrate_synthase:04,Aoshima:citrate_lyase:04}, is
further replaced by the oxidative citryl-CoA synthetase.  Finally, the
change from fumarate reduction to succinate oxidation may require a
substitution of membrane quinones~\cite{Schoepp:menaquinones:09}.  Yet
the underlying carbon skeletons over the whole pathway are completely
retained, and apart from some details of reaction ordering for
thioesters, and possibly the use of phospho-enol intermediates, the
energetic side groups are also the same.

\subsection{Reconstructed evolutionary history}
\label{sec:evol_history}

\subsubsection{Phylogeny suggests little historical contingency of
  deep evolution within the modular constraints}
 
The small number of modules that contribute to carbon fixation, and
the even smaller number of ``gateway'' molecules that serve as
interfaces between most of them, permit free recombination into many
phenotypes satisfying the constraints of autotrophy.  An important
consequence of free recombination is that a constraint of overall
autotrophy only enforces network completeness -- the existence of
\emph{some} connection between gateway molecules. Because there exist
multiple modules that can be used to satisfy these constraints,
autotrophy alone therefore does not lock in dependencies within networks over
distances longer than the modules themselves.  Homology across
intra-modular reaction sequences -- especially if it is due to
catalytic promiscuity -- further weakens any lock-in effect created by
selection for metabolic completeness.  Through these mechanisms
modularity promotes innovation-sharing~\cite{Vetsigian:code:06} and
rapid and reliable adaptation~\cite{Gerhart:facil_vary:07} to
environmental conditions, but reduces standing variation among
individuals sharing a common environment.

However, despite the potential for free recombination in
principle, distinct carbon-fixation pathway modules have very
different couplings to the chemical environment, as we reviewed in
Sec.~\ref{sec:core_carbon}.  The genome distributions reported in
Ref.~\cite{Braakman:carbon_fixation:12} show that they also have very
uneven phylogenetic distribution. For example, TCA arcs and
intermediates, as well as direct ${\mbox{C}}_1$-reduction, are nearly
universally distributed, while the 3HP arcs are restricted to specific
bacterial or archaeal clades living in alkaline environments.
Finally, we note that not all module combinations consistent with
autotrophy have been observed in extant organisms.

By combining these observations it is possible to arrange autotrophic
phenotypes on a graph according to their degree of similarity, and to
assign environmental factors as correlates of phenotypic changes over
most links.  The graph projects onto a tree with very high parsimony
and therefore requires invoking almost no horizontal gene transfer or
convergent evolution from distinct lineages. Instead, all divergences
may be interpreted as independent simple innovations driven by
environmental factors. Finally, the directionality of these links
(divergences) and the overlap of the tree with bacterial and archaeal
phylogeny motivates a natural choice of root. The lack of reticulation
in a tree of innovations in autotrophy -- at first surprising when
compared to highly-reticulated gene phylogenies~\cite{Puigbo:trees:09}
covering the same period -- becomes sensible as a record of invasion
and adaptation to new chemical environments by organisms capable of
maintaining little long-standing variation.

\subsubsection{A parsimony tree for autotrophic metabolism, and
  causation on links}
\label{sec:phylomet_tree}

The tree of autotrophic carbon-fixation phenotypes from
Ref.~\cite{Braakman:carbon_fixation:12} is shown in
Fig.~\ref{fig:metab_tree_printer_bigicon}.  All nodes in the tree
satisfy the constraint that all five universal anabolic precursors
plus glycine can be synthesized directly from ${\mbox{CO}}_2$.  We
have defined parsimony by requiring single changes over links at the
level of pathway modules, as explained above, rather than at the level
of single genes, in cases where the two criteria differ.  (This
definition separates the evolution of genetic backgrounds, such as
4-hydroxybutyrate fermentation, from the events at which organisms
came to rely on complete pathways for autotrophy.)

A complete-parsimony tree for the known phenotypes is not possible, so
we chose a tree in which the only violations are duplicate innovation
of serine synthesis from 3-phosphglycerate (3-PG), and duplication or
transfer of the short-molecule 3HP pathway.  The synthesis of serine
from 3-PG involves reactions -- the dehydrogenation of an alcohol to a
carbonyl, the transamination of this carbonyl to an amino group, and a
dephosphorylation -- that are common throughout metabolism and are
performed by highly diversified enzyme families.  We have therefore
regarded multiple occurrences of this event as not attaching a
large probability penalty to parsimony violation.  We make a similar
judgment for the 3HP pathway. This pathway contains two key
biotin-carboxylase enzymes, one of which (acetyl-CoA carboxylase) is
also part of fatty acid synthesis, which is suggested to have been
present already in the LUCA~\cite{Pereto:membrane_lipids:04}. Sequence
analysis of propionyl-CoA carboxylase has in turn been used to suggest
convergent evolution as an explanation for the multiple occurrence of
this enzyme across bacterial and archaeal
domains~\cite{Teufel:3HP_4HB:09}. The remaining reactions in this
pathway are again common metabolic reactions performed by highly
diversified enzyme families. An alternative hypothesis is
transfer: this complete pathway occurs in environments shared by the
bacteria and archaea that harbor it, and this environment also
contains a stressor (alkalinity) that may induce gene
transfer~\cite{Brazelton:biofilms:09}. Thus, both gene transfer and
convergence are plausible explanations for why this phenotype should
be paraphyletic.
  
Any tree in which either of these phenotypes was made monophyletic
would require more extensive parsimony violations than the tree
we chose, involving innovations for which convergence or transfer are
also less plausible. Such trees would require major sub-branches 
to contain both bacterial and archaeal members, and within these,
repeated divergences of major domain-specific differences.  
Common descent would fail to account for the exclusivity of
rTCA-based phenotypes within bacteria, and among archaea either
(non-THF) pterin-based one-carbon chemistry, or isoprene-related
hydroxybutarate reactions would be required to have arisen several
times. It is corroborating evidence to this parsimony argument,
that the tree we propose preserves the monophyly of bacteria and
archaea, and is consistent with the most robust signals in purely
statistical gene phylogenies, including the greatest congruence of
firmicutes with the archaea, among the bacterial
branches~\cite{Ciccarelli:phylo_tree:06,Skophammer07}.

\begin{widetext}

\begin{center}
\begin{figure}[ht]
  \begin{center} 
  \includegraphics[scale=0.6]{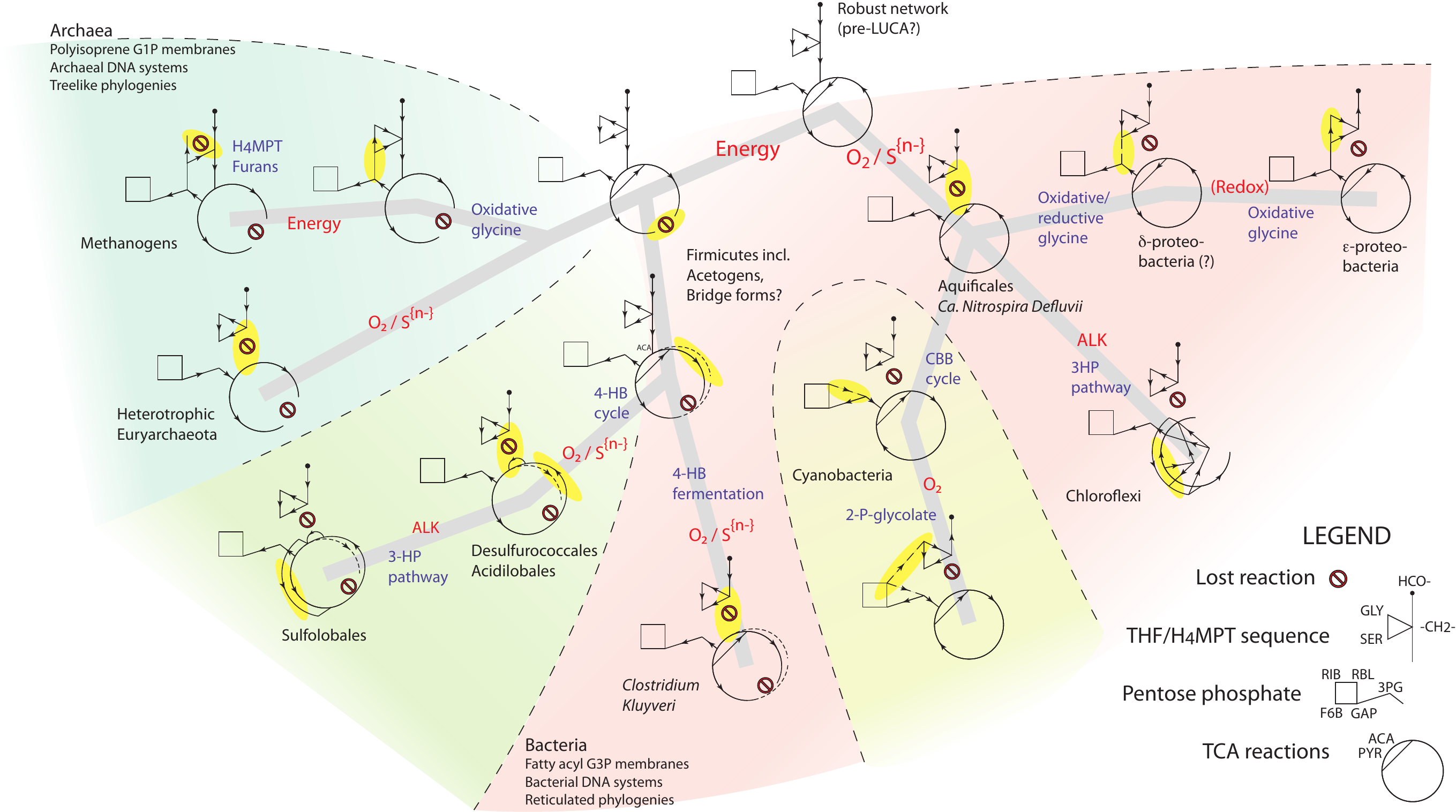}
  \caption{
  A parsimony-based reconstruction of the innovations linking the
  major carbon-fixation phenotypes, from
  Ref.~\cite{Braakman:carbon_fixation:12}. Nodes in the tree are
  autotrophic phenotypes, following the coarse-grained notation
  introduced in Fig.~\ref{fig:coarse_grained_summary}, and summarized
  in the legend.  Grey links are transitions in the maximum-parsimony
  phylometabolic reconstruction, and yellow-highlighted regions in the
  diagrams are innovations following each link.  Organism names or
  clades in which these phenotypes are found are given in black;
  fixation pathways innovated along each link are shown in blue, and
  imputed evolutionary causes are shown in red.  $\mbox{S}^{\{n-\}}$
  refers to sulfides of different oxidation states. Dashed lines
  separate regions in which the clades by phylometabolic parsimony
  follow standard phylogenetic divisions.  Abbreviations: formyl
  ($\mbox{HCO}-$); methylene ($-{\mbox{CH}}_2-$); acetyl-CoA (ACA);
  pyruvate (PYR); serine (SER); 3-phosphoglycerate (3PG);
  glyceraldehyde-3-phosphate (GAP); fructose-1,6-bisphosphate (F6B);
  ribose-phosphate (RIB); ribulose-phosphate (RBL); akalinity (ALK).
  Arrows indicate reaction directions; dashed line connecting 3PG to
  SER indicate intermittent or bi-directional reactions.
    \label{fig:metab_tree_printer_bigicon}
  }
  \end{center}
\end{figure}
\end{center}

\end{widetext}

The nodes in the tree of Fig.~\ref{fig:metab_tree_printer_bigicon} are
all phenotypes of extant organisms, with one important exception,
which is the node between the Aquificale branch and the
Firmicute/Archaea branch.  Aquificales and all phenotypes descending
from them lack the CODH/ACS enzyme, while firmicutes and archaea lack
one or more ATP-dependent acyl-CoA (citryl-CoA or succinyl-COA)
synthases.  Therefore, if we seek a connected tree of life, two
changes -- the gain of one enzyme and loss of the other -- are
required to connect these branches.  Since any organism lacking both
enzymes could not fix carbon autotrophically, we have chosen the order
of gain and loss so that the intermediate node has \emph{both} the
CODH/ACS and the acyl-CoA synthases.  It therefore has both a complete
WL pathway and an autocatalytic rTCA loop, connected through their
shared intermediate acetyl-CoA.  Losses (but not re-acquisitions) of
either of these enzymes occur at multiple points on the tree, and both
have likely explanations in either environmental chemistry or
energetics.  For this reason and several others given below, although
a parsimony tree is (\emph{a priori}) unrooted, we will regard the
joint WL/rTCA phenotype as not only a bridging node but the root of
the tree of autotrophs. 

There is one further, unproblematic exception to the assignment
of extant phenotypes to nodes in the tree, which is the insertion of
an acetogenic phenotype with a facultative oxidative pathway to serine
at the root of the Euryarchaeota.  Since methanogens use this pathway,
and since an acetogenic pathway lacking oxidative serine synthesis is
the most plausible ancestral form for all archaea as well as for
Firmicutes within the bacteria, we infer that such an intermediate
state did or does exist.  This fixation pathway is consistent with
forms observed in extant organisms, and these proposals would be
supported if such a phenotype were to be discovered or to result from
reclassification of genes in an existing organism.

In the evolution of carbon fixation from a joint WL/rTCA root, the
primary division is between the loss of the CODH/ACS, resulting in
rTCA loop-fixation phenotypes, and the loss of the acetyl-CoA or
succinyl-CoA synthetases, resulting in acetogenic phenotypes.  Very
low levels of oxygen permanently inactivate the CODH/ACS, so its loss
is probable even under microaerobic conditions.  Although the dominant
mineral buffers for oxygen in the Archaean remain a topic of
significant uncertainty~\cite{Kasting:mantel_oxygen:93,%
Kasting:anc_oxygen:06,Hazen:mineral_evol:08,Trail:Hadean_Oxygen:11},
it appears unlikely that molecular oxygen was the toxin responsible
for loss of the CODH/ACS much before the ``Great Oxidation Event''
(GOE).\footnote{The GOE is usually dated at 2.5 GYA, which may be
relevant dates to compare to genetically estimated loss events in
later branches of the Archaea or possibly in the Clostridia, but they
are not plausible as dates for the first branching in the tree of
Fig.~\ref{fig:metab_tree_printer_bigicon}. Arguments have been made
for low levels of oxygen preceding the GOE by as much as 50-100
million years~\cite{Anbar:early_whiff:07}, as well as a transient rise
in oxygen as far back as 2.9 GYA~\cite{Ono06,Ohmoto06}, but both are
actively being debated~\cite{Farquhar07,Buick08,Kump08,Sessions09}.}
Therefore the sensitivity of the CODH/ACS to sulfides or perhaps other
oxidants~(S.~Ragsdale, pers.~comm.) remains a possibly important
factor in the early divergences of carbon fixation.

Alternatively, among strict WL-anaerobes, the loss of citryl-CoA or
succinyl-CoA synthetase saves one ATP per carbon fixed, and all
acetogenic phenotypes break rTCA cycling only through the loss of one
or the other of these enzymes.  We therefore interpret the loss of
rTCA cycling as a result of selection for energy efficiency.  The
failure to regain either of these enzymes by acetogens which
subsequently also lost the CODH/ACS is perhaps surprising given the
inferred homology of the ancestral citryl-CoA and succinyl-CoA
synthetases~\cite{Aoshima:citrate_synthase:04%
  ,Aoshima:citrate_lyase:04}, but explains the absence of rTCA cycling
in either Firmicutes or any Archaea.

The remaining autotrophic phenotypes are derived from either rTCA
cycling or acetogenesis in natural stages due to plausible
environmental factors.  Oxidative serine synthesis (from 3PG) is
associated with the rise of the proteobacteria, whose differentiation
in many features tracks the rise of oxygen and the transition to
oxidizing rather than reducing environments.  RubisCO and subsequently
photorespiration arise within the cyanobacteria.  The innovation of
the 3HP bicycle from the malonate pathway arises within the
Chloroflexi.  In both Firmicutes (bacteria) and the crenarchaea,
4-hydroxybutyrate (or closely related 4-aminobutyrate) fermentation is
more or less developed.  Closure of the fermentative arcs to form a
ring, again driven by elimination of the
CODH/ACS~\cite{Braakman:carbon_fixation:12} leads to the DC/4HB
pathway in Crenarchaeota, which is then specialized in the
Sulfolobales to the alkaline 3HP/4HB pathway.  The Euryarchaeota are
distinguished by the absence of an alternative loop-fixation pathway
to rTCA, so that all members are either methanogens or heterotrophs.

Similarly, the innovation of the 3HP pathways, using biotin, emerges
as a specialization to invade extreme but relatively rare
environments. A particularly interesting case is the modification of
folates in archaea, leading from THF in ancestral nodes to
tetrahydromethanopterin in the methanogens, which enables initial
fixation of formate (formed by hydrogenation of ${\mbox{CO}}_2$) in an
ATP-free
system~\cite{Maden:folates_pterins:00,Braakman:carbon_fixation:12}.
The root position of rTCA explains the preservation of rTCA arcs both
in reductive acetyl-CoA pathways, and in anaplerotic appendages
to other fixation pathways, and the root position of direct
${\mbox{C}}_1$ reduction explains its near-universal distribution.

\subsubsection{Parsimony violation and the role of ecological
  interactions} 

A tree is by construction a summary statistic for the relations among
the phenotypes which are its leaves or internal nodes.  It is not
inherently a map of species descent, and takes on that interpretation
only when common ancestry is shown to explain the conditional
independence of branches given their (topological) parent nodes.  This
caution is especially important for the interpretation of
Fig.~\ref{fig:metab_tree_printer_bigicon}, which shows high parsimony
in the deepest branches where horizontal gene transfer is generally
believed to have been most
intense~\cite{Woese:univ_anc:98,Woese:HGTtree:00}.  We have argued
that this behavior is consistent in a tree of successive optimal
adaptations to varied environments, by organisms that could maintain
little persistent variation.  Violations of parsimony that are
improbable by evolutionary convergence contain information about
contact among historically separated lineages. Under this
interpretation the separation is primarily environmental, with the
subsequent contact identifying ecological co-habitation.  As explained
above, the possible transfer of genes for the 3HP pathway is
especially plausible, as the organisms involved may have shared the
same extreme (alkaline) environments and been under common selection
pressure, which when severe is known to accelerate rates of gene
transfer~\cite{Brazelton:biofilms:09,Brazelton:vent_ecosys:10}.

Our methods in Ref.~\cite{Braakman:carbon_fixation:12} include
flux-balance analysis of core networks, where the boundaries of
analysis are defined to be carbon input solely from ${\mbox{CO}}_2$
and the output of the universal precursors we have listed as the
interface between carbon-fixation and anabolism, as shown in
Fig.~\ref{fig:Biosphere_global}. We do not model cellular-level
mechanisms of either regulation or heredity, nor full downstream
intermediary metabolism.  Our system of metabolic flux constraints
therefore does not distinguish between individual species and
ecosystems. It does not, of course, \emph{exclude} the
possibility of representing individual organisms. The general
agreement with robust phylogenetic signatures from many different
genomic
phylogenies~\cite{Pace:tree:97,Ciccarelli:phylo_tree:06,Puigbo:trees:09}
may thus still suggest a dominant role for vertical descent among
autotrophic \emph{organisms} (and not merely consortia) in the early
evolution of carbon fixation.

\subsubsection{A non-modern but plausible form of redundancy in the
  root node}
\label{sec:character_root}

The joint WL/rTCA network was introduced into
Fig.~\ref{fig:metab_tree_printer_bigicon} to produce a connected tree
containing only autotrophic nodes.  It also gives the most
parsimonious interpretation of the nearly universal distributions of
both reductive ${\mbox{C}}_1$ chemistry and of citric acid cycle
components, and receives further circumstantial support from the
identifiability of both plausible and specific environmental driving
forces for most subsequent branches.  The constraints which jointly
required the insertion of a linked WL/rTCA network at the root have
led us to propose a kind of redundancy not found in extant fixation
pathways.  Either WL or rTCA alone is self-maintaining (in a modern
organism) so a network that incorporates both is \emph{redundantly}
autocatalytic.  This is an important and speculative departure from
known phenotypes, but it can be argued to have conferred a selective
advantage under the more primitive conditions of early cells, because
the pathway topology itself possesses a form of inherent
\emph{robustness}.  The redundant network topology of the root
phenotype would have allowed it to better cope with both internal and
external perturbation in an era when regulation and kinetic control
were probably less sophisticated and refined than they are today. In
that respect it is a more plausible phenotype for a universal ancestor
than any modern network.

The enhanced robustness of the joint network follows from the
interaction of short-loop and long-loop autocatalysis.  The threshold
for autocatalysis in the rTCA loop, fragile against parasitic side
reactions or unconstrained anabolism, is supported and given a
recovery mode when fed by an independent supply of acetyl-CoA from WL.
In turn, the production of a sufficient concentration of folates to
support direct ${\mbox{C}}_1$ reduction, fragile if the long
biosynthetic pathway is unreliable, is augmented by additional carbon
fixed in rTCA.  These arguments are topological, and do not make
specific reference to whether the catalysts for the underlying
reactions are enzymes.  They may provide context for (perhaps
multi-stage) models of transition from primordial mineral
catalysis~\cite{Martin:OrigCells:03,Russell:AcetylCoA:04} to the
eventual support of carbon fixation by biomolecules.

\begin{figure}[ht]
  \begin{center} 
  \includegraphics[scale=0.5]{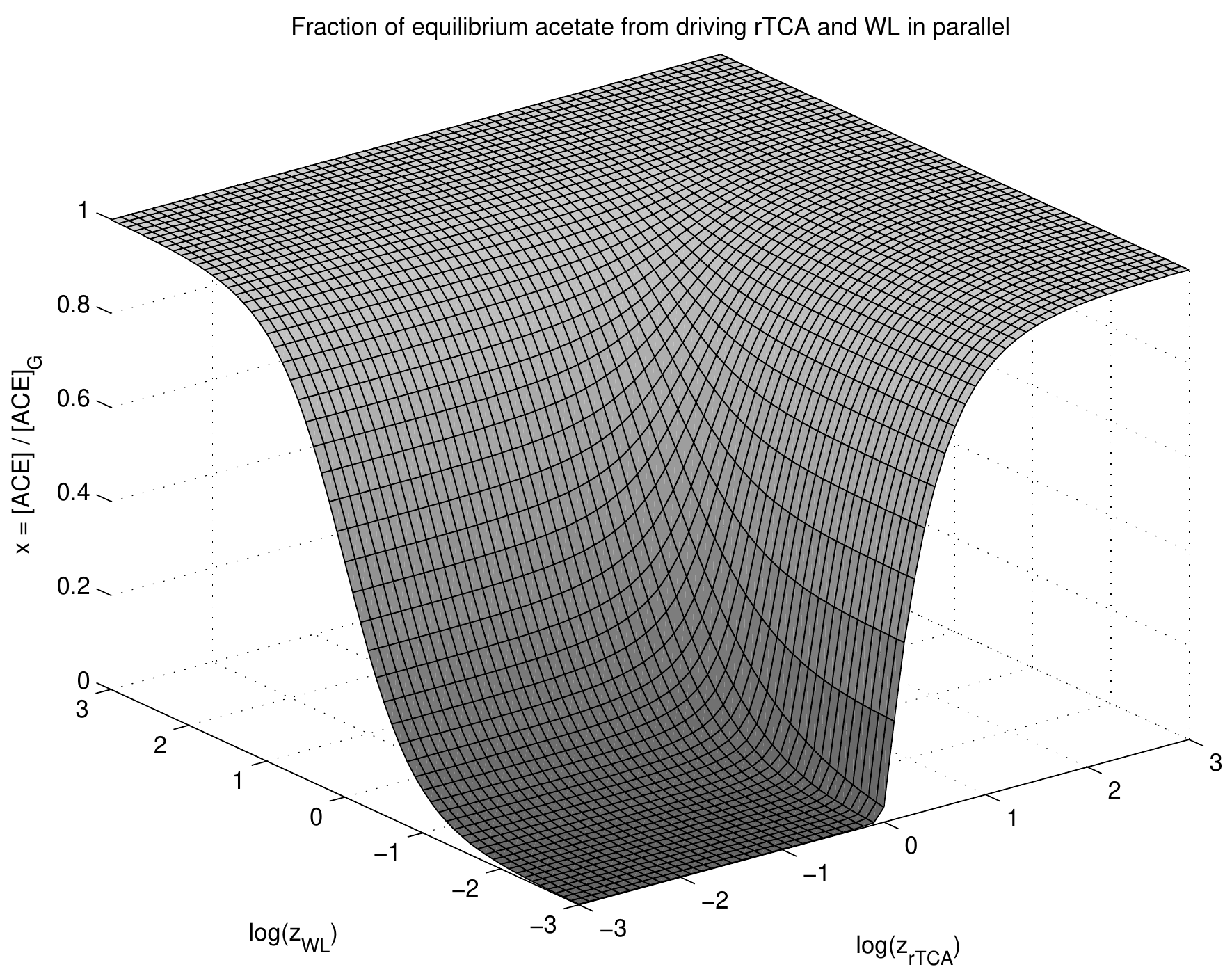}
  \caption{
  Graph of solutions to Eq.~(\ref{eq:x_quad_sol}) from
  App.~\ref{sec:graph_explanation} is shown versus base-10 logarithms
  of $z_{\mbox{\scriptsize rTCA}}$ and $z_{\mbox{\scriptsize WL}}$.
  The quantity $x$ on the z-axis is the fraction of the acetate
  concentration $\left[ \mbox{ACE} \right]$ relative to the value it
  would take in an equilibrium ensemble with carbon dioxide,
  reductant, and water.  The value $x = 1$ corresponds to an
  asymptotically zero impedance of the chemical network, compared to
  the rate of environmental drain.  The parameter
  $z_{\mbox{\scriptsize rTCA}}$ is a monotone function of the
  non-equilibrium driving chemical potential to synthesize acetate,
  and $z_{\mbox{\scriptsize WL}}$ measures the conductance of the
  ``feeder'' WL pathway.  At $z_{\mbox{\scriptsize WL}} \rightarrow
  0$, the WL pathway contributes nothing, and the rTCA network has a
  sharp catalytic threshold at $z_{\mbox{\scriptsize rTCA}} = 1$.  For
  nonzero $z_{\mbox{\scriptsize WL}}$, the transition is smoothed, so
  some excess population of rTCA intermediates occurs at any driving
  chemical potential.
    \label{fig:rTCA_WL_conc_bw_surf} 
  }
  \end{center}
\end{figure}

Fig.~\ref{fig:rTCA_WL_conc_bw_surf} shows a numerical solution for the
flux through a minimal version of the joint WL/rTCA network, with
lumped-parameter representations of parasitic side reactions and the
net free energy of formation of acetate.  (The exact rate equations
used, and their interpretation, are developed in
App.~\ref{sec:graph_explanation}.)  In the absence of a WL ``feeder''
pathway, rTCA has a sharp threshold for the maintenance of flux
through the network as a function of the free energy of formation of
its output acetate.  The existence of such a sharp threshold depending
on the rate of parasitism, below which the cycle supports \emph{no}
transport, has been one of the major sources of criticism of
network-autocatalytic pathways as models for
proto-metabolism~\cite{Orgel:cycles:08}.  When WL is added as a
feeder, however, the threshold disappears, and some nonzero flux
passes through the pathway at any negative free energy of formation of
the outputs.  The existence of a pathway that supports some
organosynthetic flux at any positive driving chemical potential --
dependent on external catalysts but not contingent on the pathway's
own internal state -- has been one of the major reasons WL has been
favored as a protometabolic pathway by molecular biologists and
chemists~\cite{Russell:AcetylCoA:04,Martin:vents:08,Fuchs11}.

The reason (beyond evidence from reconstructions) that we regard
a linear pathway modeled on the modern acetyl-CoA pathways as an
incomplete answer to the needs of incipient metabolism is that it
offers an avenue for \emph{production} of organics, but does not by
itself offer a chemical mechanism for the kind of \emph{selection} and
\emph{concentration} of fluxes that is equally central to the sparse
network of extant core metabolism~\cite{Copley:sparseness:10}.
Chemical self-amplification, if it can be demonstrated experimentally,
is the most plausible mechanism by which the biosphere can concentrate
all energy flows and material cycles through a small, stable set of
organic compounds.  It supplies the molecules that are within the loop
-- and secondarily those that are made from loop intermediates --
above the concentrations they would have in a Gibbs equilibrium
distribution, as a result of flow through the network.  The fact that
self-amplification is permitted to act in the model of
Fig.~\ref{fig:rTCA_WL_conc_bw_surf}, even below the chemical-potential
difference where the rTCA loop alone is self-sustaining, provides a
mechanism by which the loop intermediates could have been supplied in
excess in the earliest stages of the emergence of metabolism.  We
return in Sec.~\ref{sec:integration} to a related form of robustness
and selection, which applies as anabolic pathways begin to form from
loop intermediates.

A surprising observation suggested by our reconstructed history
is how conservative the biosphere has been of its intermediate stages
of innovation, as a consequence of geochemical niche diversity on
earth.  Except for the root node, we have not needed to invoke extinct
ancestral forms to explain extant diversity, an argument that even
Darwin~\cite{Darwin:origin:59} expected to be required frequently for
cases where modern, optimized forms outcompeted their more primitive
ancestors and erased direct evidence about the past.  The one case where we do invoke an essential extinct ancestral form is the root node, and its character suggests reasons why it should have become extinct that are more chemically basic and biochemically consequential
than the secondary physiological or ecological distinctions that
modern evoloutionists use to explain extinct ancestral forms.  The
topological robustness of the root node comes at the combined costs of
sub-optimal energy efficiency and oxidant sensitivity.  The fitness
advantage to shedding either of these costs would have increased
significantly as organisms obtained more sophisticated macromolecular
components and correspondingly greater control over their internal
chemistry, lessening (and ultimately removing) the selective advantage
of a redundant carbon-fixation strategy even in the absence of
external biogeochemical perturbations.

Without the ability to culture and analyze a population of LUCA
organisms, the amount we can conclude from mathematical analyses of
general network properties is of course less, but it is still within
the range commonly used to assess proposals for early metabolism.  For
example: proposals for autocatalysis in geochemical networks with
crude catalysts are routinely criticized on the basis of their
\emph{shared topological feature} of feedback and its associated
threshold fragility~\cite{Orgel:cycles:08}.  These criticisms emphasize that parasitic side-reactions are a likely problem, although the corollary that autocatalysis is generally ruled out requires the strong claim that side-reactions are a problem in \emph{all} plausible environments, which extends well beyond current experimental knowledge. In the opposite direction (and in this case based on particular and well-understood side-reactions), it is argued for the formose network~\cite{Ricardo:borate:04} that without some severe pruning mechanism, the reactions are \emph{too productive}, creating mixtures too complicated to be relevant to biochemistry.  General arguments of both kinds contribute to a
negative hypothesis behind the hope~\cite{Gesteland:RNAWorld:06} that
catalytic RNA will be a sufficient solution to both problems of
productivity and selection.  But the larger points in both arguments
are important, and should be applied at many other places in
hypotheses about the emergence and early elaboration of metabolism.
They emphasize that both robustness \emph{and} selectivity are needed
features of any mechanism responsible for the earliest cellular
organosynthesis.  At the same level of generality as they are raised
as criticisms, these criteria create meaningful distinctions between
topologies for early carbon-fixation pathways, to which the
lumped-parameter model of our root node gives a quantative form.

\subsection{The rise of oxygen, and changes in the evolutionary
  dynamics of core metabolism}

The limits of the phylometabolic tree we show in
Fig.~\ref{fig:metab_tree_printer_bigicon} fall on a horizon that
coincides with the rise of oxygen.  More precisely: we do not show
branches that phylogenetically trace lineage divisions later than this
horizon, because no known divisions in carbon fixation distinguish
such later branches.  Many of the late branches contain only
heterotrophs, and to the extent that post-oxygen lineage divisions
follow divisions in metabolism, they are divisions in forms of
heterotrophy.  The rise of oxygen seems to have put an end to
innovation in carbon fixation, and led to a florescence of innovation
in carbon sharing. By ``sharing'' we refer to general exchanges in
which organic compounds are re-used without \emph{de novo} synthesis;
we do not intend only symbiotic associations.  At the level of
aggregate-ecosystem net primary production, the exchange of organics
with incomplete catabolism may, however, reduce the free energy cost
of the \emph{de novo} synthesis of biomass that supports a given level
of phenotypic diversity or specialization, allowing ecologies of
complementary specialists to partially displace ecologies of
generalist autotrophs.

On the same horizon, the high parsimony of the tree we have shown
ends, and it becomes necessary to explain complex metabolisms as a
consequence of transfer of metabolic modules among clades in which
they had evolved separately.  We no longer expect that it would be
possible to explain -- and to some extent to predict -- these
innovations given only constraints of chemistry and invasion of new
geochemical environments.  Instead, they rely chemically on
ecologically determined carbon flows, and genetically on opportunities
for transfer of genes or pathway segments.  Therefore any explanation
will require some explicit model of ecological dynamics, and may
require invoking some accidents of historical contingency.  This
contrast of phylometabolic reconstructions, between later and earlier
periods, illustrates our association of parsimony violation with the
role of ecosystems and explicit contributions of multilevel dynamics
to evolution.

It is perhaps counterintuitive, but we believe consistent, that the
phylometabolic tree is more tree-like in the earlier era of more
extensive single-gene lateral transfers, and becomes \emph{less}
tree-like and more reticulated, in the era of complex ecosystems
enabled by oxygenic metabolisms, which may have come as much as 1.5
billion years later.  For reticulation to appear in a tree of
reconstructed metabolisms, it is necessary that variants which evolved
independently -- as we have argued, under distinct selection pressures
-- be maintained in new environments where they can be brought into
both contact and interdependence.  The maintenance of standing
variation is facilitated both by the evolution of more advanced
mechanisms to integrate genomes and limit horizontal transfer, and by
the greater power density of oxygenic metabolisms.

\begin{widetext}

\begin{center}
\begin{figure}[ht]
  \begin{center} 
   \includegraphics[scale=0.68]{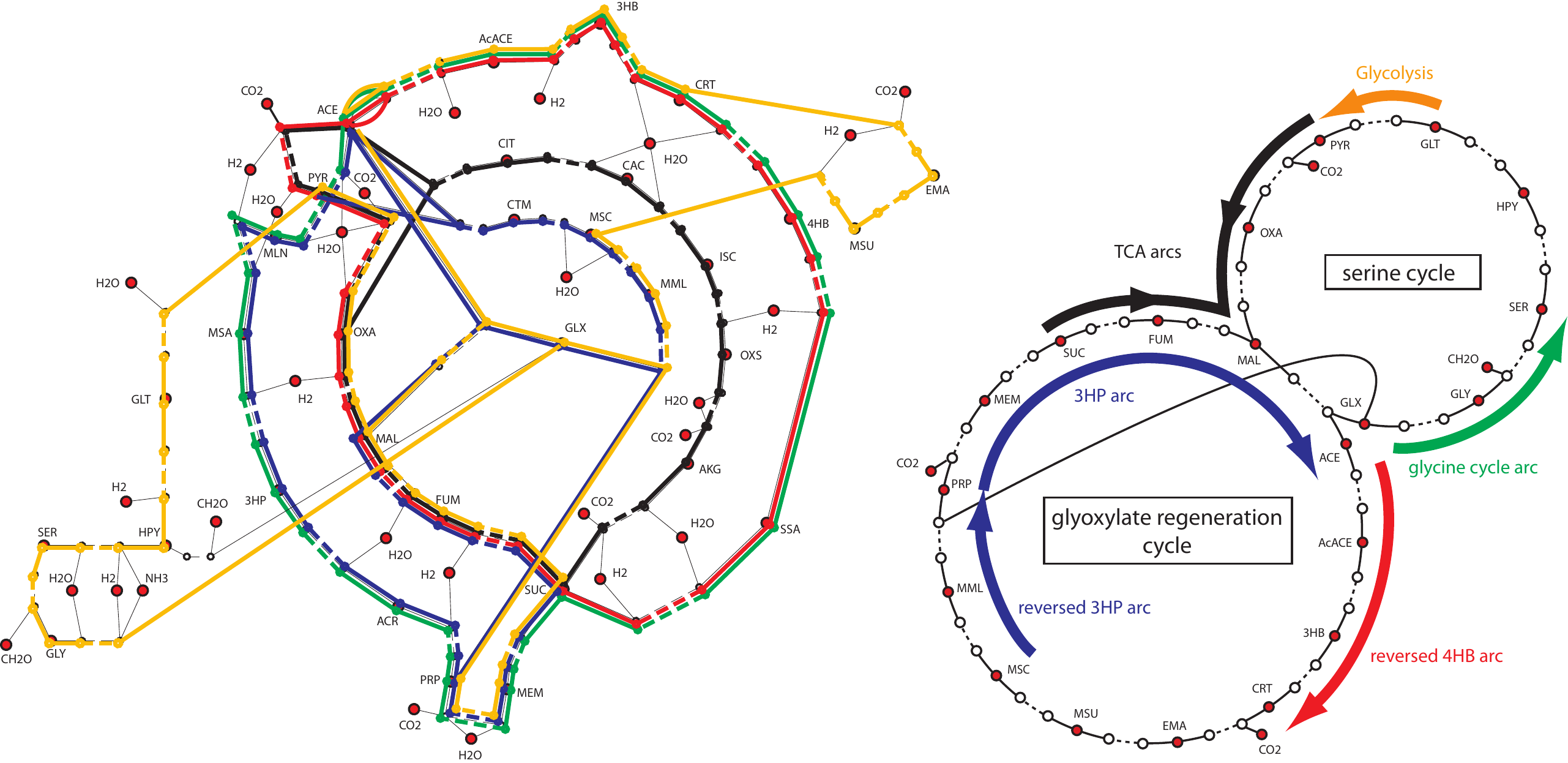}
  \caption{
  The serine cycle/glyoxylate-regeneration cycle of methylotrophy.
  Left panel shows the stoichiometric pathway overlaid on the
  autotrophic loop pathways from
  Fig.~\ref{fig:pathways_compare_ring_4col}.  Right panel gives a
  projection of the serine cycle and glyoxylate regeneration cycle
  showing pathway directions; overlaps with the predecessor
  autotrophic pathways are labeled. Abbreviations: hydroxy pyruvate (HPY); ethyl-malonate (EMA); methyl-succinate (MSU), others as in Fig.~\ref{fig:carbon_anabolism_CHO}.
    \label{fig:pathways_compare_ring_4col_ser} 
  }
  \end{center}
\end{figure}
\end{center}

\end{widetext}

The \emph{serine cycle} used by some methylotrophic proteobacteria,
shown in Fig.~\ref{fig:pathways_compare_ring_4col_ser}, provides an
example of the structure and complex inheritance of a post-oxygen,
heterotrophic pathway.  Methylotrophs possess both an ${\mbox{H}}_4
\mbox{MPT}$ system transferred from methanogenic
archaea~\cite{Chistoserdova:C1_transfer:98,Vorholt:methylotrophy:99},
and a conserved THF system ancestral to the proteobacteria (and we
argue, to the universal common ancestor).  In methylotrophs,
${\mbox{H}}_4 \mbox{MPT}$ is primarily used for the oxidation of
formaldehyde to formate, while THF can be used in both the oxidative
direction as part of the demethylation of various reduced one-carbon
compounds, and in the reduction of formate. ${\mbox{C}}_1$ compounds
are then assimilated either as ${\mbox{CO}}_2$ in the CBB cycle, as
methylene-groups and ${\mbox{CO}}_2$ in the serine cycle or as
formaldehyde in the ribulose monophosphate (RuMP) cycle, in which
formaldehyde is attached to ribulose-5-phosphate to produce
fructose-6-phosphate~\cite{Chistoserdova:methylotrophy:09,Chistoserdova11}.

The full substrate network of the most complex assimilatory pathway of
methylotrophy is a bicycle in which the serine cycle is coupled to the
\emph{glyoxylate regeneration cycle.}  This full network employs
segments of all four loop-autotrophic pathways, as well as reactions
in glycolysis, and part of the ``glycine cycle''.  Carbon enters the
pathway at several points. Methylene groups enter through the glycine
cycle, combining with glycine to form serine.  Serine is then
deaminated and reduced to pyruvate, which is combined with a
${\mbox{CO}}_2$ in a carboxylation to enter the core of TCA reactions.
TCA arcs are performed reductively from pyruvate to malate, and
oxidatively from succinate to malate, following the pattern of the 3HP
pathway plus anaplerotic reactions from its output pyruvate.  The
short-molecule arc of 3HP is run as in the autotrophic carbon-fixation
pathway starting from propionate, but part of the long- molecule arc
of 3HP is reversed in the glyoxylate regeneration cycle.  The 4HB
pathway arc, transferred from archaea, is also reversed to feed this
glyoxylate cycle, and is followed by a final additional carboxylation
unique to this pathway~\cite{Erb07,Erb09}.

The serine/gyoxylate cycle of methylotrophy is a remarkable
``Frankenstein's monster'' of metabolism, stitched together from parts
of all pre-existing pathways, but requiring almost nothing new in its
own local chemistry.  Notably, the modules in this bacterial pathway
which have been inherited from archaea are all reversed from the
archaeal direction.

\subsection{Summary: Catalytic control as a central source of
  modularity in metabolism}

Focusing on the metabolic foundation of the biosphere --
carbon-fixation and its interface with anabolism -- we have seen many
examples of ways in which catalytic control is a central organizing
principle in metabolism. The most complex and conserved reaction
mechanisms in carbon-fixation often have unique and very elaborate
metal centers and cofactors associated to them, reflecting the
difficulty (or at least unique requirements) of the catalytic problem
being solved. Not surprisingly, these reactions form the boundaries at
which the various modules making up carbon-fixation are connected.
Carbon fixation is the precursor to all biosynthesis, and in the
context of a fan/bowtie network where it is part of the core, it is
therefore also a strong constraint.  Finally, under comparative
analysis we find the maximum-parsimony assignment of innovation events
in the compositional structure of the network of possibilities to
coincide closely with robust signals from genomic phylogenies.  We
interpret the convergence of these diverse observations to mean that
innovations in carbon fixation were at least a large factor in the
major early evolutionary divergences of bacteria and archaea.  The
preservation of this evolutionary signal over very long periods and
the very small diversity of pathway innovations suggests that they
have also been some of the strongest long-term constraints on
evolution. The fixation-module boundaries act as ``turnstiles'' along
which the flow of carbon into the biosphere is redirected upon
biogeochemical perturbations, and they are preconditions for
higher-level diversification.

The catalytic control of classes of organic reactions also leads to a
secondary source of modularity, the locking in of various core
pathways by the elaboration of downstream intermediary metabolism.
The most striking example of lock-in is the origination of all
anabolic pathways in only a very small number of molecules, mostly
within the TCA cycle, even when different carbon-fixation strategies
are used.  The suggested interpretation is that much of intermediary
metabolism had elaborated prior to the divergences in carbon-fixation.
A related, but slightly different form of lock-in is found in the
construction of methylotrophic pathways, which circumvents innovations
in the catalytic control of difficult chemistry by re-using a wide
range of parts from pre-existing carbon-fixation pathways.

\section{Cofactors, and the emergence and centralization of
    metabolic control}
\label{sec:cofactors}

Cofactors form a unique and essential class of components within
biochemistry, both as individual molecules and as a distinctive level
in the control over metabolism.  In synthesis and structure they tend
to be among the most complex of the metabolites, and unlike amino
acids, nucleotides, sugars and lipids, they are not primary structural
elements of the macromolecular components of cells.  Instead,
cofactors provide a limited but essential inventory of functions,
which are used widely and in a variety of macromolecular contexts.  As
a result they often have the highest connectivity (forming topological
``hubs'') within metabolic networks, and are required in conjunction
with key inputs or
enzymes~\cite{Handorf:met_expand:05,Raymond:oxy_networks:06,%
  Schutte:met_coevol:10} to complete the most elaborate metabolisms.

  Cofactor chemistry is in its own right an essential component of the
  logic underlying metabolic architecture and evolution.  We argued in
  Sec.~\ref{sec:network_autocat} that part of the structure of the
  small-molecule substrate network is explained by reaction mechanisms
  and autocatalysis in short-loop pathways, which may once have been
  supported by external mineral catalysts.  At least since the first
  cells, however, all such pathways have been realized only with the
  essential participation of intermediates from the hierarchically and
  functionally distinct cofactor class, which add a second layer of
  network-catalytic feedback.  The more structurally complex cofactors
  tend to be associated with more catalytically complex functions
  within carbon-fixation.  Their long synthetic pathways result in
  long feedback loops, creating new needs for pathway stabilization
  and control.  Because cofactors often mediate kinetic bottlenecks in
  metabolism, their inventory of functions may constrain the
  evolutionary possibilities for new pathways, so innovations in
  cofactor synthesis can have dramatic consequences for the
  large-scale structure of evolution.

  As we note below, cofactors are among the less well-understood
  components of metabolism.  Our ability to decompose cofactor
  functions and reconstruct the likely history of their elaboration is
  therefore less comprehensive than the analysis we have given of the
  small-molecule substrate.  However, many functions that divide the
  cofactors into groups, which seem also to have been responsible for
  cases of convergent evolution and have perhaps stabilized major
  functional categories, relate directly to properties of particular
  chemical elements.  Others are molecular properties shared as a
  consequence of derivation from a common precursor.  In this section
  we select aspects of cofactor chemistry that seem to us most
  essential to overall metabolic architecture and evolution, with the
  goal of framing as much as of answering questions.  As our
  understanding of cofactor chemistry improves through laboratory
  studies, so will our ability to integrate the observations in this
  section into a more complete theory of metabolic architecture and
  evolution.

\subsection{Introduction to cofactors as a group, and why they define
  an essential layer in the control of metabolism}

\subsubsection{Cofactors as a class in extant biochemistry}

The biosynthesis of cofactors involves some of the most elaborate and
least understood organic chemistry used by organisms. The pathways
leading to several major cofactors have only very recently been
elucidated or remain to be fully described, and their study continues
to lead to the discovery of novel reaction mechanisms and enzymes that
are unique to cofactor synthesis~\cite{Graham02,Begley08,Jurgenson09}.
While cofactor biosynthetic pathways often branch from core metabolic
pathways, their novel reactions may produce special bonds and
molecular structures not found elsewhere in metabolism. These novel
bonds and structures are generally central in their catalytic
functions.

Structurally, many cofactors form a class in transition between the
core metabolites and the oligomers.  They contain some of the largest
directly-assembled organic monomers (pterins, flavins, thiamin,
tetrapyrroles), but many also show the beginnings of polymerization of
standard amino acids, lipids or ribonucleotides.  These may be joined
by the same phosphate ester bonds that link RNA oligomers or
aminoacyl-tRNA, or they may use distinctive bonds
(\emph{e.g.}~$5^{\prime}$-$5^{\prime}$ esters) found only in the
cofactor class~\cite{Huang:RNACofactors:00}.

The polymerization exhibited within cofactors is distinguished from
that of oligomers by its heterogeneity.  Srinivasan and
Morowitz~\cite{Srinivasan:aquifex_analysis:09} have termed cofactors
``chimeromers'', because they often include monomeric components from
several molecule classes.  Examples are coenzyme-A, which includes
several peptide units and an ATP; folates, which join a pterin moiety
to para-aminobenzoic acid (PABA); quinones, which join a PABA
derivative to an isoprene lipid tail; and a variety of cofactors
assembled on phosphoribosyl-pyrophosphate (PRPP) to which RNA
``handles'' are esterified.

We may understand the border between small and large molecules, where
most cofactors are found, as more fundamentally a border between the
use of heterogeneous organic chemistry to encode biological
information in covalent structures, and the transition to homogeneous
phosphate chemistry, with information carried in sequences or
higher-order non-covalent structures.  The chemistry of the metabolic
substrate is mostly the chemistry of organic reactions.  Phosphates
and thioesters may appear in intermediates, but their role generally
is to provide energy for leaving groups, enabling formation of the
main structural bonds among C, N, O, and H.  One of the striking
characteristics scales in metabolism is that its organic reactions,
the near-universal mode of construction for molecules of 20 to 30
carbons or less, cease to be used in the synthesis of larger
molecules. Even siderophores, among the most complex of widely-used
organic compounds, are often elaborations of functional centers that
are small core metabolites, such as citrate~\cite{Crosa02,Butler05}.
Large oligomeric macromolecules are almost entirely synthesized using
the dehydration potential of
phosphates~\cite{Westheimer:phosphates:87} to link monomers drawn from
the inventory~\cite{Srinivasan:aquifex_chart:09} of small core
metabolites.  Many cofactors have structure of both kinds, and they
are the smallest molecules that as a class commonly use phosphate
esters as permanent structural elements~\cite{White:coenzymes:76}.

Finally, cofactors are distinguished by structure-function relations
determined mostly at the single-molecule scale.  The monomers that are
incorporated into macromolecules are often distinguished by general
properties, and only take on more specific functional roles that
depend strongly on location and
context~\cite{Petsko:proteins:03,Gutteridge:catalysts:05}.  In
contrast, the functions of cofactors are specific, often finely tuned
by evolution~\cite{Maden:folates_pterins:00}, and deployable in a wide
range of macromolecular contexts.  Usually they are carriers or
transfer agents of functional groups or reductants in intermediary
metabolism~\cite{Fischer:cofactors:10}.  Nearly half of enzymes
require cofactors as
coenzymes~\cite{White:coenzymes:76,Fischer:cofactors:10}. If we extend
this grouping to include chelated
metals~\cite{Andreini:bio_metals:08,Andreini:metal_MACiE:09} and
clusters, ranging from common iron-sulfur centers to the elaborate
metal centers of gas-handling
enzymes~\cite{Fontecilla:metalloenzymes:09,Bender11}, more than half
of enzymes require coenzymes or metals in the active site.

The universal reactions of intermediary metabolism depend on only
about 30 cofactors \cite{Fischer:cofactors:10} (though this number
depends on the specific definition used).  Major functional roles
include 1) transition-metal-mediated redox reactions (heme, cobalamin,
the Nickel tetrapyrrole ${\mbox{F}}_{430}$, chlorophylls\footnote{It
is natural in many respects to include Ferredoxins (and related
flavodoxins) in this list.  Although not cofactors by the criteria of
size and biosynthetic complexity, these small, widely-diversified,
ancient, and general-purpose ${\mbox{Fe}}_2{\mbox{S}}_2$,
${\mbox{Fe}}_3{\mbox{S}}_4$, and ${\mbox{Fe}}_4{\mbox{S}}_4$-binding
polypeptides are unique low-potential (high-energy) electron donors.
Reduced Ferredoxins are often generated in reactions involving radical
intermediates in iron-sulfur enzymes, described below in connection
with electron bifurcation.}), 2) transport of one-carbon groups that
range in redox state from oxidized (biotin for carboxyl groups,
methanofurans for formyl groups) to reduced (lipoic acid for methylene
groups, S-adenosyl methionine, coenzyme-M and cobalamin for methyl
groups), with some cofactors spanning this range and mediating
interconversion of oxidation states (the folate family interconverting
formyl to methyl groups), 3) transport of amino groups (pyridoxal
phosphate, glutamate, glutamine), 4) reductants (nicotinamide
cofactors, flavins, deazaflavins, lipoic acid, and coenzyme-B), 5)
membrane electron transport and temporary storage (quinones), 6)
transport of more complex units such as acyl and amino-acyl groups
(panthetheine in CoA and in the acyl-carrier protein (ACP), lipoic
acid, thiamine pyrophosphate), 7) transport of dehydration potential
from phosphate esters (nucleoside di- and tri-phosphates), and 8)
sources of thioester bonds for substrate-level phosphorylation and
other reactions (panthetheine in CoA).

\subsubsection{Roles as controllers, and consequences for the
  emergence and early evolution of life}

Cofactors fill roles in network or molecular catalysis below the level
of enzymes, but they share with all catalysts the property that they
are not consumed by participating in reactions, and therefore are key
loci of control over metabolism.  Cofactors as transfer agents are
essential to completing many network-catalytic loops.  In association
with enzymes, they can create channels and active sites, and thus they
facilitate molecular catalysis.  An example of the creation of
channels by cofactors is given by the function of cobalamin as a
${\mbox{C}}_1$ transfer agent to the Nickel reaction center in the
acetyl-CoA synthase from a corrinoid iron-sulfur
protein~\cite{Ragsdale:ACS_enzymes:91,Ragsdale:CODH_ACS:96,%
Ragsdale:acetogenesis:08}.  An example of cofactor incorporation in
active sites is the role of TPP as the reaction center in the
pyruvate-ferredoxin oxidoreductase (PFOR), which lies at the end of a
long electron-transport channel formed by Fe-S
clusters~\cite{Chabriere:oxidoreductases:99}.  Through the limits in
their own functions or in the functional groups they transport through
networks, they may impose constraints on chemical diversity or create
bottlenecks to evolutionary innovation.  The previous sections have
shown that many module boundaries in carbon fixation and core
metabolism are defined by idiosyncratic reactions, and we have noted
that many of these idiosyncrasies are associated with specific
cofactor functions.

Cofactors, as topological hubs, and participants in reactions at
high-flux boundaries in core and intermediary metabolism, are focal
points of natural selection.  The adaptations available to key atoms
and bonds include altering charge or pKa, changing energy level
spacing through non-local electron transport, or altering orbital
geometry through ring strains.  Divergences in low-level cofactor
chemistry may alter the distribution of functional groups and thereby
change the global topology of metabolic networks, and some of these
changes map onto deep lineage divergences in the tree of life. A
well-understood example is the repartitioning of ${\mbox{C}}_1$ flux
from methanopterins versus
folates~\cite{Maden:folates_pterins:00,Braakman:carbon_fixation:12}.
The same adaptation that enables formylation of methanopterins within
an exclusively thioester system, where the homologous folate reaction
requires ATP, reduces the potential for methylene-group transfer, and
necessitates the oxidative formation of serine from 3PG in
methanogens, which is not required of acetogens.

Most research on the origin of life has focused either on the
metabolic substrate~\cite{Morowitz:BCL:92,Fry:ELE:00} or catalysis by
RNA~\cite{Gesteland:RNAWorld:06}, but we believe the priority of
cofactors deserves (and is beginning to receive) greater
consideration~\cite{Copley:PMRNA:06,Yarus:pre_RNA:11}.  In the
expansion of metabolic substrates from inorganic inputs, the pathways
to produce even such complex cofactors as folates \emph{et alia} are
comparable in position and complexity to those for purine RNA, while
some for functional groups such as nicotinamide~\cite{Copley:PMRNA:06}
or chorismate are considerably simpler.  Therefore, even though it is
not known what catalytic support or memory mechanisms enabled the
initial elaboration of metabolism, any solutions to this problem
should also support the early emergence of at least the major redox
and C- and N-transfer cofactors.  Conversely, the pervasive dependence
of biosynthetic reactions on cofactor intermediates makes the
expansion of protometabolic networks most plausible if it was
supported by contemporaneous emergence and elaboration of cofactor
groups.  In this interpretation cofactors occupy an intermediate
position in chemistry and complexity, between the small-metabolite and
oligomer levels~\cite{Copley:PMRNA:06}. They were the transitional
phase when the reaction mechanisms of core metabolism came under
selection and control of organic as opposed to mineral-based
chemistry, and they provided the structured foundation from which the
oligomer world grew.

  We argue next that a few properties of the elements have governed
  both functional diversification and evolutionary optimization of
  many cofactors, especially those associated with core
  carbon-fixation.  We focus on heterocycles with conjugated double
  bonds incorporating nitrogen, and on the groups of functions that
  exploit special properties of bonds to sulfur atoms.  The
  recruitment of elements or special small-molecule contexts
  constitutes an additional distinct form of modularity within
  metabolism.  Like the substrate network, cofactor groups often share
  or re-use synthetic reaction sequences.  However, unlike the
  small-molecule network, cofactors can also be grouped by criteria of
  catalytic similarity that are independent of pathway recapitulation.
  For example, alkyl-thiol cofactors, which comprise diverse groups of
  molecules, all make essential use of distinctive properties of the
  sulfur bond to carbon, which appear nowhere else in
  biochemistry.  As an example involving elements in specific
  contexts, a large group of cofactors employing C-N heterocycles all
  arise from a single sub-network whose reactions are catalyzed by
  related enzymes, and the transport and catalytic functions performed
  by the heterocycles are distinctive of this cofactor group.

\subsection{The cofactors derived from purine RNA}
\label{sec:purine_cofactors}

Most of the cofactors that use heterocycles for their primary
functions have biosynthetic reactions closely related to those for
purine RNA.  These reactions are performed by a diverse class of
cyclohydrolase enzymes, which are responsible for the key
ring-formation and ring-rearrangement steps.  The cyclohydrolases can
split and reform the ribosyl ring in PRPP, jointly with the 5- and
6-membered rings of guanine and adenine.  Five biosynthetically
related cofactor groups are formed in this way.  Three of these -- the
folates, flavins and deazaflavins -- are formed from GTP, while one --
thiamin -- is formed from a direct precursor to GTP, as shown in
Fig.~\ref{fig:Purine_Cofactors}.

{\bf \noindent Folates:} The folates are structurally most similar to
GTP, but have undergone the widest range of secondary specializations,
particularly in the Archaea.  They are primarily responsible for
binding ${\mbox{C}}_1$ groups during reduction from formyl to
methylene or methyl oxidation states, and their secondary
diversifications are apparently results of selection to tune the
free-energy landscape of these oxidation states.

{\bf \noindent Flavins and deazaflavins:} The flavins are tricyclic
compounds formed by condensation of two pterin groups, while
deazaflavins are synthesized through a modified version of this
pathway, in which one pterin group is replaced by a benzene ring
derived from chorismate.  Flavins are general-purpose reductants,
while deazaflavins are specifically associated with methanogenesis.

{\bf \noindent Thiamin:} Thiamine combines a C-N heterocycle common to
the GTP-derived cofactors with a thiazole group (so incorporating
sulfur), and shares functions with both the purine cofactor group and
the alkyl-thiol group reviewed in the next subsection.

\begin{widetext}

\begin{center}
\begin{figure}[ht]
  \begin{center} 
  \includegraphics[scale=0.6]{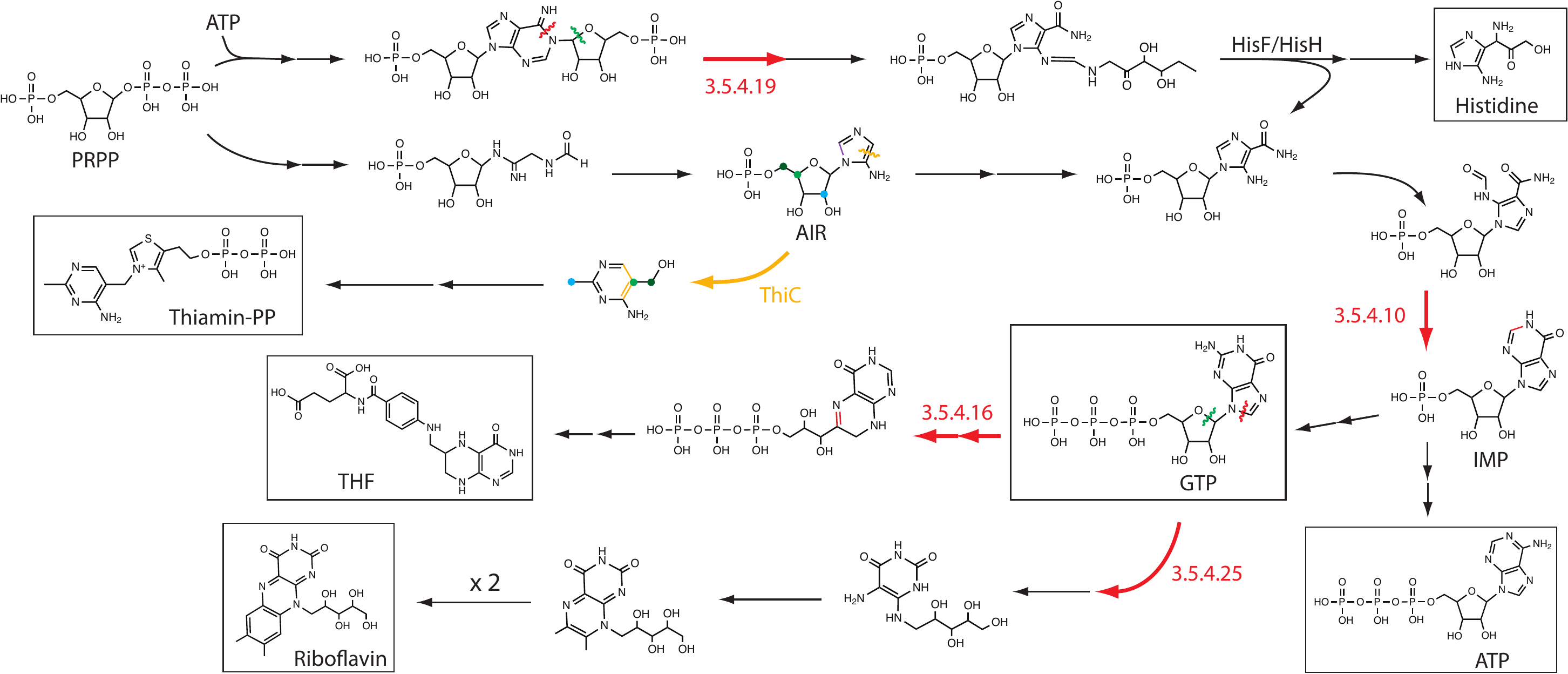}
  \caption{
  Key molecular re-arrangements in the network leading from AIR to
  purines and the purine-derived cofactors.  The 3.5.4 class of
  cyclohydrolases (red) convert FAICAR to IMP (precursor to purines),
  and subsequently convert GTP to folates and flavins by opening the
  imidazole ring.  Acting on the 6-member ring of ATP and on a second
  attached PRPP, the enzyme 3.5.4.19 initiates the pathway to
  histidinol.  The thiamine pathway, which uses the unclassified
  enzyme ThiC to hydrolize imidazole and ribosyl moieties, is the most
  complex, involving multiple group rearrangements (indicated by
  colored atoms).  This complexity, together with the subsequent
  attachment of a thiazole group, lead us to place thiamine latest in
  evolutionary origin among these cofactors.
    \label{fig:Purine_Cofactors} 
  }
  \end{center}
\end{figure}
\end{center}

\end{widetext}

{\bf \noindent Histidine:} The last ``cofactor'' in this group is the
amino acid histidine, synthesized from ATP rather than GTP but using
similar reactions.  Histidine is a general acid-base catalyst with
unique pKa, which in many ways functions as a ``cofactor in amino acid
form''~\cite{Srinivasan:aquifex_analysis:09}.

We will first describe in detail the remarkable role of the folate
group in the evolutionary diversification of the Wood-Ljungdahl
pathway, and then return to general patterns found among the
purine-derived cofactors, and their placement within the elaboration
of metabolism and RNA chemistry.

\subsubsection{Folates and the central superhighway of ${\mbox{C}}_1$
  metabolism}

\begin{widetext}

\begin{figure}[ht]
  \begin{center} 
  \includegraphics[scale=0.6]{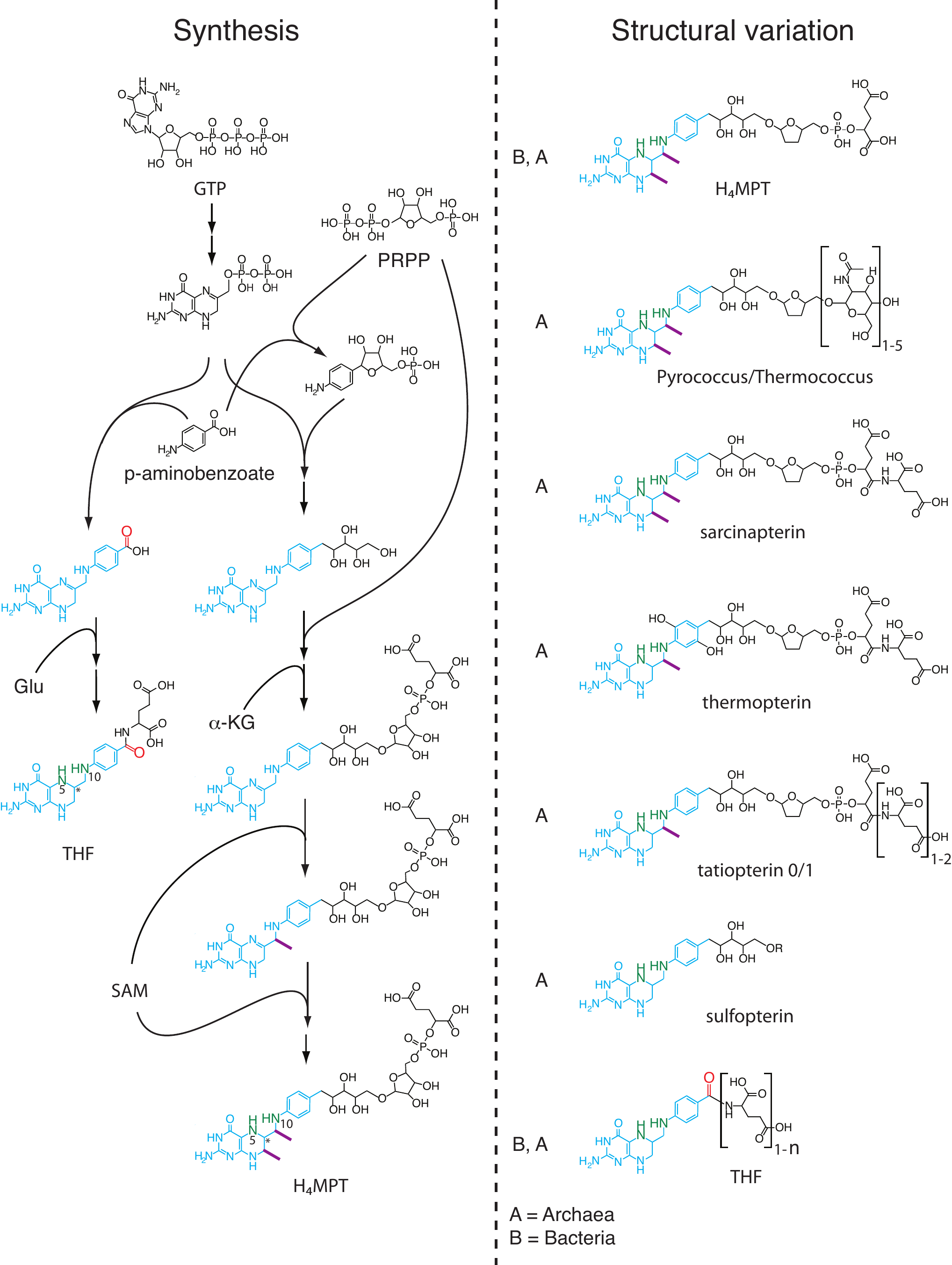}
  \caption{
  Structural variants among cofactors in the folate family, shown with the biosynthetic pathways that produce these variations.  Pteridine and benzene groups are shown in blue, active nitrogens are shown in green, electron-withdrawing carbonyl groups are shown in red, and methyl groups that regulate steric hindrance are shown in purple.  
    \label{fig:folate_variants} 
  }
  \end{center}
\end{figure}

\end{widetext}

Members of the folate family carry ${\mbox{C}}_1$ groups bound to
either the ${\mbox{N}}^5$ nitrogen of a heterocycle derived from GTP,
an exocyclic ${\mbox{N}}^{10}$ nitrogen derived from a
para-aminobenzoic acid (PABA), or both.  The two most common folates
are tetrahydrofolate (THF), ubiquitous in bacteria and common in many
archaeal groups, and tetrahydromethanopterin (${\mbox{H}}_4
\mbox{MPT}$), essential for methanogens and found in a small number of
late-branching bacterial clades.  Other members of this family are
exclusive to the archaeal domain and are structural intermediates
between THF and ${\mbox{H}}_4 \mbox{MPT}$.  Two kinds of structural
variation are found among folates, as shown in
Fig.~\ref{fig:folate_variants}.  First, only THF retains the carbonyl
group of PABA, which shifts electron density away from
${\mbox{N}}^{10}$ \emph{via} the benzene ring, and lowers its pKa
relative to ${\mbox{N}}^5$ of the heterocycle.  All other members of
the family lack this carbonyl.  Second, all folates besides THF
incorporate one or two methyl groups that impede rotation between the
pteridine and aryl-amine planes, changing the relative entropies of
formation among different binding states for the attached
${\mbox{C}}_1$~\cite{MacKenzie:folates:84,Maden:folates_pterins:00,%
  Braakman:carbon_fixation:12}.

Folates mediate a diverse array of ${\mbox{C}}_1$ chemistry, various
parts of which are essential in the biosynthesis of all
organisms~\cite{Maden:folates_pterins:00}.  The collection of
reactions, summarized in Fig.~\ref{fig:C1_metabolism}, has been termed
the ``central superhighway'' of one-carbon metabolism.  Functional
groups supplied by folate chemistry, connected by interconversion of
${\mbox{C}}_1$-oxidation states along the superhighway, include 1)
formyl groups for synthesis of purines, formyl-tRNA, and formylation
of methionine (fMet) during translation, 2) methylene groups to form
thymidilate, which are also used in many deep-branching organisms to
synthesize glycine and serine, forming the ancestral pathway to these
amino acids \cite{Braakman:carbon_fixation:12}, and 3) methyl groups
which may be transferred to S-adenosyl-methionine (SAM) as a general
methyl donor in anabolism, to the acetyl-CoA synthase to form
acetyl-CoA in the Wood-Ljungdahl pathway, or to coenzyme-M where the
conversion to methane is the last step in the energy system of
methanogenesis.

The variations among folates, shown in Fig.~\ref{fig:folate_variants},
leave the charge, pKa and resulting C-N bond energy at ${\mbox{N}}^5$
roughly unaffected, while the the ${\mbox{N}}^{10}$ charge, pKa, and
C-N bond energy change significantly across the family.  This charge
effect, together with entropic effects due to steric hindrance from
methyl groups, can sharply vary the functional roles that different
folates play in anabolism.

\vfill
\eject

The biggest difference lies between THF and ${\mbox{H}}_4 \mbox{MPT}$.
In THF, the ${\mbox{N}}^{10}$ pKa is as much as 6.0 natural-log units
lower than that of ${\mbox{N}}^5$~\cite{Kallen:THF_pKa:66}. The
resulting higher-energy C-N bond cannot be formed without hydrolysis
of one ATP, either to bind formate to ${\mbox{N}}^{10}$ of THF, or to
cyclize ${\mbox{N}}^5$-formyl-THF to form
${\mbox{N}}^5$,${\mbox{N}}^{10}$-methenyl-THF (see
Fig.~\ref{fig:C1_metabolism}).  This latter reaction is the mirror
image of the cyclization of ${\mbox{N}}^{10}$-formyl-THF, and as we
will argue below a plausibly conserved evolutionary intermediate in
the attachment of formate onto folates.  After further reduction, the
resulting methylene is readily transferred to lipoic acid to form
glycine and serine, in what we have termed the ``glycine
cycle''~\cite{Braakman:carbon_fixation:12} (the lipoyl-protein based
cycle on the right in Fig.~\ref{fig:C1_metabolism}.

In contrast, in ${\mbox{H}}_4 \mbox{MPT}$ the difference in pKa
between ${\mbox{N}}^{10}$ and ${\mbox{N}}^5$ is only 2.4 natural-log
units.  The lower ${\mbox{C-N}}^{10}$ bond energy permits spontaneous
cyclization of ${\mbox{N}}^5$-formyl-${\mbox{H}}_4 \mbox{MPT}$,
following (also ATP-independent) transfer of formate from a
formyl-methanofuran cofactor.  Through this sequence, methanogens fix
formate in an ATP-independent system using only redox chemistry.  The
initial free energy to attach formate to methanofuran is provided by
the terminal methane released in methanogenesis (the Co-M/Co-B cycle
in Fig.~\ref{fig:C1_metabolism}).  The resulting downstream methylene
group, however, has too little energy as a leaving group to transfer
to an alkyl-thiol cofactor, so methanogens sacrifice the ability to
form glycine and serine by direct reduction of formate.

The reconstructed ancestral use of the 7-9 reactions in
Fig.~\ref{fig:C1_metabolism} is to reduce formate to acetyl-CoA or
methane.  However, the reversibility of many reactions in the
sequence, possibly requiring substitution of reductant/oxidant
cofactors, allows folates to accept and donate ${\mbox{C}}_1$ groups
in a variety of oxidation states, from and into many pathways
including salvage pathways.  Methylotrophic proteobacteria which have
obtained ${\mbox{H}}_4 \mbox{MPT}$ through horizontal gene
transfer~\cite{Vorholt:methylotrophy:99,Chistoserdova:methylotrophy:09}
may run the full reaction sequence in reverse. They may use either
${\mbox{H}}_4 \mbox{MPT}$ to oxidize formaldehyde or THF to oxidize
various methylated ${\mbox{C}}_1$ compounds, in both cases leading to
formate, or other intermediary oxidation states (from THF) as inputs
to anabolic pathways. In many late-branching bacteria, some archaea,
and eukaryotes, the THF based pathway may run in part oxidatively and
in part reductively, through connections to either
gluconeogenesis/glycolysis or glyoxylate metabolism.  In these
organisms serine (derived through oxidation, amination and
dephosporylation from 3-phosphoglycerate) or glycine (derived through
amination of glyoxylate) become the sources of transferable methyl
groups in anabolism.  This versatility has preserved the folate
pathway as an essential module of biosynthesis in all domains of life,
and at the same time has made it a pivot of evolutionary variation.

\subsubsection{Refinement of folate-${\mbox{C}}_1$ chemistry maps onto
  lineage divergence of methanogens}

The structural and functional variation within the folate family
illustrates the way that selection, acting on cofactors, can create
large-scale re-arrangements in metabolism, enabling adaptations that
are reflected in lineage divergences.  The free-energy cascade
described in the last section, linking ATP hydrolysis, the charge and
pKa of the ${\mbox{N}}^{10}$ nitrogen, and the leaving-group activity
of the resulting bound carbon for transfer to alkyl-thiol cofactors or
other anabolic pathways, is a fundamental long-range constraint of
folate-${\mbox{C}}_1$ chemistry.  A comparative analysis of gene
profiles in pathways for glycine and serine synthesis, explained in
Ref.~\cite{Braakman:carbon_fixation:12}, shows that while the
constraint cannot be overcome, its impact on the form of metabolism
can vary widely depending on the structure of the mediating folate
cofactor.

The annotated role for ATP hydrolysis in WL autotrophs is to attach
formate to ${\mbox{N}}^{10}$ of THF, initiating the reduction
sequence.  However, many deep-branching bacteria and archaea show no
gene for this reaction, while multiple lines of evidence indicate that
THF nonetheless functions as a carbon-fixation cofactor in these
organisms~\cite{Braakman:carbon_fixation:12}.  In almost all cases
where an ATP-dependent ${\mbox{N}}^{10}$-formyl-THF synthase is
absent, an ATP-dependent ${\mbox{N}}^5$-formyl-THF cycloligase
\cite{Stover93,Huang95} is found. This is another case where a broad
evolutionary context allows an alternate interpretation.
${\mbox{N}}^5$-formyl-THF cycloligase was originally discovered in
mammalian systems, where its function has been highly uncertain and
hypothesized to be the salvage mechanism as part of a futile
cycle~\cite{Stover93,Huang95}, before being found to be widespread
across the tree of life~\cite{Braakman:carbon_fixation:12}. If we
deduce by reconstruction, however, that ancestral folate chemistry
operated in the fully reductive direction, and that in ${\mbox{H}}_4
\mbox{MPT}$ systems formate is attached at the ${\mbox{N}}^5$
position, while in THF systems formate is attached at the
${\mbox{N}}^{10}$ position, the widespread distribution of the
cycloligase takes on a different possible meaning. It is plausible
that the ${\mbox{N}}^5$-formyl-THF cycloligase allows a formate
incorporation pathway that is an evolutionary intermediate between the
commonly recognized pathway using THF and its evolutionary derivative
using ${\mbox{H}}_4 \mbox{MPT}$ (see Fig.~\ref{fig:C1_metabolism}).
The ATP-dependent cycloligase produces
${\mbox{N}}^5$,${\mbox{N}}^{10}$-methenyl-THF from
${\mbox{N}}^5$-formyl-THF, which may potentially form spontaneously
due to the higher ${\mbox{N}}^5$-pKa~\cite{Huang95}.  ATP hydrolysis
is thus specifically linked to the ${\mbox{N}}^{10}$-carbon bond which
is the primary donor for carbon groups from folates.  Methanogens, in
contrast, escape the dependence on ATP hydrolysis by decarboxylating
PABA before it is linked to pteridine to form methanopterin (see
Fig.~\ref{fig:folate_variants}), but they sacrifice methyl-group
donation from ${\mbox{H}}_4 \mbox{MPT}$ to most anabolic pathways,
making methanogenesis viable only in clades that evolved the oxidative
pathway to serine from 3-phosphoglycerate.

We noted in Sec.~\ref{sec:evol_history} that the elimination of one
ATP-dependent acyl-CoA synthase in acetogens reduces the free energy
cost of carbon fixation relative to rTCA cycling.  The decoupling of
the formate-fixation step on methanopterins from ATP hydrolysis is a
further significant innovation, lowering the ATP cost for uptake of
${\mbox{CO}}_2$.  This divergence of ${\mbox{H}}_4 \mbox{MPT}$ from
THF, and a related divergence of deazaflavins from flavins (see
Fig.~\ref{fig:Pterins_Flavins}), follow phylogenetically (and we
believe, were responsible for) the divergence of the methanogens from
other euryarcheota~\cite{Braakman:carbon_fixation:12}.

We regard this example as representative of the way that innovations
in cofactor chemistry more generally mediated large-scale
rearrangements in metabolism, and corresponding evolutionary (and
ecological) divergences of clades. Another similar example comes from
the quinones, a diverse family of cofactors mediating membrane
electron transport~\cite{Collins81}.
Ref.~\cite{Schoepp:menaquinones:09} found that the synthetic
divergence of mena- and ubiquinone follows the pattern of phylogenetic
diversification within proteobacteria. $\delta$- and
$\epsilon$-proteobacteria use menaquinone, $\gamma$-proteobacteria use
both mena- and ubiquinone, and $\alpha$- and $\beta$-proteobacteria
use only ubiquinone. Because mena- and ubiquinone have different
midpoint potentials, it was suggested that their distribution reflects
changes in environmental redox state as the proteobacteria diversified
during the rise of oxygen~\cite{Schoepp:menaquinones:09,Nitschke95}.

\subsubsection{Relation of the organic superhighway to minerals}

An interpretive frame for many of these observations is the proposal
that metabolism is an outgrowth of
geochemistry~\cite{Russell:FeS:06,Russell:AcetylCoA:04,Morowitz:EFOoL:07},
which came under the control of living
organisms~\cite{Smith:auto_hetero:10} (see Sec.~\ref{sec:origins} for
dedicated discussion).  If we wish to judge this proposal, then it is
informative to look for parallels and differences between biochemical
and plausible geochemical reaction sequences.  The distinctive
features of biochemical ${\mbox{C}}_1$ reduction are the attachment of
formate to tuned heterocyclic or aryl-amine nitrogen atoms for
reduction, and the transfer of reduced ${\mbox{C}}_1$ groups to
sulfhydryl groups (of SAM, lipoic acid, or CoM).  In the
mineral-origin hypothesis for direct reduction, the ${\mbox{C}}_1$
were adsorbed at metals and either reduced through crystal
oxidation~\cite{Wachtershauser:FeS_world:92} or by reductant in
solution.  The transfer of reduced ${\mbox{C}}_1$ groups to
alkyl-thiol cofactors may show continuity with reduction on
metal-sulfide minerals.  However, the mediation of reduction by
nitrogens appears to be a distinctively biochemical innovation.

\subsubsection{Cyclohydrolases as the central enzymes in the family,
  and the resulting structural homologies among cofactors}

The common reaction mechanism unifying the purine-derived cofactors is
an initial hydrolysis of both purine and ribose rings performed by
cyclohydrolases assigned EC numbers 3.5.4 (see
Fig.~\ref{fig:Purine_Cofactors}).  All cyclohydrolases within this EC
family are used for biosynthesis or conversions within this class of
molecules. They are responsible for the synthesis of
inosine-monophosphate (IMP, precursor to AMP and GMP) from
5-formamidoimidazole-4-carboxamide ribonucleotide (FAICAR), for the
first committed steps in the syntheses of both folates and flavins
from GTP, and for the initial ring-opening step in the synthesis of
Histidine from ATP and PRPP.  Fig.~\ref{fig:Purine_Cofactors} shows
the key steps in the network synthesizing both purines and the
pterins, folates, flavins, thiamine, and histidine.

The common function of the 3.5.4 cyclohydrolases is hydrolysis of
rings on adjacent nucleobase and ribose groups, or the formation of
cycles by ligation of ring fragments.  In all cases, the ribosyl
moieties come from phosphoribosyl-pyrophosphate (PRPP).  In the
synthesis of pterins from GTP and of histidinol from ATP, both a
nucleobase cycle and a ribose are cleaved.  In pterin synthesis, the
imidazole of guanine and the purine ribose are cleaved.  In histidine
synthesis, the six-membered ring of adenine is cleaved (at a different
bond than the one synthesized from FAICAR), and the ribose comes from
a secondary PRPP.

By far the most complex synthesis in this family is that of thiamin
from aminoimidazole ribonucleotide (AIR). This sequence begins with an
elaborate molecular rearrangement, performed in a single step by the
enzyme ThiC~\cite{Jurgenson09}. (Eukaryotes use an entirely different
pathway, in which the pyrimidine is synthesized from histidine and
pyridoxal-5-phosphate~\cite{Tazuya95}.) While the ThiC enzyme is
unclassified, and its reaction mechanism incompletely understood, it
shares apparent characteristics with members of the 3.5.4
cyclohydrolases. As in the first committed steps in the synthesis of
folates and flavins from GTP, both a ribose ring and a 5-member
heterocycle are cleaved and subsequently (as in folate synthesis)
recombined into a 6-member heterocycle. The complexity of this
enzymatic mechanism makes a pre-enzymatic homologue to ThiC difficult
to imagine, and suggests that thiamin is both of later origin, and
more highly derived, than other cofactors in this family. This derived
status is supported by the fact that the resulting functional role of
thiamin is not performed on the pyrimidine ring itself, but rather on
the thiazole ring to which it is attached, and which is likewise
created in an elaborate synthetic sequence~\cite{Jurgenson09}.  

\begin{figure}[ht]
  \begin{center} 
  \includegraphics[scale=0.45]{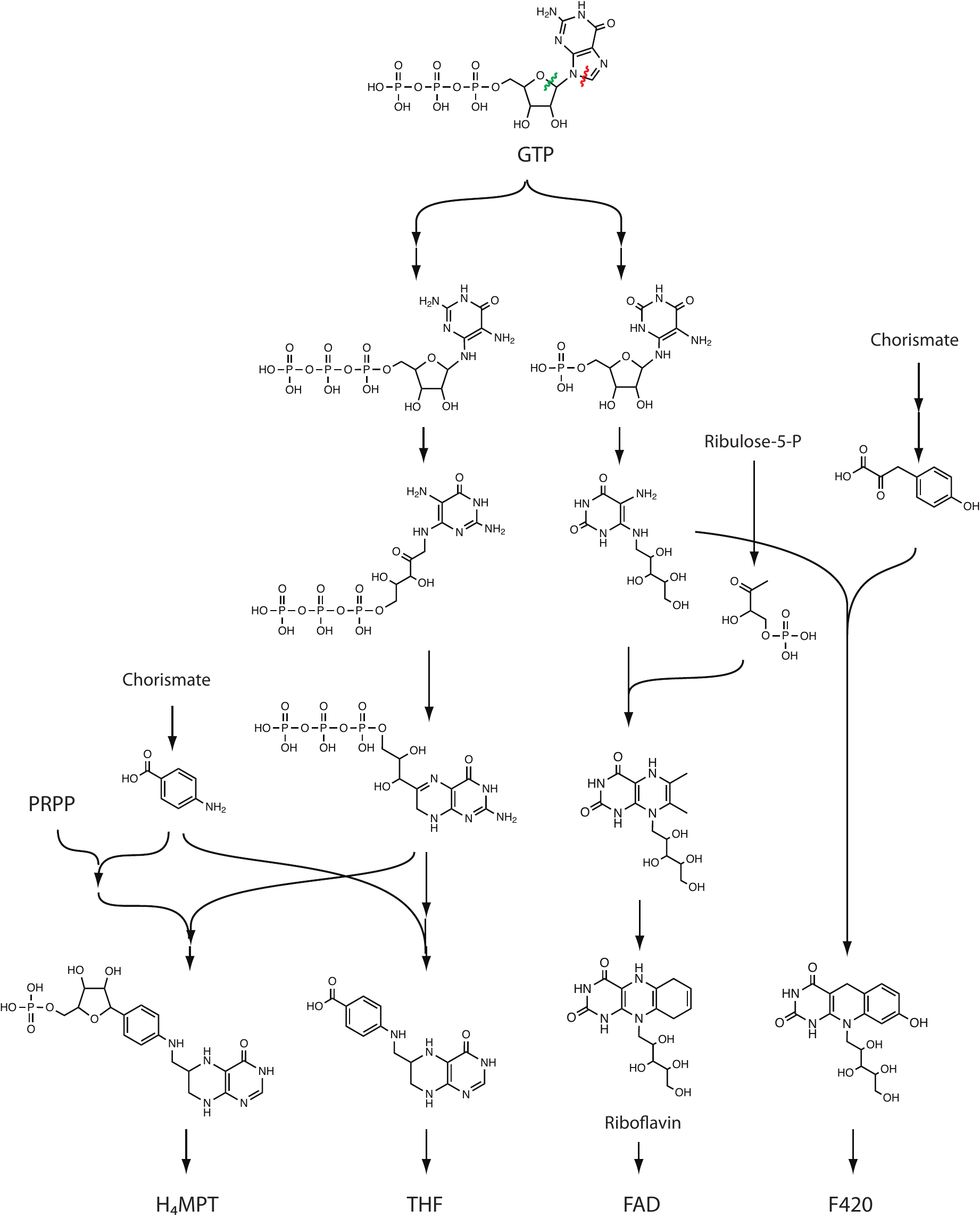}
  \caption{
  The substrate modifications leading from GTP to the four major
  cofactors ${\mbox{H}}_4 \mbox{MPT}$, THF, riboflavin (in FAD) and
  the archaeal homologue deazaflavin ${\mbox{F}}_{420}$.  The branches
  indicating substrate diversification may also reflect an
  evolutionary lineage.
    \label{fig:Pterins_Flavins} 
  }
  \end{center}
\end{figure}

Fig.~\ref{fig:Pterins_Flavins} shows the detailed substrate
re-arrangement in the sub-network leading from GTP to methanopterins,
folates, riboflavin, and the archaeal deazaflavin ${\mbox{F}}_{420}$.
In the pterin branch, both rings of neopterin are synthesized directly
from GTP, and an aryl-amine originating in PABA provides the second
essential nitrogen atom.  PABA is either used directly (in folates) or
decarboxylated with attachment of a PRPP (in methanopterins) to vary
the pKa of the amine.  In contrast, the flavin branch is characterized
by the integration of either ribulose (in riboflavin) or chorismate
(in ${\mbox{F}}_{420}$) to form the internal rings.  Two
6,7-dimethyl-8-(D-ribityl)lumazine are condensed to form riboflavin,
whereas a single GTP with chorismate forms ${\mbox{F}}_{420}$.

The cyclohydrolase reactions can be considered the key innovation
enabling the biosynthesis of this whole family of cofactors, and
importantly, of purine RNA itself.  The heterocycles that are formed
or cleaved by these reactions provide the central structural
components of the active parts of the final cofactor molecules. In
this sense, except for TPP, the distinctions among purine-derived
cofactors can be considered secondary modifications on a background
structured by PRPP and C-N heterocycles. If we consider sub-networks
of metabolism as producing key structural or functional components, in
this case for the synthesis of cofactors, then this family draws on
only two such developed sub-networks. The first of these is purine
synthesis, and the other is synthesis of chorismate, the precursor to
PABA and the unique source of single benzene rings in
biochemistry~\cite{Bentley90}.  Flexibility in the ways that
chorismate is modified to control electron density, and the way the
benzene ring is combined with other heterocycles, contributes to the
combinatorial elaboration within the family.

\subsubsection{Placing the members of the class within the network
  expansion of metabolism}

The following observations suggest to us the possibility that most of
the purine-derived cofactors (perhaps excepting thiamin) were
available contemporaneously with monomer purine RNA.

For some reactions, the abstraction of enzyme mechanisms is advanced
enough to identify small-molecule organocatalysts that could have
provided similar
functions~\cite{Barbas:organocat:08,MacMillan:organocat:08}. The
current understanding of cyclohydrolase proteins, however, does not
suggest other simpler mechanisms by which similar reactions might
first have been catalyzed, leaving us almost wholly uncertain about
how RNA was first formed. Unless non-enzymatic mechanisms are
discovered which are both plausible and \emph{selective}, our previous
arguments about the permissiveness of crude catalysts lead us to
expect that, at whatever stage catalysts capable of interconverting
AIR, AICAR, FAICAR, and IMP first became available, pteridines would
have been formed contemporaneously and possibly played a role in the
elaboration of the metabolic network.  (See Sec.~\ref{sec:innovation}
for further discussion on promiscuous vs. selective enzymes.)  If the chorismate pathway (which begins in
the sugar-phosphate network) had also arisen by that stage, the same
arguments suggest that folates and flavins may also have been
available.  In this supposition we are treating the first three EC
numbers as an appropriate guide to reaction mechanism without
restriction of the molecular substrate.  Whether the first RNA were
produced in this way, or through structurally very dissimilar stages,
is a currently active question~\cite{Powner:RNA_denovo:09}.

As in our discussion of the root node in
Sec.~\ref{sec:character_root}, we consider it important to apply
\emph{ubiquitously} the premise that enabling network throughput and
pruning network diversity were concurrent ongoing requirements in the
co-evolution of substrate reactions and their catalysts.
Most-often~\cite{Ricardo:borate:04,Benner:RNA:06,Orgel:cycles:08} the
inability to prune networks is recognized as a problem for the early
formation of order.  In the case of the purine-derived cofactors, it
it may offer both clues to help explain the structure of the
biosynthetic network, and a way to break down the problem of early
metabolic evolution into simpler steps with intermediate criteria for
selection.

The patterns that characterize current metabolism as a
recursive network
expansion~\cite{Handorf:met_expand:05,Raymond:oxy_networks:06} about
inorganic inputs are most easily understood as a reflection of the
organic-chemical possibilities opened by cofactors.  Pterins, as
donors of activated formyl groups, support (among other reactions) the
synthesis of purines, forming a short autocatalytic loop.  Similarly,
flavins would have augmented redox reactions.  Finally, it has long
been recognized that acid/base catalysis is uniquely served by
histidine, which has a $\mbox{pKa} \approx 6.5$ on the
$\varepsilon$-nitrogen, a property not found among any biological
ribonucleotides (though possible for some substituted adenine
derivatives)~\cite{Decout:RNA_cat:93}.

Within the class of GTP-derived cofactors, a sub-structure may perhaps
be suggested: the dimer condensation that forms riboflavin is a
hierarchical use of building blocks formed from GTP.  Although simple
and consisting of a single key reaction, this could reflect a later
stage of refinement.  It is recognized~\cite{Stryer:BC:81} that
flavins are somewhat specialized reductants, both biosynthetically and
functionally more specific than the much simpler nicotinamide
cofactors, which plausibly preceded them~\cite{Copley:PMRNA:06}.

\subsubsection{Purine-derived cofactors selected before RNA itself, as
  opposed to having descended from an RNA world defined through base
  pairing?}

The overlap between RNA and cofactor biosynthesis, and the
incorporation of AMP in several cofactors (where is serves primarily
as a ``handle'' for docking), has been noticed and given the
interpretation that cofactors are a degenerated relic of an
\emph{oligomer} RNA world~\cite{White:coenzymes:76}.  While
\emph{monomer} RNA is of comparable complexity to small-molecule
cofactors, oligomer RNA is significantly more complex. The only
significant logical motivation to place oligomer RNA prior to
small-molecule cofactors, is therefore the premise that RNA base
pairing and replication is the \emph{least}-complex plausible
mechanism supporting (specifically, Darwinian) selection and
persistence of catalysts that are hypothesized to have been required
for the elaboration of biosynthesis.

This is still a complex premise, however, as it requires not only
organosynthesis of oligomer RNA, but also chiral selection and
mechanisms to enable base pairing and (presumably template-directed)
ligation~\cite{Lincoln:RNA_evol:09}. A particular problem for RNA
replication is the steric restriction to $3^{\prime}$-$5^{\prime}$
phosphate esters, over the kinetically favored
$2^{\prime}$-$5^{\prime}$ linkage.  In comparison, small-molecule
catalysis by either RNA~\cite{Copley:codes:05} or related cofactors
may be considered in any context that supports their
synthesis.\footnote{The relative importance of synthesis and selection
depends on whether opening access to a space of reactions, or
concentrating flux within a few channels in that space, is the primary
limit on the emergence of order at each phase in the elaboration of
metabolism.  Following our earlier arguments about the need for
autocatalysis, selection will be essential in some stages, and this
remains an important problem for metabolism-first
premises~\cite{Orgel:cycles:08}.  Chemical selection criteria derived
from differential growth rate pose no problem in the domain of
small-molecule organocatalysis, but the identification of plausible
mechanisms to preserve selected differences remains an important area
of work.  Most mechanisms that do not derive from RNA base pairing
involve separation by spatial geometry or material phases, including
porous-medium processes akin to invasion
percolation~\cite{Martin:OrigCells:03}, or more general proposals for
compositional
inheritance~\cite{Segre:StatChem:99,Segre:CompGenomes:00,%
Segre:CompInherit:01}, abstracted from models of coascervate
chemistry.}  If chemical mechanisms are found which support structured
organosynthesis and selection -- a requirement for any
metabolism-first theory of the origin of life -- the default premise
may favor simplicity: that heterocycles were first selected as
cofactors, and that purine RNA, only one among many species maintained
by the same generalized reactions, was subsequently selected for
chirality, base-pairing, and ligation.

\subsection{The alkyl-thiol cofactors}

The major chemicals in this class include the sulfonated alkane-thiols
coenzyme-B (CoB) and coenzyme-M (CoM), cysteine and homocysteine
including the activated forms S-adenosyl-homocysteine (which under
methylation becomes SAM), lipoic acid, and pantetheine or pantothenic
acid, including pantetheine-phosphate.  The common structure of the
alkyl-thiol cofactors is an alkane chain terminated by one or more
sulfhydryl (SH) groups.  In all cases except lipoic acid, a single SH
is bound to the terminal carbon; in lipoic acid two SH groups are
bound at sub-adjacent carbons.  Differences among the alkyl-thiol
cofactors arise from their biosynthetic context, the length of their
alkane chains, and perhaps foremost the functional groups that
terminate the other ends of the chains.  These may be as simple as
sulfones (in CoB) or as complex as peptide bonds (in CoA).

Cofactors in this class serve three primary functions, as reductants
(cysteine, CoB, pantetheine, and one sulfur on lipoic acid), carriers
of methyl groups (CoM, SAM, one sulfur on lipoic acid), and carriers
of larger functional groups such as acyl groups (lipoic acid in lipoyl
protein, phosphopantotheine in acyl-carrier protein).  A highly
specialized role in which H is a leaving group is the formation of
thioesters at carboxyl groups (pantethenic acid in CoA, lipoic acid in
lipoyl protein) This function is essential to substrate-level
phosphorylation~\cite{deDuve:BC:91}, and appears repeatedly in the
deepest and putatively oldest reactions in core metabolism.  A final
function closely related to reduction is the formation and cleavage of
$\mbox{S}-\mbox{S}$ linkages by cysteine in response to redox state,
which is a major controller of both committed and plastic tertiary
structure in proteins.  The sulfur atoms on cysteine often form
coordinate bonds to metals in metallo-enzymes, a function that we may
associate with protein ligands, in contrast to the more common
nitrogen atoms that coordinate metals in pyrrole cofactors.

The properties of the alkyl-thiol cofactors derive largely from the
properties of sulfur, which is a ``soft'' period-3
element~\cite{Gray:chemistry:94} that forms relatively unstable
(usually termed ``high-energy'') bonds with the hard period-2 element
carbon.  For the alkyl-thiol cofactors in which sulfur plays direct
chemical roles, three main bonds dictate their chemistry:
$\mbox{S}-\mbox{C}$, $\mbox{S}-\mbox{S}$, and $\mbox{S}-\mbox{H}$.
Sulfur can also exist in a wide range of oxidation states, and for
this reason often plays an important role in energy
metabolism~\cite{Wald:periods:62}, particularly for chemotrophs, and
due to its versatility has been suggested to precede oxygen in
photosynthesis~\cite{Hohmann:photosynth_rev:11}.  The electronic
versatility of sulfur and the high-energy $\mbox{C}-\mbox{S}$ bonds
combine with the large atomic radius of sulfur to give access to
additional geometrical, electronic and ring-straining possibilities
not available to CHON chemistry.

Although not alkyl-thiol compounds as categorized above, two
additional cofactors that make important indirect use of sulfur are
thiamin and biotin.  In neither case is sulfur the element to which
transferred ${\mbox{C}}_1$ groups are bound. For reactions involving
TPP the ${\mbox{C}}_1$-unit is bound to the carbon between sulfur and
the positively charged nitrogen, while in biotin ${\mbox{C}}_1$-units
are bound to the carboxamide nitrogen in the (non-aromatic)
heterocycle opposite the sulfur-containing ring. It seems likely,
however, that the sulfur indirectly contributes to the properties of
the binding carbon or nitrogen, through some combination of
electrostatic, resonance, or possibly ring-straining interactions.
The importance of the sulfur to the focal carbon or nitrogen atom is
suggested by the complexity of the chemistry and enzymes involved in
its incorporation into these two cofactors
\cite{Jurgenson09,Berkovitch04}.

\subsubsection{Biochemical roles and phylogenetic distribution}

{\bf Transfer of methyl or methylene groups:} The S atoms of CoM,
lipoic acid, and S-adenosyl-homocysteine accept methyl or methylene
groups from the nitrogen atoms of pterins.  Considering that
transition-metal sulfide minerals are the favored substrates for
prebiotic direct-${\mbox{C}}_1$
reduction~\cite{Huber:Ace_CoA:00,Cody:ACF:01,Russell:FeS:06}, a
question of particular interest is how, in mineral scenarios for the
emergence of carbon fixation, the distinctive relation between tuned
nitrogen atoms in pterins as carbon carriers, and alkyl-thiol
compounds as carbon acceptors, would have formed.

{\bf Reductants and co-reductants:} CoB and CoM act together as methyl
carrier and reductant to form methane in methanogenesis. In this
complex transfer~\cite{Fontecilla:metalloenzymes:09}, the
fully-reduced (${\mbox{Ni}}^{+}$) state of the Nickel tetrapyrrole
${\mbox{F}}_{430}$ forms a dative bond to $-{\mbox{CH}}_3$ displacing
the CoM carrier, effectively re-oxidizing ${\mbox{F}}_{430}$ to
${\mbox{Ni}}^{3+}$.  Reduced ${\mbox{F}}_{430}$ is regenerated through
two sequential single-electron transfers.  The first, from CoM-SH,
generates a ${\mbox{Ni}}^{2+}$ state that releases methane, while
forming a radical ${\mbox{CoB}}^{\cdot}-\mbox{S}-\mbox{S}-\mbox{CoM}$
intermediate with CoB.  The radical then donates the second electron,
restoring ${\mbox{Ni}}^{+}$.  The strongly oxidizing heterodisulfide
CoB-S-S-CoM is subsequently reduced with two NADH, regenerating CoM-SH
and CoB-SH.

A similar role as methylene carrier and reductant is performed by the
two SH groups in lipoic acid.  CoM is specific to methanogenic
archaea~\cite{Balche:CoM:79}, while lipoic acid and
S-adenosyl-homocysteine are found in all three
domains~\cite{Danson:archaea_metab:93,Braakman:carbon_fixation:12}.
Lipoic acid is formed from octanoyl-CoA, emerging from the
biotin-dependent malonate pathway to fatty acid synthesis, and along
with fatty acid synthesis~\cite{Lombard:biotin:11}, may have been
present in the universal common ancestor. The previously noted
universal distribution of the glycine cycle supports this hypothesis.

{\bf Role in the reversal of citric-acid cycling:} Lipoic acid becomes
the electron acceptor in the oxidative decarboxylation of
$\alpha$-ketoglutarate and pyruvate in the oxidative Krebs cycle,
replacing the role taken by reduced ferredoxin in the rTCA cycle.
Thus the prior availability of lipoic acid was an enabling
precondition for reversal of the cycle in response to the rise of
oxygen.

{\bf Carriers of acyl groups:} Transport of acyl groups in the
acyl-carrier protein (ACP) proceeds through thioesterification with
pantetheine phosphate, similar to the thioesterification in fixation
pathways.  In fatty acid biosynthesis acyl groups are further
processed while attached to the panthetheine phosphate prosthetic
group.

{\bf Electron bifurcation:} The heterodisulfide bond of CoB-S-S-CoM
has a high midpoint potential ($E_0^{\prime} = -140 \mbox{mV}$),
relative to the $H^{+}/{\mbox{H}}_2$ couple ($E_0^{\prime} = -414
\mbox{mV}$), and its reduction is the source of free energy for the
\emph{endergonic} production of reduced Ferredoxin
(${\mbox{Fd}}^{2-}$, $E_0^{\prime}$ \emph{in situ} unknown but between
$-520 \mbox{mV}$ and $-414 \mbox{mV}$)~\cite{Kaster:bifurcation:12},
which in turn powers the initial uptake of ${\mbox{CO}}_2$ on
${\mbox{H}}_4\mbox{MPT}$ in methanogens.  The remarkable direct
coupling of exergonic and endergonic redox reactions through splitting
of binding pairs into pairs of radicals, which are then directed to
paired high-potential/low-potential acceptors, is known as
\emph{electron bifurcation}~\cite{Herrmann:flavoproteins:08}.  Variant
forms of bifurcation are coming to be recognized as a widely-used
strategy of metal-center enzymes, either consuming oxidants as energy
sources to generate uniquely biotic low-potential reductants such as
${\mbox{Fd}}^{2-}$~\cite{Gerhardt:C_aminobutyr:00,Schutz:Naphthoquinol:03,%
Li:Crotonyl_bifurc:08,Kaster:bifurcation:12}, or to ``titrate'' redox
potential to minimize dissipation and achieve reversibility of redox
reactions involving reductants at diverse potentials, \emph{e.g.} by
combining low-potential (${\mbox{Fd}}^{2-}$, $E_0^{\prime} = -420
\mbox{mV}$) and high-potential (NADH, $E_0^{\prime} = -300 \mbox{mV}$)
reductants to produce intermediate-potential reductants (NADPH,
$E_0^{\prime} = -360 \mbox{mV}$)~\cite{Wang:bifurcation:10}.  Together
with substrate-level phosphorylation (SLP), electron bifurcation may
be the principal chemical mechanism (contrasted with membrane-mediated
oxidative phosphorylation) for interconverting biological energy
currencies, and along with SLP~\cite{deDuve:BC:91}, a mechanism of
central importance in the origin of
metabolism~\cite{Martin:bifurcation:12}.  Small metabolites including
such heterodisulfides of cofactors, which can form radical
intermediates exchanging single electrons with Fe-S clusters
(typically \emph{via} flavins) are essential sources and repositories
of free energy in pathways using bifurcation.  Both electron
bifurcation and the stepwise reduction of ${\mbox{F}}_{430}$ (above)
illustrate the central role of metals as mediators of single-electron
transfer processes in metabolism.

\subsubsection{Participation in carbon fixation pathway modules}

The similarity between the glycine cycle and methanogenesis in
Fig.~\ref{fig:C1_metabolism} emphasizes the convergent roles of
alkyl-thiol cofactors.  In the glycine cycle, methylene groups are
accepted by the terminal sulfur on lipoic acid, and the subadjacent SH
serves as reductant when glycine is produced, leaving a disulfide bond
in lipoic acid.  The disulfide bond is subsequently reduced with NADH.
In methanogenesis, a methyl group from ${\mbox{H}}_4 \mbox{MPT}$ is
transferred to CoM, with the subsequent transfer to
${\mbox{F}}_{430}$, and the release from ${\mbox{F}}_{430}$ as methane
in the methyl-CoM reductase, coupled to formation of
$\mbox{CoB}-\mbox{S}-\mbox{S}-\mbox{CoM}$.  The heterodisulfide is
again reduced with NADH, but employs a pair of electron bifurcations
to retain the excess free energy in the production of
${\mbox{Fd}}^{2-}$ rather than dissipating it as
heat~\cite{Kaster:bifurcation:12}.  Methanogenesis is thus associated
with 7 distinctive cofactors beyond even the set known to have
diversified functions within the archaea~\cite{Lengeler:BP:99}, again
suggesting the derived and highly optimized nature of this
Euryarchaeal phenotype.  The striking similarity of these two
methyl-transfer systems, mediated by independently evolved and
structurally quite different cofactors, suggests evolutionary
convergence driven specifically by properties of alkyl thiols.

\begin{widetext}

\begin{figure}[ht]
  \begin{center} 
  \includegraphics[scale=0.7]{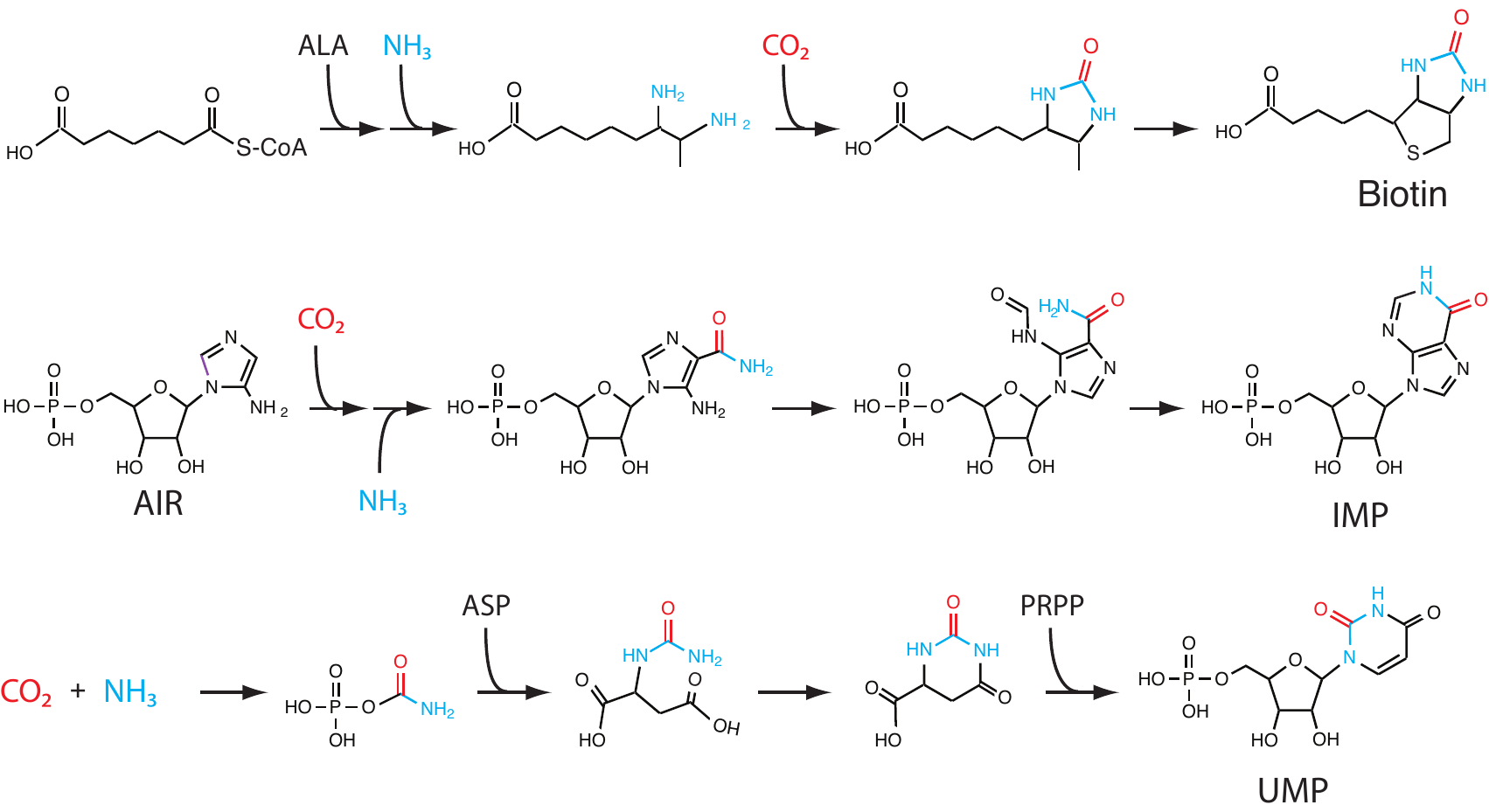}
  \caption{
  Carboxylation reactions in the synthesis of cofactors. The sequences
  show the immediate amination of the carboxyl group to a carboxamide
  group, which is then preserved into the final heterocyclic
  structure. As the only carboxylations not used in core carbon
  uptake, these reaction sequences form a distinct class of chemistry.
  Amination reactions are shown as net additions of ammonia, which may
  be derived from other sources (such as glutamine, aspartate or
  S-adenosyl-methionine). Abbreviations: Alanine (ALA); Aspartate
  (ASP); phosphoribosyl pyrophosphate (PRPP).
    \label{fig:cofactor_carboxylations} 
  }
  \end{center}
\end{figure}

\end{widetext}

A curious pattern, which we note but do not attempt to interpret, is
the association of non-sulfur, nitrogen-heterocycle cofactors with WL
carbon fixation, contrasted with the use of sulfur-containing
heterocycles in carboxylation reactions of the rTCA cycle.  The
non-sulfur cofactors THF and ${\mbox{H}}_4 \mbox{MPT}$ are used in the
reactions of the WL pathway, while the biosynthetically-related but
sulfur-containing cofactor Thiamin mediates the carbonyl insertion (at
a thioester) in rTCA~\cite{Ma:PFOR:97,Chabriere:oxidoreductases:99}.
Biotin -- which has been generally associated with malonate synthesis
in the fatty-acid pathway (and derivatives such as propionate
carboxylation to methyl-malonate in 3HP~\cite{Lombard:biotin:11}) --
mediates the subsequent $\beta$-carboxylation of pyruvate and of
$\alpha$-ketoglutarate~\cite{Cronan02,Zhang03,Aoshima:AKG_carbox:04}.
Thus the two cofactors we have identified as using sulfur indirectly
to tune properties of carbon or nitrogen ${\mbox{C}}_1$-bonding atoms
mediate the two chemically quite different sequential carboxylations
in rTCA.

\subsection{Carboxylation reactions in cofactor synthesis}

Carboxylation reactions can be classified as falling into two general
categories: those used in core carbon ``uptake", and those used
exclusively in the synthesis of specific cofactors. In addition to
carboxylation reactions in carbon-fixation pathways, the former
category includes the carboxylation of crotonyl-CoA in the glyoxylate
regeneration cycle. This cycle is a mixotrophic rather than an
autotrophic pathway, but this reaction does form a distinct entry
point for ${\mbox{CO}}_2$ into the biosphere. The carboxylation of
acetyl-CoA to malonyl-CoA further serves a dual purpose, in being both
the starting point for fatty acid synthesis, as well as a key step in
the 3HP pathway used in several carbon-fixation pathways. All these
carboxylation reactions thus have in common that they are used at
least in some organism as the central source for cellular carbon. All
other carboxylation reactions that are not used as part of core carbon
uptake, are used in the synthesis of the biotin cofactor, and the
purine and pyrimidine nucleotides (see
Fig. \ref{fig:cofactor_carboxylations}).

If we consider the sequences in which these carboxylation reactions
are used to synthesize biotin, purine and pyrimidine, they also form a
distinct class of chemistry. In all three cases the resulting carboxyl
group is immediately aminated, either as part of the carboxylation
reaction, or in the following reaction, and the carboxamide group is
subsequently maintained into the final heterocyclic structure. In
addition we previously saw that IMP becomes the source for the folate
and flavin family (through GTP).  Carboxylation reactions are thus
either a general source for cellular carbon in core metabolism, or a
specific source of carboxamide groups in the synthesis of cofactors
that are part of the catalytic control of core metabolism.

\subsection{The chorismate pathway in both amino acid and cofactor
  synthesis}  

Chorismate is the sole source of single benzene rings in
biochemistry~\cite{Bentley90}.  The non-local $\pi$-bond resonance is
used in a variety of charge-transfer and electron transfer and storage
functions, in functional groups and cofactors derived from chorismate.
We have noted the charge-transfer function of PABA in tuning
${\mbox{N}}^{10}$ of folates, and its impact on ${\mbox{C}}_1$
chemistry.  The para-oriented carbonyl groups of quinones may be
converted to partially- or fully-resonant orbitals in the benzene
ring, enabling fully oxidized (quinone), half-reduced (semiquinone),
or fully reduced (hydroquinone) states~\cite{Stryer:BC:81}.  Finally,
the aromatic ring in tryptophan (a second amino acid which behaves in
many ways like a cofactor) has at least one function in the active
sites of enzymes as a mediator of non-local
electron-transfers~\cite{Kuki:e_tunnel:00}.

\section{Innovation: promiscuous catalysis, serendipitous pathways}
\label{sec:innovation}

The previous sections argued for the existence of low-level chemical
and cofactor/catalyst constraints on metabolic innovations, and
presented evolutionary divergences that either respected these as
constraints, or were enabled by the diversification of cofactor and
catalytic functions.  In this section we consider the dynamics by
which innovation occurs, and its main organizing principles.
Innovation in modern metabolism occurs principally by duplication and
divergence of enzyme
function~\cite{Jensen:recruitment:76,Copley:promiscuity:03,%
Khersonsky:promiscuity:10}.  Often it relies on similarity of
functions among paralogous enzymes, but in some cases may exploit more
distant or accidental overlap of functions.

Innovation always requires some degree of enzymatic
promiscuity~\cite{Copley:promiscuity:03}, which may be the ability to
catalyze more than one reaction (catalytic promiscuity) or to admit
more than one substrate (substrate ambiguity).  Pathway innovation
also requires \emph{serendipity}~\cite{Kim:serendip_paths:10}, which
refers to the coincidence of new enzymatic function with some avenue
for pathway completion that generates an advantageous phenotype from
the new reaction.  Although most modern enzymes are highly specific,
broad substrate-specificity is no longer considered rare, and is even
explained as an expected outcome in cases where costs of refinement
are higher than can be supported by natural selection, and in other
cases by positive selection for phenotypic
plasticity~\cite{Khersonsky:promiscuity:10,Nam12}. However, when
enzymes are specific -- whether due to structure or due to evolved
regulation -- they are of necessity diversified in order to cover the
broad range of metabolic reactions used in the modern biosphere.
Serendipitous pathways assembled from a diversified inventory of
specific enzymes will in most cases be strongly historically
contingent as they depend on either overlap of narrow affinity domains
or on ``accidental'' enzyme features not under selection from
pre-existing functions.  Such pathways therefore seem unpredictable
from first principles; whether they are rare will depend on the degree
to which the diversity of enzyme substrate-affinities compensates for
their specificity.

A key question for early metabolic evolution is whether the trade-off
between specificity and diversity was different in the deep past than
in the present, either in degree or in structure, in ways that
affected either the discovery of pathway completions or the likelihood
that new metabolites could be retained within existing networks.
These structural aspects of promiscuity and serendipity determine the
regulatory problem faced by evolution in balancing the elaboration of
metabolism with its preservation and selection for function.

\subsubsection{Creating reaction mechanisms and restricting
  substrates, while evolving genes} 

Metabolism is characterized at all levels by a tension between
creating reaction mechanisms that introduce new chemical
possibilities, and then pruning those possibilities by selectively
restricting reaction substrates.  Whether this tension creates a
difficult or an easy problem for natural selection to solve depends at
any time on whether the accessible changes in catalytic function,
starting from integrated pathways, readily produce new integrated
pathways whose metabolites can be recycled in autocatalytic loops.  We
argue that the conservation of pathway mechanisms, particularly when
these are defined by generic functional groups such as carboxyls,
ketones, and enols, with promiscuity coming from substrate ambiguity
with respect to molecular properties away from the reacting functional
group, favors the kind of orderly pathway duplication that we observe
in the extant diversity of core metabolism.  Therefore we expect that
serendipitous pathway formation was both facile in those instances in
the early phases of metabolic evolution where innovations in
radical-based mechanisms for carbon incorporation occurred, and
structured according to the same local-group chemistry around which
the substrate network is organized.

Modern enzymes both create reaction mechanisms and restrict
substrates, but the parts of their sequence and structure that are
under selection for these two categories of function may be quite
different, so the two functions can evolve to a considerable degree
independently.  Active-site mechanisms in enzymes for organic
reactions will often depend sensitively on a small number of highly
conserved catalytic residues in a relatively fixed geometry, while
substrate selection can depend on a wide range of properties of enzyme
shape or conformation dynamics~\cite{Khersonsky:promiscuity:10}, on
local functional-group properties of the substrate that have been
termed ``chemophores''~\cite{Khersonsky:enzyme_families:11}, as well
as (in some cases) on detailed relations between the substrate and
active-site geometry or residues.  An extreme example of the potential
for separability between reaction mechanism and substrate selection is
found in the polymerases.  A stereotypical reaction mechanism of
attack on activating phosphoryl groups requires little more than
correct positioning of the substrates.  In the case of DNA
polymerases, at least six known categories (A, B, C, D, X, and Y) with
apparently independent sequence origin have converged on a geometry
likened to a ``right hand'' which provides the required
orientation~\cite{Braithwaite:DNA_polym:93,Bailey:DNA_polym:06}.

At the same time as evolving enzymes needed to provide solutions to
the biosynthetic problem of enabling and regulating metabolic network
expansion, they were themselves dependent on the evolving capabilities
of genomic and translation systems for maintaining complexity and
diversity.  Jensen~\cite{Jensen:recruitment:76} originally argued that
high enzymatic specificity was no more plausible in primitive cells
than highly diversified functionality,\footnote{This argument was
largely a rebuttal of an earlier proposal by
Horowitz~\cite{Horowitz:retrograde:45} for ``retrograde evolution'' of
enzyme functions.  The 1940s witnessed the rise of an overly-narrow
interpretation of ``one gene, one enzyme, one substrate, one
reaction'' (a rigid codification of what would become Crick's
\emph{Central Dogma}~\cite{Crick:dogma:70}), which in the context of
complex pathway evolution appeared to be incompatible with natural
selection for function of intermediate states.  The Horowitz solution
was to depend on an all-inclusive ``primordial
soup''~\cite{Haldane:OOL:67}, in which pathways could grow backward
from their final products, propagating selection stepwise downward in
the pathway until a pre-existing metabolite or inorganic input was
found as a pathway origin.} and that enzymatic promiscuity was both
evolutionarily necessary and consistent with what was known at that
time about substrate ambiguity and catalytic promiscuity.  Modern
reviews~\cite{Copley:promiscuity:03,Khersonsky:promiscuity:10,%
Khersonsky:enzyme_families:11} of the mechanisms underlying functional
diversity, promiscuity, and serendipity confirm that substrate
ambiguity is the primary source of promiscuity that has led to the
diversification of enzyme families.  It is striking that, even in
cases where substrate affinity has been the conserved property while
alternate reaction mechanisms or even alternate active sites have been
exploited, it is often local functional groups on one or more
substrates that appear to determine much of this
affinity~\cite{Khersonsky:promiscuity:10}.

\subsubsection{Evidence in our module substructure that early
  innovation was governed principally by local chemistry}

The substructure of modules, and the sequence of innovations, we have
sketched in Sec.~\ref{sec:core_carbon} appears to be dominated by
substrate ambiguity in enzymes or enzyme families with conserved
reaction mechanisms.  The key reactions in carbon fixation are of two
types: Crucial reactions typically involve metal centers or cofactors
that could have antedated enzymes, and it is primarily reaction sites,
not molecular selectivity, that distinguishes pathways at the stage of
these reactions.  Recall that the enzymes that have been argued to be
the ancestral forms of both the acetyl- and succinyl-CoA ligases and
the pyruvate and $\alpha$-ketoglutarate biotin-carboxylases show very
close sequence
homology~\cite{Aoshima:citrate_synthase:04,Aoshima:AKG_carbox:04},
suggesting shared ancestral enzymes for both.  The shared internal
sequence of reductions and isomerizations common to modules
(Fig.~\ref{fig:pathways_compare_1_annot}) are very broadly duplicated,
and the molecular specificity in their enzymes today is not correlated
with significant reaction-sequence changes in the internal structure
of pathways.  These pathways could plausibly function much as they do
today with less-specific hydrogenases and aconitases.

A quantitative reconstruction of early evolutionary \emph{dynamics}
will require merging probability models for networks and metabolic
phenotypes with those for sequences and structure of enzyme families.
The goal is a consistent model of the temporal sequence of ancestral
states of catalyst families, and of the substrate networks on which
they acted.

\section{Integration of cellular systems}
\label{sec:integration}

The features of metabolism that display a ``logic'' of composition,
which is then reflected in their evolutionary history, are those with
few and robust responses to environmental conditions that can be
inferred from present diversity.  These are the subsystems whose
evolution has been simplified and decoupled by modularity.  Their
relative immunity from historical contingency, resulting in more
``thermodynamic'' modes of evolution, results from rapid,
high-probability convergence in populations that can share
innovations~\cite{Vetsigian:code:06}.

The larger roles for standing variation and historical contingency
that are so often emphasized~\cite{Gould:SET:02} in accounts of
evolutionary dynamics are made possible by longer-range correlations
that link modules, creating mutual dependencies and restricting viable
changes~\cite{Davidson:kernels:06,Erwin:kernels:09}.  The most
important source of such linkage in extant life is the unification of
metabolic substrates and control processes within
cells~\cite{Wilson:biol_cell:25}.  Cellular death or reproduction
couples fitness contributions from many metabolic-phenotype traits,
together with genome replication systems.  This enables the
accumulation of diversity as genomes capture and exploit gains from
metabolic control, complementary
specialization~\cite{Poisot:specialization:11}, and the emergence of
ecological assemblies of specialists as significant mediators of
contingent aspects of evolutionary innovation (as we illustrated with
the example of methylotrophy).

We consider in this section several important ways in which
aggregation of metabolic processes within cells follows its own
orderly hierarchy and progression.  We note that even a single cell
does not impose only one type of aggregation, but at least three
types, and that these are the bases for different selection pressures
and could have arisen at different times.  Within cellular subsystems,
the coupling of chemical processes is often mediated by coupling of
their energy systems, which has probably developed in stages that we
may be able to identify.  Finally, even where molecular replication is
coupled to cellular physiology, in the genetic code, strong and
perhaps surprising signatures of metabolic modularity are
recapitulated.

\subsection{Cells provide at least three functionally distinct forms of
  compartmentation} 

Under even the coarsest functional abstraction, the cell provides not
one form of compartmentation, but at least
three~\cite{Nelson:biochem:04,Alberts:MBC:02}.  The geometry and
topology of closed spheres or shells, and the capacitance and proton
impermeability of lipid bilayers, permit the buildup of pH and voltage
differences, and thus the coupling of redox and phosphate energy
systems through intermediate proton-motive (or in many cases,
sodium-motive) force~\cite{Morowitz:FB:78}.  The concentration of
catalysts with substrates enhances reactions that are second-order in
organic species, and the equally important homeostatic control of the
cytoplasm regulates metabolic reaction rates and precludes parasitic
reactions.  Finally, the cell couples genetic variations to internal
biochemical and physiological variations much more exclusively than
they are coupled to shared resources such as biofilms or siderophores,
leading to the different evolutionary dynamics of development from
niche construction~\cite{OdlingSmee:NC:03}.\footnote{For an argument
that somatic development and niche construction are variants on a
common process, distinguished by the genome's level of control and
exploitation of the constructed resources, see
Ref.~\cite{Bershad:soma_niche:08}.} The perspective that this is an
active coupling, which defines one of the forms of
\emph{individuality} rather than providing a complete characterization
of the nature of the living state, is supported by the complex
ecosystems including viral RNA and DNA that are partly autonomous of
the physiology of particular
cells~\cite{Claverie:virus_cell_evol:06,Forterre:virus_view:10}. Each
of these different forms of coupling affects the function and
evolution of the modules we have discussed.

\subsubsection{Coupling of redox and phosphate energy systems may have
  been the first form of compartmentation selected}

Biochemical subsystems driven, respectively, by redox potential or
phosphoanhydride-bond dehydration potential, cannot usually be
directly coupled to one another due to lack of ``transducer''
reactions that draw on both energy systems. In addition to the
ultimate physical constraint of limits to free energy, biochemistry
also operates under additional proximate constraints from the chemical
and quantum-mechanical substrates in which that free energy it is
carried. The notable exception to the general lack of direct coupling
between energy systems is the exchange of phosphate and sulfur groups
in substrate-level phosphorylation~\cite{Stryer:BC:81} from thioesters
(which may proceed in either direction depending on conditions).
Although it provides a less flexible mode of coupling than
membrane-mediated oxidative phosphorylation, this crucial reaction
type, which occurs in some of the deepest reactions in biochemistry
(those employing CoA, including all those in the six carbon fixation
pathways), has been proposed as the earliest coupling of redox and
phosphate~\cite{deDuve:BC:91}, and the original source of
phosphoanhydride potential~\cite{Martin:OMP:07} enabling pathways that
require both reduction and dehydration reactions.

Phosphate concentration limits growth of many biological systems
today, and phosphate concentrations appear to be even lower in vent
fluids~\cite{Kelley:VFMOR:02} than on average in the ocean, making it
difficult to account for the emergence of many metabolic steps in
hydrothermal vent scenarios for the origin of life.  Serpentinization
and other rock-water interactions that produce copious reductants 
-- and are believed to have been broadly similar at least from the
early archean to the
present~\cite{Russell:FeS:97,Russell:AcetylCoA:04} -- also scavenge
phosphates into mineral form. Unless new mechanisms are
discovered that could have produced an increased amount of phopshate
for early vents, it thus appears doubtful that phosphates were abundant in
the environments otherwise most favorable to geochemical
organosynthesis.  What little phosphate is found in water is primarily
orthophosphate, because the phosphoanhydride bond is unstable to
hydrolysis.  Therefore the retention of orthophosphate, and the
continuous regeneration of pyrophosphate and
polyphosphates~\cite{Baltscheffsky:PPi:66,Yamagata:PolyP:91,%
Kornberg:PolyP:99,Baltscheffsky:PPi:99,Brown:PolyP:04}, may have been
essential to the spread of early life beyond relatively rare
geochemical environments.

The membrane-bound ATP-synthetase, which couples phosphorylation to a
variety of redox reactions~\cite{Lengeler:BP:99} through proton or
sodium pumping, is therefore essential in nearly all biosynthetic
pathways, and must have been among the first functions of the
integrated cell.  Without a steady source of phosphate esters, none of
the three oligomer families could exist.  The ATP synthetase itself is
homologous in all organisms, providing one strong argument (among
many~\cite{Pereto:membrane_lipids:04,Lombard:biotin:11}) for a
membrane-bound last common ancestor.  Proton-mediated phosphorylation
(best known through oxidative phosphorylation in the respiratory
chain~\cite{Stryer:BC:81}) requires a topologically enclosed space to
function as a proton capacitor~\cite{Morowitz:FB:78}.  However, as
shown by gram-negative bacteria~\cite{Lengeler:BP:99} and their
descendants mitochondria and plastids, which acidify the periplasmic
space or thylakoid lumen, the proton capacitor need not be (and
generally is not) the same compartment as the cytoplasm containing
enzymatic reactions.  Because the coupling of energy systems is a
different function from regulating reaction rates catalytically, the
phosphorylation system should not generally have been subject to the
same set of evolutionary pressures and constraints as other cellular
compartments, and need not have arisen at the same time.  We note
that, because it may have lower osmotic pressure than the cytoplasm,
the acidified space required for proton-driven phosphorylation may not
have required a cell wall, greatly simplifying the number of
concurrent innovations required for compartmentalization, compared to
those for the cytoplasm.  Therefore we conjecture that proton-mediated
phosphorylation could have been the first function leading to
selection for lipid-bilayer compartmentalization, allowing other
cellular functions to accrete at later times.

\subsubsection{Regulation of biosynthetic rates may have been
  prerequisite for the optimization of loop-autocatalytic cycles}
\label{sec:rate_regulation}

The second function of cellular compartments, and the one most
emphasized in vesicle theories of the origin of
life~\cite{Morowitz:BCL:92,Hanczyk:protocells:03,Luisi:OOL_text:06},
is the enhancement of second-order reactions by collocation of
catalysts and their substrates.  Here we note another role that we
have not seen mentioned, which is more closely related to the
functions of the cell that \emph{inhibit} reactions.  Organisms
employing autocatalytic-loop carbon fixation pathways must reliably
limit their anabolic rates to avoid drawing off excess network
catalysts into anabolism, resulting in passage below the autocatalytic
threshold for self-maintenance, and collapse of carbon fixation and
metabolism.  Regulating anabolism to maintain viability and growth may
have been an early function of cells.

We noted in Sec.~\ref{sec:character_root} the fragility of
autocatalytic-loop pathways to parasitic side-reactions, and the way
the addition of a linear pathway such as WL stabilizes loop
autocatalysis in the root node of
Fig.~\ref{fig:metab_tree_printer_bigicon}. For proto-metabolism,
spontaneous abiotic side-reactions may be hazardous, if catalysts in
the main fixation pathway do not sufficiently accelerate their
reaction rates, creating a separation of timescales relative to the
uncatalyzed background.  Within the first cells, the same hazard is
posed by secondary anabolism, as its reaction rates become enhanced by
catalysts similar to those in the core.  This fact was clearly noted
already in Ref.~\cite{Huber:Ace_CoA:00}.  It may thus be that the
optimizations in either branch of the carbon-fixation tree were not
possible until rates of anabolism were sufficiently well-regulated to
protect supplies of loop intermediates or essential cofactors.
Therefore, while the root node is plausible as a
pre-cellular~\cite{Huber:Ace_CoA:00} or an early cellular (but
non-optimized) form, either branch from it may have required the
greater control afforded by quite refined cellular regulation of
reaction rates.  {It is here that we envision a crucial role for
feedback regulation at the genomic level~\cite{Savageau74,Savageau75}
as a support for the \emph{architectural stability} of the underlying
substrate network, prior even to its service in homeostasis in complex
environments or in phenotypic plasticity.}

\subsection{Coupling of metabolism to molecular replication, and
  signatures of chemical regularity in the genetic code}

Among the subsystems coupled by modern cells, perhaps none is more
elaborate than the combined apparatus of amino acid and nucleotide
biosynthesis and protein coding.  The most remarkable chemical aspect
of the protein-coding system is that it is an \emph{informational}
system: a sophisticated machinery of transcription, tRNA formation and
aminoacylation, and ribosomal translation separates the chemical
properties of DNA and RNA from those of proteins, permitting almost
free selection of sequences in both alphabets in response to
requirements of heredity and protein function.\footnote{The
  observation that enzymes acting on DNA have evolved to actively
  mitigate chemical differences in the bases, to enable a more nearly
  neutral combinatorial alphabet, is due to Peter Schuster
  (pers.~comm.)}  The interface at which this separation occurs is the
genetic code.  From the informational suppression of chemical details
that defines the coding system, the code itself might have been
expected to be a random map, but empirically the code is known to
contain many very strong regularities related to amino acid
biosynthesis and chemical properties, and perhaps to the evolutionary
history of the aminoacyl-tRNA synthetases.

Many explanations have been advanced for redundancy in the genetic
code, as a source of robustness of protein properties against
single-point
mutations~\cite{Freeland:code:98,Knight:codes:99,Lu:AA_alphabet:06,%
  Vetsigian:code:06}, but in all of these the source of selection
originates in the elaborate and highly evolved function of coding
itself.  In many cases the redundancy of amino acids at adjacent
coding positions reflects chemical or structural similarities,
consistent with this robustness-criterion for selection, but in nearly
all cases redundancy of bases in the code correlates even more
strongly with shared elements of \emph{biosynthetic pathways} for the
amino acids.  The co-evolutionary hypothesis of
Wong~\cite{Wong:code_coevol:75} accounts for the correlation of the
first base-position with amino-acid backbones as a consequence of
duplication and divergence of amino acid biosynthetic enzymes together
with aminoacyl-tRNA synthetases (aaRS).  The stereochemical hypothesis
of Woese~\cite{Woese:stereochem:66} addresses a correlation of the
second coding position with a measure of hydrophobicity called the
\emph{polar requirement}.  The remarkable fact that both correlations
are highly significant relative to random assignments, but that they
are \emph{segregated} between first and second codon bases, is not
specifically addressed in either of these accounts.  Copley \emph{et
  al.}~\cite{Copley:codes:05} address the same regularities as both
the Wong and Woese hypotheses, but link them to much more striking
redundancies in biosynthetic pathways, which they propose are
consequences of small-molecule organo-catalytic roles of dimer RNA in
the earliest biosynthesis of amino acids.

We note here a further chemical regularity in the genetic code, which
falls outside the scope of the previous explanations, and possibly
relates to the biosynthetic pathways of the purine cofactors as
discussed in Sec.~\ref{sec:purine_cofactors}. This regularity concerns
triplet codons with purines at the second position, and takes one of
two forms. Several amino acids that use GTP-derived cofactors in their
biosynthetic pathways are associated with triplet codons containing
Guanine at the second position, while another amino acid (Histidine)
that in its synthesis is directly derived from ATP is associated with
triplet codons containing Adenine at the second position.  This
association is much more comprehensive for G-second codons than for
A-second codons, and it does not suggest the same kinds of mechanistic
relations in the two cases.  However, it further compresses the
description of patterns in the code that were not addressed in
Ref.~\cite{Copley:codes:05}, in terms of similar chemical and
biosynthetic associations.

The correlation between the glycine cycle for amino acid biosynthesis
from ${\mbox{C}}_1$ groups on folate cofactors, and codons XGX, where
X is any base and G is guanine is strong.  (In what follows we
abbreviate wobble-base positions y for pyrimidines U and C, or u for
purines A or G.)  This group includes glycine (GGX), serine (AGy),
cysteine (UGy), and tryptophan (UGu).\footnote{Both purines are used
in the mitochondrial code and only UGG is used in the nuclear code.}
We do not propose a specific mechanism for such an association here,
but our earlier argument that folates would have been contemporaneous
with GTP suggests that biosynthesis through the glycine cycle was the
important source of these amino acids at the time they became
incorporated into the code.  Some of these amino acids satisfy
multiple regularities, as in the correlation of glycine with
$\mbox{GXX} \leftrightarrow \mbox{reductive transamination}$, or
cysteine with $\mbox{UXX} \leftrightarrow \mbox{pyruvate backbone}$,
proposed in Ref.~\cite{Copley:codes:05}.

The position (CAy) of histidine, synthesized from ATP, is the only
case we recognize of a related correlation in XAX codons.  For this
position, the availability of ATP seems to have been associated with
the synthesis of histidine directly through the cyclohydrolase
function (rather than through secondary cofactor functions), at the
time this amino acid became incorporated into the code.

Much more than correlation is required to impute causation, and all
existing theories of cause for regularities in the genetic code are
either highly circumstantial or require additional experimental
support.  Therefore we limit the aspects of these observations that we
regard as significant to the following three points:

{\bf \noindent The existence of a compression:} The idealized adaptive
function of coding is to give maximum evolutionary plasticity to
aspects of phenotype derived from protein sequence, uncoupled from
constraints of underlying biosynthesis.  The near-wholesale transition
from organic chemistry to polymer chemistry around the
${\mbox{C}}_{20}$ scale suggests that this separation has been
effectively maintained by evolution.  Strong regularities which make
the description of the genetic code \emph{compressible} relative to a
random code reflect \emph{failures} of this separation which have
transmitted selection pressure across levels, during either the
emergence or maintenance of the code.  These include base-substitution
errors, whether from mutation or in the transcription and translation
processes, but also apparently chemical relations between nucleobases
and amino acids.

{\bf \noindent The segregation of the roles of different base
  positions and in some cases different bases in terms of their
  biochemical correlates:} The genetic code is like a ``rule book''
for steps in the biosynthesis of many amino acids, but the chemical
correlations which are its rules are of many kinds. The correlations in
the code may be understood as rules because the biosynthetic pathways
may be placed on a decision tree, with branches labeling alternative
reactions at several stages of synthesis, and branching directions
indicated by the position-dependent codon
bases~\cite{Copley:codes:05}. Beyond the mere existence of those
rules, and their collective role as indices of regularity threading
the code, we must explain why rules of different kinds are so neatly
\emph{segregated} over different base positions and sometimes over
different bases (as in the XGX and XAX codons).

{\bf \noindent A compression that references process rather than
  property:} The role of biosynthetic pathways as correlates of
regularities makes this compression of the genetic code a reference to
the \emph{process} and metabolic network context within which amino
acids are produced, and not merely to their \emph{properties}.  (Many
of the chemical properties recognized as criteria of selection,
whether size or hydrophobicity, are shared at least in part because
they result from shared substrates or biosynthetic steps.)  We think
of the function of coding as separating biosynthetic process from
phenotype: transcription and translation are ``Markovian'' in the
sense that the only information from the biosynthetic process which
survives to affect the translated protein is what is inherent in the
structure of the amino acid. In technical terms, one says that the
phenotype is \emph{conditionally independent} of the biosynthetic
pathway, \emph{given} the amino acid.  Thus selection on
post-translation phenotypes should only be responsive to the finished
amino acids.  The existence of regularities in the genetic code which
show additional correlation with intermediate steps in the
biosynthetic process therefore requires either causes other than
selection on the post-coding phenotype (including its robustness), or
a history-dependence in the formation of the code that reflects
earlier selection on intermediate pathway states.  If they reflect
causal links to metabolic chemistry, these ``failures'' of the
separation between biosynthetic constraint and selection of polymers
for phenotype may have broken down the emergence of molecular
replication into a sequence of simpler, more constrained, and
therefore more attainable steps.

\section{The extrapolation of metabolic logic to questions of
  emergence} 
\label{sec:origins}

Comparative analysis and its formal extension to diachronic
reconstruction simultaneously estimate two properties of systems: a
model for a \emph{generating process} including constraints or laws
that have operated over the system's history, and a collection of
\emph{idiosyncratic or historically contingent events} that make each
history distinct and are not assumed to be reproducible or
predictable.  In our review up to this point the reconstructed period
has been the genomic era, and snapshots of ancestral states in this
era do not directly carry information about pre-LUCA or pre-cellular
forms except through constraints that we can argue were common both
pre- and post-LUCA.  The surprising feature of the compositional and
evolutionary logic of metabolism that we have sketched is how much of
it apparently reflects constraints from low-level organic or
organometallic chemistry that are not distinctively biological, or
homologies and energetic contacts with geochemistry that were arguably
broadly continuous through the emergence of
life~\cite{Russell:FeS:97,Fenchel:OEL:02,Reysenbach:vents:02}.  In
several places prebiotic scenarios have been mentioned as interpretive
frames for our observations, so here we summarize which features of
the logic of metabolism we think are strong constraints on theories of
the emergence of life.

\subsection{Autotrophic versus heterotrophic origins}

In Sec.~\ref{sec:ana_cata} we summarized reasons to regard carbon
fixation from ${\mbox{CO}}_2$ and anabolism as both the ancestral
pathways of cellular metabolism and the set of prior constraints
around which catabolic pathways, and the diverse array of
heterotrophic metabolisms they enable, subsequently evolved.  The
evidence in favor of this view is highly multi-factorial, including
historical reconstruction, inferred geochemical context, pathway
chemistry, and degree of universality.  The most striking property of
this evidence is that it leads to quite specific inferences about
ancestral autotrophic phenotypes, whereas we do not know of comparable
proposals about ancestral heterotrophy that are similarly specific and
that unify a similar diversity of observations.

It is then natural to ask: was the early role of autotrophy and
anabolism in the cellular era a continuation of geochemical processes
of similar character, or was it the outcome of a reversal of earlier
pre-LUCA or pre-cellular metabolisms fed by organics from pathways
unrelated to those in extant biochemistry?  The distinction is not
quite the same as that between autotrophic and heterotrophic organisms
(though these terms are often borrowed)~\cite{Smith:auto_hetero:10}. Rather,
it is a distinction between a hypothesis of continuity with
geochemistry which was gradually brought under autonomous control of
bio-organic chemistry, and a hypothesis of discontinuity requiring
that early organisms have evolved the mechanisms and networks of
biochemistry \emph{de novo}.  

The arguments for geochemical
continuity~\cite{Wachtershauser:SM:88,Wachtershauser:rTCA:90,%
Morowitz:BCL:92,Smith:universality:04,Martin:vents:08,Fuchs11} are
founded first on detailed accounts of the capacity of a range of
geochemical energy systems to to support extant
life~\cite{Amend:Energetics:01,Shock:mineral_energy:09}.  A subset of
the entries in Table~1 of Ref.~\cite{Amend:Energetics:01}, involving
${\mbox{Fe}}^{2+}$ reduction or autotrophic methanogenesis, can be
applied directly to early-earth environments.  (Note, however, that
many entries in their table of environments involve sulfates,
nitrates, ferric iron, or small amount of molecular oxygen (the
Knallgas reaction) as terminal electron acceptors.  The breadth of
organic conversions detailed in the paper is meant to provide a basis
for habitability analysis today, so plausible pathways in the Hadean
must be understood as having been limited by the available terminal
electron acceptors.)  Where the continuity hypothesis supposes that
extant life has ``enfolded'' prior geochemical mechanisms, it cites
detailed similarities between transition-metal/sulfide mineral unit
cells and metallo-enzyme active
sites~\cite{Russell:FeS:97,Russell:AcetylCoA:04,Russell:FeS:06}, which
may reflect~\cite{Martin:bifurcation:12} mineral precursors to the
widespread use of radical mechanisms in reactions catalyzed by metal-center
enzymes~\cite{Fontecilla:metalloenzymes:09}, as we have mentioned
previously.  The richness of hydrothermal vent environments in
particular, in geometry, surface
catalysis~\cite{Wachtershauser:SM:88,Wachtershauser:rTCA:90,%
Wachtershauser:FeS_world:92}, thermal and pH gradients, and the
overall similarity of the aqueous redox environment of hydrothermal
fluids to biochemistry~\cite{Shapiro:SmallMolecules:06,%
Shapiro:MetabFirst:07,Martin:vents:08,Berg:carbon_fixn:10}, provides
specific locations where catalysis and also other functions such as
containment or selective diffusion would have been provided.  Finally,
the geochemical hypothesis has been circumstantially supported by
experimental evidence that minerals can catalyze reactions in the
citric-acid cycle~\cite{Cody:pyruvate:00}, and an extensive range of
reductions~\cite{Heinen96,Cody:ACF:01}, including synthesis of
acetyl-thioesters~\cite{Huber:Ace_CoA:00}, which for a variety of
reasons we have noted in this review are among the most-central
compounds of core metabolism.

The specificity of the links which the continuity hypothesis is
in a sense \emph{required} to propose derive from the very restrictive
boundary condition of ${\mbox{CO}}_2$ as sole carbon source, the same
constraint that permitted specific claims in our reconstruction of
cellular autotrophy.  Our approach of gathering formal evidence about
the structure and strength of constraints, and of testing these for
consistency within both organism physiology and ecology, is very
similar in spirit to the approach of Ref.~\cite{Benner:palimpsest:89}
to a ``breakout organism'' from the RNA world. However, we aim at fewer and
chemically lower-level facts that plausibly reach further back to
pre-cellular or pre-RNA times.  (The details reconstructed in
Ref.~\cite{Benner:palimpsest:89} are also very compatible with our
reconstructions of early carbon fixation, and we regard our proposal
of an even earlier role for some cofactors than for RNA base pairing
to be very much in the general spirit of their arguments.)

In contrast, heterotrophic-origins stories are largely objections
to problems with geochemical organosynthesis \emph{and selection}
requiring ``something else'' in its place. They may be quite
unrestrictive about what the original organic inputs were, as in the
original proposals of Oparin~\cite{Oparin:OOL:67} and
Haldane~\cite{Haldane:OOL:67}. Their most restrictive quantitative
constraints (such as pathway minimization) may not directly determine
pathway direction~\cite{Melendez96}, and may presume an optimization
problem different from the one performed if the molecular inventory
was not pre-fixed. They may also show only limited overlap with extant
biochemistry~\cite{Miller:AA:53}, without suggestions for how missing
components were filled in or abiotic components were pruned.

\subsection{The joint WL/rTCA network as a pre-cellular form}

The importance of balancing considerations of accessibility and
robustness with selectivity in incipient and early-cellular
biochemistry were mentioned in Sec.~\ref{sec:character_root} and
Sec.~\ref{sec:purine_cofactors}.  As a solution to the problem of
reconstructing history, the root WL/rTCA network was put forward as a
quantitative example in which the consequences of topology for both
robustness and selectivity could be analyzed.  That treatment was
essentially backward-looking, asking how well our proposed root node
meets multiple criteria required by inference from the present, such
as pathway distribution, plausible causes for innovations, and
selection of the extant biosynthetic precursors.  The converse
question in the pre-cellular era is dynamical and echos Leibniz's
question ``Why is there something instead of nothing?''.

In its earliest forms, such a joint network would be presumed to have
mineral or perhaps soluble metal-ligand catalysts for both direct
${\mbox{C}}_1$ reduction and rTCA cycling, perhaps already showing the
distinctions between the functions of the nitrogenous cofactors for
${\mbox{C}}_1$ reduction and the functions of sulfur-containing
cofactors in rTCA.  We may ask, would a hybrid pathway out-compete
alternatives \emph{chemically} as a kinetic channel for carbon
reduction by ${\mbox{H}}_2$ (or perhaps directly by reduced iron)?  To
this we argue that a feeder augmented by a loop outcompetes an unaided
feeder on average by virtue of autocatalytic self-amplification~\cite{Hordijk12}.  A
loop with a feeder outcompetes a bare loop in the context of loss or
fluctuations because of greater robustness and recovery
(self-re-ignition).  The important observation is that chemical
selection already shows features common to Darwinian selection:
fitness can come both from average behavior and from stability under
perturbations, and different components of a pathway may provide
different elements of fitness.

\subsection{A synthetic description: geochemistry, the metabolic
  substrate, and catalysis} 

While many mechanisms and components -- particularly
catalysts~\cite{Copley:PMRNA:06} -- must have been replaced in a
sequence leading from prebiotic geochemistry to the earliest cellular
biochemistry, the three elements we have emphasized of a logic of
metabolism should have remained continuous across the transition.  We
favor scenarios in which chemical networks at the aggregate scale of
the biosphere originated in an abundant supply of ${\mbox{CO}}_2$ and
${\mbox{H}}_2$, containing driving redox potential resulting in an
accumulation of stress, which became coupled to a robust concentration
mechanism within organic chemistry, forming networks that in turn
became increasingly stable with the emergence of intermediary
metabolism and the appearance of complex cofactors and additional
long-loop feedback mechanisms.

\section{Conclusions}

We have argued that the fundamental problem of electron transfer in
aqueous solution leads to a qualitative division between catalytically
``hard'' and ``easy'' chemistry, and that this division in one form or
another has led to much of the architecture and long-term evolution of
metabolism and the biosphere.  Hard chemistry involves electron
transfers whose intermediate states would be unstable or energetically
inaccessible in water if not mediated by transition-metal centers in
metal-ligand complexes and/or elaborate and structurally complex organic cofactors.
Easy chemistry involves hydrogenations and
hydrations, intramolecular redox reactions, and a wide array of
acid-base chemistry.  Easy chemistry is promiscuously re-used and
provides the internal reactions within modules of core metabolism.
Hard chemistry defines the module boundaries and the key constraints
on evolutionary innovation.  These simple ideas underlie a modular
decomposition of carbon fixation that accounts for all known
diversity, largely in terms of unique adaptations to chemically simple
variations in the abiotic environment.  On the foundation of core
metabolism laid by carbon fixation, the remainder of biosynthesis is
arranged as a fan of increasingly independent anabolic pathways.  The
unifying role of the core permits diverse anabolic pathways to
independently reverse and become catabolic, and the combinatorics of
possible reversals in communities of organisms determines the space of
evolutionary possibilities for heterotrophic ecology.

We have emphasized the role of feedback in biochemistry, which takes
different forms at several levels.  Network autocatalysis, if we take
as a separate question the origin of external catalytic and cofactor
functions, is found as a property internal to the small-molecule
substrate networks for many core pathways.  A qualitatively different
form of feedback is achieved through cofactors, which may act either
as molecular or as network catalysts.  As network catalysts they
differ from small metabolites because their internal structure is not
changed except at one or two bonds, over the reactions they enable.
The cofactors act as ``keys'' that incorporate domains of organic
chemistry within biochemistry, and this has made them both
extraordinarily productive and severely limiting.  No extant core
pathways function without cofactors, and cofactor diversification
appears to have been as fundamental as enzyme diversification in some
deep evolutionary branches.  We have therefore argued for a closely
linked co-evolution of cofactor functions with the expansion of the
universal metabolic network from inorganic inputs, and attempted to
place key cofactor groups within the dependency hierarchy of
biosynthetic pathways, particularly in relation to the first ability
to synthesize RNA.

The most important message we hope to convey is the remarkable imprint
left by very low-level chemical constraints, even up to very high
levels of biological organization.  Only seven carbon fixation
modules, mostly determined by distinctive, metal-dependent
carboxylation reactions, cover all known phylogenetic diversity and
provide the building blocks for both autotrophic and heterotrophic
metabolic innovation.  A similar, small collection of organic or
organometallic cofactor families have been the gateways that determine
metabolic network structure from the earliest cells to the present.
The number of these cofactors that we consider distinct may be
somewhat further reduced if we recognize biosynthetic relatedness that
leads to functional relatedness (as in the purine-derived or
chorismate-derived cofactors), or cases of evolutionary convergence
dominated by properties of elements (as for lipoic acid and the
CoB-CoM system).

We believe that these regularities should be understood as laws of
biological organization. In a proper, geochemically-embedded theory of
the emergence of metabolism, such regularities should be predictable
from the properties of the underlying organic chemistry.  As our
understanding of relevant organic chemistry continues to expand,
particular forms, such as distinctive metal chemistry or convergent
uses of nitrogen and sulfur, should become predictable from their
distinct catalytic properties.  Properties of distributions, as in the
use of network modules or the diversity of cofactors, should in turn
be predictable from asymmetries in catalytic constraints that are
likely to arise within a large and diverse possibility space of
organic chemistry.  Moreover, this lawfulness should have been
expected: the factors that reduce (or encrypt) the role of laws in
biology, and lead to unpredictable historical contingencies, arise
from long-range correlation.  Correlation of multiple variables leads
to large spaces of possibility and entangles the histories of
different traits, making the space difficult to sample uniformly.  But
correlation in biology is in large part a \emph{constructed} property;
it has not been equally strong in all eras and its persistence depends
on timescales.  Long-term evolution permits recombination even in
modern integrated cells and genomes. Early life, in contrast, with its
less-integrated cells and genomes, and its more loosely-coupled
traits, had constructed less long-range correlation. These are the
domains where the simpler but invariant constraints of underlying
chemistry and physics should show through.

\subsection*{Acknowledgments}

This work was completed as part of the NSF FIBR grant nr. 0526747:
From geochemistry to the genetic code.  DES thanks Insight Venture
Partners for support.  RB is further supported by an Omidyar
Fellowship at the Santa Fe Institute. We are grateful to Harold
Morowitz, Shelley Copley, and Charles McHenry for critical
conversations of these ideas and essential references, and to two
anonymous referees for many helpful suggestions and references.

\newpage
\appendix

\begin{widetext}

\section{Glossary of some terms used in the text}
\label{sec:glossary}

\begin{table}[ht!]
\begin{tabular}{l l}
  \parbox[t]{1.5in}{term} & \parbox[t]{5.5in}{usage} \\
\hline
  \parbox[t]{1.5in}{\raggedright
  carbon fixation
} & \parbox[t]{5.5in}{\raggedright\mbox{ }
  Any process by which organisms convert ${\mbox{CO}}_2$ (or another
  inorganic one-carbon source such as bicarbonate or formic acid) into
  molecules possessing $\mbox{C}-\mbox{C}$ bonds.  All biosynthesis
  rests ultimately on carbon fixation, because the biosphere does not
  rely on organic carbon from abiotic sources.
} \\
  \parbox[t]{1.5in}{\raggedright
  anabolism
} & \parbox[t]{5.5in}{\raggedright\mbox{ }
  Biochemical processes that build up molecule size.  We will be
  concerned particularly with buildup by reactions of organic
  chemistry (as opposed to phosphate-driven polymerization, which is
  chemically simple and homogeneous).  Anabolism is a net consumer of
  reductants, as biomass is more reduced than its input
  ${\mbox{CO}}_2$.
} \\
  \parbox[t]{1.5in}{\raggedright
  catabolism
} & \parbox[t]{5.5in}{\raggedright\mbox{ }
  Biochemical processes that break down organic molecules taken as
  inputs.  The breakdown may provide energy or biosynthetic precursors
  to other anabolic reactions.
} \\
  \parbox[t]{1.5in}{\raggedright
  autotrophy
} & \parbox[t]{5.5in}{\raggedright\mbox{ }
  A self-sufficient mode of metabolism in which all biomolecules can
  be synthesized using ${\mbox{CO}}_2$ as sole carbon source.  In
  strict usage the term denotes self-sufficiency of the metabolic
  network of an \emph{organism}.  For purposes of understanding the
  constraints implied by flux-balance analysis, we will extend the
  scope to include appropriate consortia of organisms.
} \\
  \parbox[t]{1.5in}{\raggedright
  heterotrophy
} & \parbox[t]{5.5in}{\raggedright\mbox{ }
  A mode of metabolism in which the focal network must draw carbon
  from some organic source, because it lacks necessary reactions to
  synthesize some essential metabolites starting from ${\mbox{CO}}_2$.
} \\
  \parbox[t]{1.5in}{\raggedright
  compositional logic
} & \parbox[t]{5.5in}{\raggedright\mbox{ }
  Principles of assembly of metabolic networks in organisms or
  consortia which capture regularities in the structure of the
  resulting networks, or of their dynamics which are responsible for
  phenotype or ecological role.
} \\
  \parbox[t]{1.5in}{\raggedright
  evolutionary logic
} & \parbox[t]{5.5in}{\raggedright\mbox{ }
  Principles of selection or constraints which compactly express
  regularities in evolutionary branching and relate these to aspects
  of phenotype which may have constrained innovations or determined
  fitness.
} \\
  \parbox[t]{1.5in}{\raggedright
  autocatalysis
} & \parbox[t]{5.5in}{\raggedright\mbox{ }
  A property of reaction networks, that intermediates or outputs of
  the reaction system act to catalyze earlier reactions in the system
  (their own biosynthetic pathways or others), leading to
  self-amplification of the reaction fluxes.  Autocatalysis may be
  provided by individual molecules such as enzymes, or may result from
  completing network cycles that connect inputs to outputs while
  regenerating network intermediates.  For formalizations
  see~\cite{Zachar10,Hordijk:autocat:12} 
} \\
  \parbox[t]{1.5in}{\raggedright
  catalytic control
} & \parbox[t]{5.5in}{\raggedright\mbox{ }
  The bringing-into-existence, or the regulation, of particular
  input-output characteristics of a reaction system through
  introduction, or control of the concentrations, specificities, or
  activities, of catalysts for its reactions.  Control may be through
  mutation, concentration, physical location, multiple-unit
  interactions, or allosteric regulation.  Cofactors as well as
  enzymes may be control elements over catalysis.  Less directly,
  assembly of catalysts for several reactions to form a network may
  lead to new input-output characteristics through the formation of
  network-catalytic pathways.
} \\
  \parbox[t]{1.5in}{\raggedright
  topological modularity
} & \parbox[t]{5.5in}{\raggedright\mbox{ }
  Used interchangeably with ``network modularity'' in this review.  A
  property of the connectivity in a network that permits its
  decomposition into a collection of \emph{clusters} or
  \emph{communities}, with greater link density among members within a
  community than between members in distinct communities.  Many
  measures such as \emph{network modularity}~\cite{Newman06} or
  Girvan-Newman community detection based on betweenness
  centrality~\cite{Girvan:modularity:02} may be used.  Examples: a
  link whose removal separates the pentose-phosphate network from the
  core network containing universal biosynthetic precursors is the
  synthesis of 3-phosphoglycerate from phosphoenolpyruvate in
  gluconeogenesis; a node whose removal decomposes many loop
  carbon-fixation pathways is succinyl-CoA.
} \\
  \parbox[t]{1.5in}{\raggedright
  robustness
} & \parbox[t]{5.5in}{\raggedright\mbox{ }
  Preservation of some property of structure or function under
  incident perturbations.  These may be external, such as
  concentration fluctuations, or internal, such as removal of a
  reaction or reduction in its flux due to fluctuation in
  concentration of a catalyst.  A property that autotrophic systems
  require to be robust is the ability to \emph{produce} all members in
  a key set of metabolites with ${\mbox{CO}}_2$ as the only carbon
  source.
} \\
  \parbox[t]{1.5in}{\raggedright
  maximum parsimony
} & \parbox[t]{5.5in}{\raggedright\mbox{ }
  A criterion for constructing trees of relatedness that minimizes
  repeated instances of the same innovation over links.  Strict
  parsimony is well-defined, but ranking among solutions that do
  require some repeated instances is not defined by the parsimony
  criterion alone.  In practice, ordering of solutions by parsimony is
  often accompanied by judgments of the probability penalty that a
  richer method such as maximum-likelihood or Bayesian reconstruction
  would attach to repeated innovations.
} \\
  \parbox[t]{1.5in}{\raggedright
  hypergraph
} & \parbox[t]{5.5in}{\raggedright\mbox{ }
  (Used in the appendix) A generalization from the concept of a graph.
  The edges in hypergraphs (called \emph{hyperedges}) possess sets of
  nodes as their boundary, rather than pairs of nodes as for simple
  graphs.  Directed hypergraphs are necessary to capture the
  stoichiometric relations of chemical reaction networks.  See
  Ref.~\cite{Berge:hypergraphs:73}.
} \\
\hline
\end{tabular}
\label{tab:social_makeup}
\end{table}

\clearpage

\end{widetext}

\section{Bipartite graph representations for chemical reaction networks}
\label{sec:graph_explanation}

The stoichiometry of a chemical reaction may be represented by a
\emph{directed hypergraph}~\cite{Berge:hypergraphs:73}.  A hypergraph
differs from a simple graph in that, where each edge of a simple graph
has two points as its boundary, in a hypergraph, a hyper-edge may have
a set of points as its boundary.  In a directed hypergraph, the input
and output sets in the boundary are distinguished.  For the
application to chemistry presented here, each hyper-edge corresponds
to a reaction, and its input and output boundary sets correspond to
moles of the reactant and product molecules.

It is possible to display the hypergraphs representing chemical
reactions as \emph{doubly-bipartite} simple graphs, meaning that both
nodes and edges exist in two types, and that well-formed graphs permit
only certain kinds of connections of nodes to edges.  The bipartite
graph representation of a reaction has an intuitive similarity to the
conventional chemical-reaction notation (shown in
Fig.~\ref{eq:one_reaction} below), but it makes more explicit
reference to the chemical mass-action law as well as to the reaction
stoichiometry.  For appropriately constructed graphs, graph-rewrite
rules correspond one-to-one with evaluation steps of mass-action
kinetics, permitting simplification of complex reaction networks to
isolate key features, while retaining correspondence of the visual and
mathematical representations.

We use graph representations of reaction networks in the text where we
need to show relations among multiple pathways that may connect the
same inputs and outputs (such as acetyl-CoA and succinyl-CoA), and may
draw from the same input and output species (such as ${\mbox{CO}}_2$,
reductant, and water).  Parallel input and output sequences appear as
``ladder'' topology in these graphs, and for the particular pathways
of biological carbon fixation, this is due to the recurrence of
identical functional-group reaction sequences in multiple pathways, as
discussed in Sec.~\ref{sec:modularity}.

In this appendix we define the graph representation used in the text,
introduce graph-reduction procedures and prove that they satisfy the
mathematical property of associativity, and provide solutions for the
particular simplification of interacting rTCA and Wood-Ljungdahl
pathways in a diluting environment. 

All examples in this appendix use the same simplified projection onto the CHO sector that is used in diagrams in the main text.  Actual reaction free energies will be driven by coupled energies of hydrolysis of ATP or oxidation of thiols to thioesters.  The graph-reduction methods described in the next section may be used to include such effects into lumped-parameter representations of multi-reagent reaction sequences that regenerate energetic intermediates such as ATP or CoA in a network where these are made explicit.

\subsection{Definition of graphic elements}

\subsubsection{Basic elements and well-formed graphs}

The elements in a bipartite-graph representation of a chemical
reaction or reaction network are defined as follows:

\begin{itemize}

\item Filled dots dots represent concentrations of chemical species.
  Each such dot is given a label indicating the species, such as
  \begin{displaymath}
    \mbox{\includegraphics[scale=0.65]{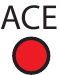}} 
    \leftrightarrow
    \left[ \mbox{ACE} \right] , 
  \end{displaymath}
  used to refer to acetate in the text.

\item Dashed lines represent transition states of reactions.  Each is
  given a label indicating the reaction, as in
  $\mbox{\includegraphics[scale=0.65]{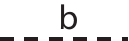}}$.  

\item Hollow circles indicate inputs or outputs between molecular
  species and transition states, as in
  \begin{displaymath}
    \mbox{\includegraphics[scale=0.65]{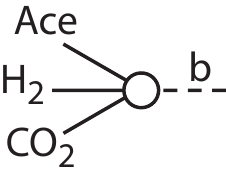}} . 
  \end{displaymath} 
  Each circle is associated with the complex of reactants or products
  to the associated reaction, indicated as labeled line stubs.  

\item Hollow circles are tied to molecular concentrations with solid
  lines $\mbox{\includegraphics[scale=0.65]{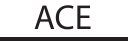}}$;
  one line per mole of reactant or product participating in the
  reaction.  (That is, if $m$ moles of a species A enter a reaction
  $b$, then $m$ lines connect the dot corresponding to $\left[
    \mbox{A} \right]$ to the hollow circle leading into reaction $b$.
  This choice uses graph elements to carry information about
  stoichiometry, as an alternative to labeling input- or output-lines
  to indicate numbers of moles.)

\item Full reactions are defined when two hollow circles are connected
  by the appropriate transition state, as in
  \begin{displaymath}
    \mbox{\includegraphics[scale=0.65]{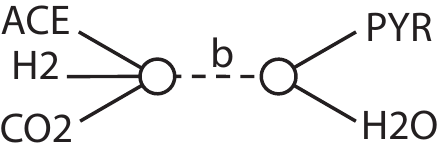}} , 
  \end{displaymath}
  describing the reductive carboxylation of acetate to form pyruvate.

\item The bipartite graph for a fully specified reaction takes the form
  \begin{equation}
    \mbox{\includegraphics[scale=0.65]{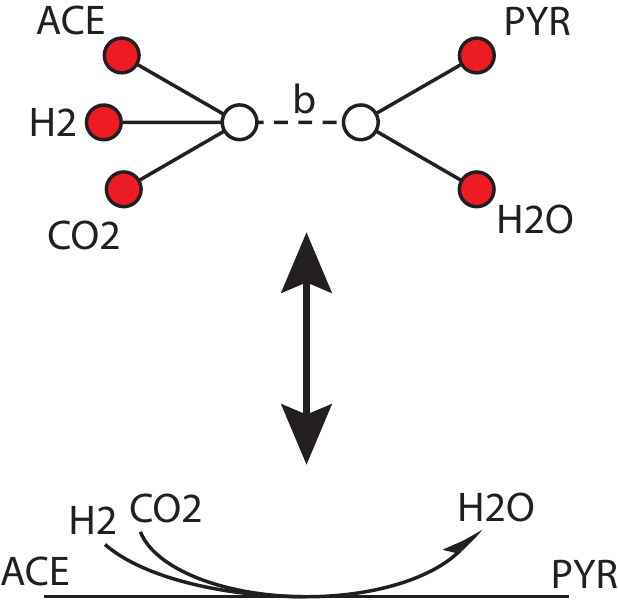}} , 
  \label{eq:one_reaction}
  \end{equation}
  where labeled stubs are connected to filled circles by mole-lines.
  The bipartite-graph corresponds to the standard chemical notation
  for the same reaction as shown. 

\end{itemize}

\subsubsection{Assignment of graph elements to terms in the
  mass-action rate equation}

The mass-action kinetics for a graph such as the reductive
carboxylation of acetate is given in terms of two
half-reaction currents, which we may denote with the reaction label
and an arbitrary sign as
\begin{eqnarray}
  j_b^{+} 
& = & 
  k_b
  \left[ \mbox{ACE} \right]
  \left[ {\mbox{CO}}_2 \right]
  \left[ {\mbox{H}}_2 \right]
\nonumber \\
  j_b^{-} 
& = & 
  {\bar{k}}_b
  \left[ \mbox{PYR} \right]
  \left[ {\mbox{H}}_2 \mbox{O} \right] . 
\label{eq:react_b_half_currents}
\end{eqnarray}
$k_b$ and ${\bar{k}}_b$ denote the forward and reverse half-reaction
rate constants.  The total reaction current $J_b \equiv j_b^{+} -
j_b^{-}$ is related to the contribution of this reaction to the
changes in concentration as
\begin{eqnarray}
  \dot{\left[ \mbox{ACE} \right]} = 
  \dot{\left[ {\mbox{CO}}_2 \right]} = 
  \dot{\left[ {\mbox{H}}_2 \right]}
& = & 
  -J_b
\nonumber \\
  \dot{\left[ \mbox{PYR} \right]} = 
  \dot{\left[ {\mbox{H}}_2 \mbox{O} \right]}
& = & 
  J_b , 
\label{eq:react_b_conc_changes}
\end{eqnarray}
where the overdot denotes the time derivative.  Reaction currents on
graphs do not have inherent directions, reflecting the microscopic
reversibility of reactions.  All sources of irreversibility are to be
made explicit in the chemical potentials that constitute the boundary
conditions for reactions.

Each term in the mass-action rate equation may be identified with a
specific graphical element in the bipartite representation.  The
half-reaction rate constants $k_b$, ${\bar{k}}_b$ are associated with
the hollow circles, and the current $J_b$ (which is the
time-derivative of the coordinate giving the ``extent of the
reaction'') is associated with the transition-state dashed line.
Concentrations, as noted, are associated with filled dots, and
stoichiometric coefficients are associated with the multiplicities of
solid lines.

\subsection{Graph reduction for reaction networks in steady state}

Networks of chemical reactions in steady state satisfy the constraints
that the input and output currents to each chemical species (including
any external sources or sinks) sum to zero.  These constraints are the
basis of stoichiometric flux-balance
analysis~\cite{Price:ext_pathways:02,Famili:FBA_SVD:03,%
  Palsson:systems_bio:06,Feist:networks:09}, but they can also be used
to eliminate internal nodes as explicit variables, leading to
lumped-parameter expressions for entire sub-networks as ``effective''
vertices or reactions.  With appropriate absorption of externally
buffered reagents into rate constants, this network reduction can be
done exactly, without loss of information.  An example of such a
reduction is the Michaelis-Menton representation of multiple substrate
binding at enzymes.  Systematic methods for network reduction were one
motivation behind Sinan$\breve{\mbox{o}}$glu's graphic
methods~\cite{Sinanoglu:ChemNetworks:75,Sinanoglu:AlgChem:84}.  More
sophisticated stochastic approaches have recently been used to include
fluctuation properties in effective vertices, generalizing the
Michaelis relation beyond mean field~\cite{Sinitsyn:CG_chem_nets:09}.

The map we have given of mass-action rate parameters to graphic
elements allows us to represent steady-state network reduction in
terms of graph reduction.  In this approach, rewrite rules for the
removal of graph elements are mapped to composition rules for
half-reaction rate constants and stoichiometric coefficients.  These
composition rules can be proved to be associative, leading to an
algebra for graph reduction.  Here we sketch the rewrite rules
relevant to reduction of the citric-acid cycle graph.  In the next
subsection we will reduce the graph, to the form used in the text.

\subsubsection{The base composition rule for removal of a single
  internal species}
\label{sec:base_case}

The simplest reduction is removal of an intermediate chemical species
that is the sole output to one reaction, and the sole input to
another, in a linear chain.  Examples in the TCA cycle include malate
(MAL) and isocitrate (ISC), produced by reductions and consumed by
dehydrations.  They also include citrate (CIT) itself, produced by the
hydration of aconitate and consumed by retro-aldol cleavage. 

\begin{figure}[ht]
  \begin{center} 
  \includegraphics[scale=0.5]{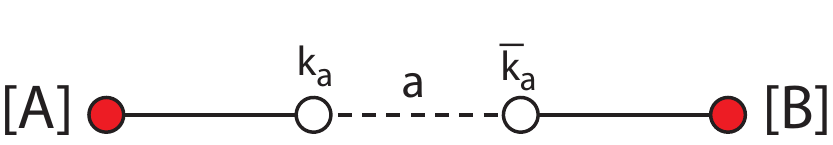}
  \caption{
  Basic reaction graph.  $\left[ \mbox{A} \right]$ and $\left[
    \mbox{B} \right]$ are concentrations associated with the two
  colored nodes.  Forward and backward rate constants $k_a$ and
  ${\bar{k}}_a$ are associated with the two unfilled circles.  The
  associated reaction state current is $J_a$.
    \label{fig:basic_reaction} 
  }
  \end{center}
\end{figure}

For a single linear reaction as shown in
Fig.~\ref{fig:basic_reaction}, the mass-action law is 
\begin{equation}
  \left[ \mbox{A} \right]
  k_a - 
  \left[ \mbox{B} \right]
  {\bar{k}}_a = 
  J_a , 
\label{eq:rate_const_def}
\end{equation}
and concentrations change as 
\begin{eqnarray}
  \dot{\left[ \mbox{A} \right]}
& = & 
  - J_a
\nonumber \\
  \dot{\left[ \mbox{B} \right]}
& = & 
  J_a .
\label{eq:J_a_dots_def}
\end{eqnarray}
The equilibrium constant for the reaction $\mbox{A} \rightarrow
\mbox{B}$ is
\begin{equation}
  K_{
    \mbox{\scriptsize A} \rightarrow \mbox{\scriptsize B}
  } = 
  \frac{k_a}{{\bar{k}}_a} . 
\label{eq:eqm_const_def}
\end{equation}

\begin{figure}[ht]
  \begin{center} 
  \includegraphics[scale=0.5]{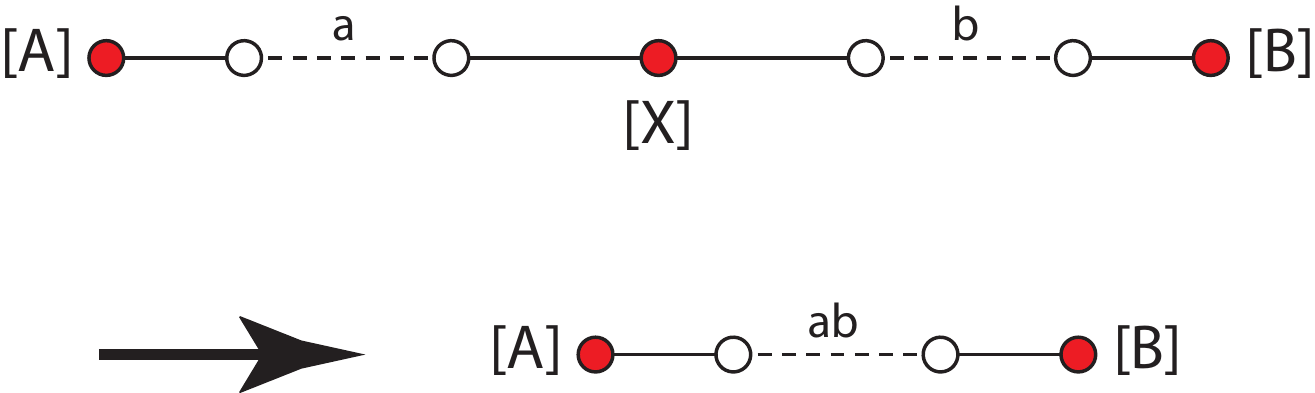}
  \caption{
  Removal of an internal species $\mbox{X}$ from a diagram with
  elementary reactions.  Rate constant pairs $\left( k_a , {\bar{k}}_a
  \right)$, $\left( k_b , {\bar{k}}_b \right)$ are used to define new
  rate constants $\left( k_{ab} , {\bar{k}}_{ab} \right)$ for the
  effective transition state $ab$.
    \label{fig:two_basic_reactions} 
  }
  \end{center}
\end{figure}

For two such reactions in a chain, as shown in
Fig.~\ref{fig:two_basic_reactions}, the mass-action laws are 
\begin{eqnarray}
  \left[ \mbox{A} \right]
  k_a - 
  \left[ \mbox{X} \right]
  {\bar{k}}_a 
& = & 
  J_a 
\nonumber \\
  \left[ \mbox{X} \right]
  k_b - 
  \left[ \mbox{B} \right]
  {\bar{k}}_b 
& = & 
  J_b , 
\label{eq:two_rate_const_def}
\end{eqnarray}
and the conservation laws become
\begin{eqnarray}
  \dot{\left[ \mbox{A} \right]}
& = & 
  - J_a
\nonumber \\
  \dot{\left[ \mbox{X} \right]} 
& = &  
  J_a - J_b 
\nonumber \\
  \dot{\left[ \mbox{B} \right]}
& = & 
  J_b .  
\label{eq:J_ab_dots_def}
\end{eqnarray}

Under the steady-state condition $\dot{\left[ \mbox{X} \right]} = 0$,
we wish to replace the
equations~(\ref{eq:two_rate_const_def},\ref{eq:J_ab_dots_def}) with a
rate equation 
\begin{equation}
  \left[ \mbox{A} \right]
  k_{ab} - 
  \left[ \mbox{B} \right]
  {\bar{k}}_{ab} = 
  J_{ab} 
\label{eq:bound_rate_const_def}
\end{equation}
and a conservation law expressed in terms of $J_a = J_{ab} = J_b$.
The rate constants in Eq.~(\ref{eq:bound_rate_const_def}) are to be
specified through a composition rule 
\begin{equation}
  \left( k_a , {\bar{k}}_a \right) \circ
  \left( k_b , {\bar{k}}_b \right) = 
  \left( k_{ab} , {\bar{k}}_{ab} \right) 
\label{eq:composition_notation}
\end{equation}
derived from the graph rewrite.  Removing $\left[ \mbox{X} \right]$
from the mass-action equations using $\dot{\left[ \mbox{X} \right]} =
0$, we derive that the rate constants satisfying
Eq.~(\ref{eq:bound_rate_const_def}) are given by 
\begin{eqnarray}
  k_{ab} 
& = & 
  \frac{
    k_a k_b
  }{
    {\bar{k}}_a + k_b
  }
\nonumber \\ 
  {\bar{k}}_{ab} 
& = & 
  \frac{
    {\bar{k}}_a {\bar{k}}_b
  }{
    {\bar{k}}_a + k_b
  } . 
\label{eq:elem_comp_rule}
\end{eqnarray}
The associated equilibrium constant correctly satisfies the relation
\begin{equation}
  \frac{k_{ab}}{{\bar{k}}_{ab}} = 
  \frac{k_a}{{\bar{k}}_a} 
  \frac{k_b}{{\bar{k}}_b} . 
\label{eq:eqm_const_compound}
\end{equation}

\subsubsection{Associativity of the elementary composition rule}

\begin{figure}[ht]
  \begin{center} 
  \includegraphics[scale=0.5]{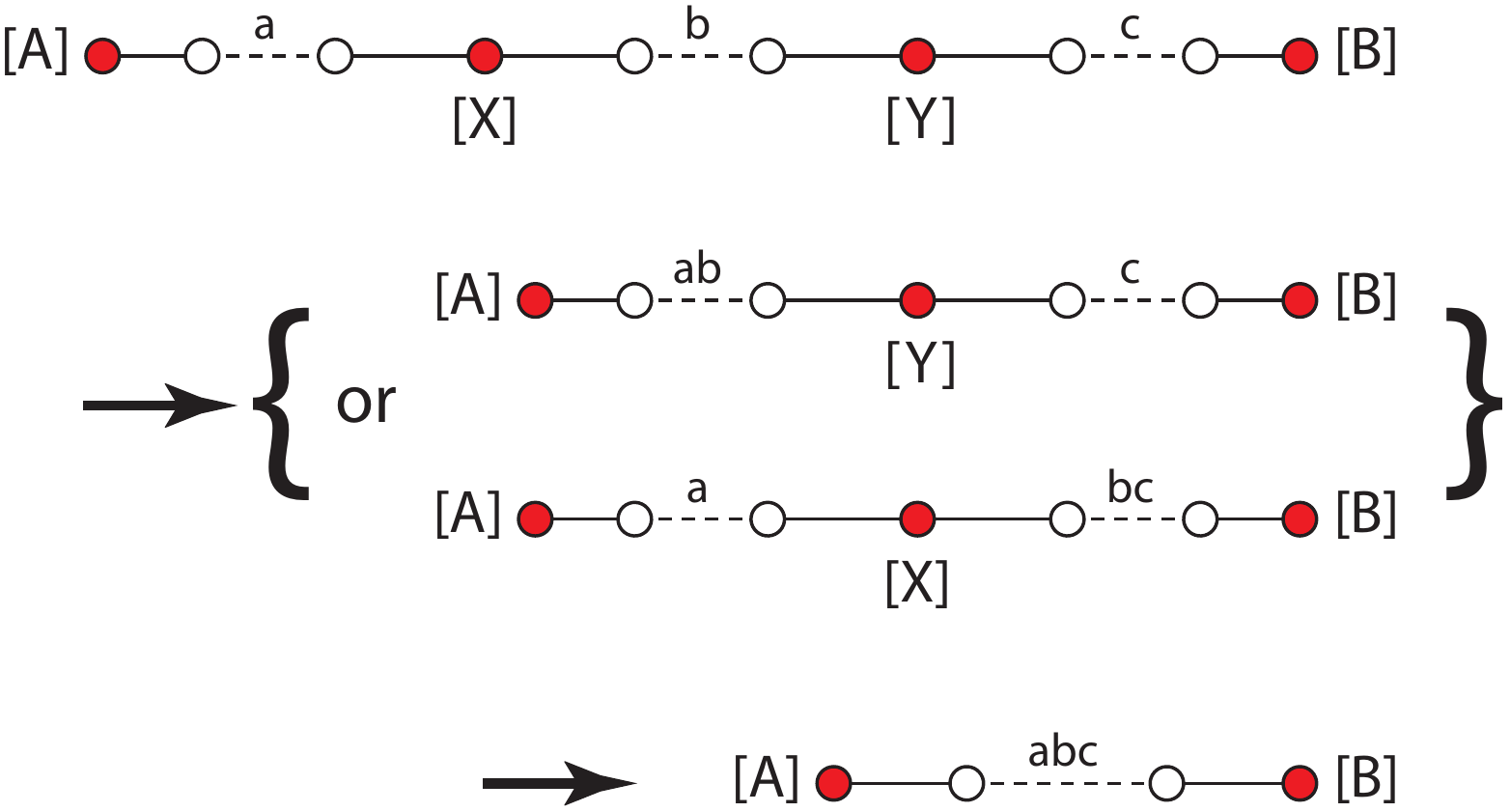}
  \caption{
  Composition of three reactions $a$, $b$, $c$ can proceed by
  elimination of either $\mbox{X}$ or $\mbox{Y}$ first.
    \label{fig:three_basic_reactions} 
  }
  \end{center}
\end{figure}

The composition rule~(\ref{eq:eqm_const_compound}) is associative,
meaning that internal nodes may be removed from chains of reactions in
any order, as shown in Fig.~\ref{fig:three_basic_reactions}.  All
composition rules derived in the remainder of this appendix will be
variants on the elementary rule (with additional buffered
concentration variables added), so we demonstrate associativity for
the base case as the foundation for other cases.

From Eq.~(\ref{eq:eqm_const_compound}) for $\left( k_a , {\bar{k}}_a
\right) \circ \left( k_b , {\bar{k}}_b \right)$, followed by the
equivalent expressions for $\left( k_{ab} , {\bar{k}}_{ab} \right)
\circ \left( k_c , {\bar{k}}_c \right)$, $\left( k_a , {\bar{k}}_a
\right) \circ \left( k_{bc} , {\bar{k}}_{bc} \right)$, and $\left( k_b
, {\bar{k}}_b \right) \circ \left( k_c , {\bar{k}}_c \right)$, we
derive the sequence of reductions 
\begin{eqnarray}
  k_{abc}
& = & 
  \frac{
    k_{ab} k_c
  }{
    {\bar{k}}_{ab} + k_c
  }
\nonumber \\
& = & 
  \frac{
    k_a k_b k_c
  }{
    {\bar{k}}_a {\bar{k}}_b + 
    \left( {\bar{k}}_a + k_b \right) k_c
  }
\nonumber \\
& = & 
  \frac{
    k_a k_b k_c
  }{
    {\bar{k}}_a  \left( {\bar{k}}_b + k_c \right) + 
    k_b k_c 
  }
\nonumber \\
& = & 
  \frac{
    k_a k_{bc}
  }{
    {\bar{k}}_a + k_{bc}
  } , 
\label{eq:elem_comp_assoc}
\end{eqnarray}
and a similar equation follows for ${\bar{k}}_{abc}$.  Thus we have 
\begin{equation}
  \left[
    \left( k_a , {\bar{k}}_a \right) \circ 
    \left( k_b , {\bar{k}}_b \right)
  \right] \circ
  \left( k_c , {\bar{k}}_c \right) = 
  \left( k_a , {\bar{k}}_a \right) \circ 
  \left[
    \left( k_b , {\bar{k}}_b \right) \circ 
    \left( k_c , {\bar{k}}_c \right)
  \right] . 
\label{eq:notation_comp_assoc}
\end{equation}

\subsubsection{Removal of internal nodes that require other inputs or
  outputs} 
\label{sec:composite_case}

Next we consider the elimination of an internal node $\left[ \mbox{X}
  \right]$ that is produced or consumed together with other products
or reactants.  Conservation $\dot{\left[ \mbox{X} \right]} = 0$
implies relations among the currents of these other species as well.
All remaining graph reductions that we will perform for the TCA cycle
are of this kind.  In some cases both the secondary product and
reactant are the solvent (water), as in the aconitase reactions
(repeated in TCA, 3HB, 4HB, and bicycle pathways).  In other cases
they are reductants or inputs such as ${\mbox{CO}}_2$ that we consider
buffered in the environment.

The pair of mass action equations we wish to reduce are\footnote{In
  this and the following examples, we consider single additional
  species $\left[ \mbox{C} \right]$ and $\left[ \mbox{D} \right]$.
  These may readily be generalized to a variety of cases in which the
  additional reagents are $\prod_{k=1}^p \left[ {\mbox{C}}_k \right]$
  and $\prod_{l=1}^q \left[ {\mbox{D}}_l \right]$.}
\begin{eqnarray}
  \left[ \mbox{A} \right]
  k_a - 
  \left[ \mbox{X} \right]
  \left[ \mbox{C} \right]
  {\bar{k}}_a 
& = & 
  J_a 
\nonumber \\
  \left[ \mbox{X} \right]
  \left[ \mbox{D} \right]
  k_b - 
  \left[ \mbox{B} \right]
  {\bar{k}}_b 
& = & 
  J_b , 
\label{eq:two_rate_const_comp_def}
\end{eqnarray}
and the desired reduced form is 
\begin{equation}
  \left[ \mbox{A} \right]
  \left[ \mbox{D} \right]
  k_{ab} - 
  \left[ \mbox{C} \right]
  \left[ \mbox{B} \right]
  {\bar{k}}_{ab} = 
  J_{ab} .  
\label{eq:bound_rate_const_comp_def}
\end{equation}

\begin{figure}[ht]
  \begin{center} 
  \includegraphics[scale=0.5]{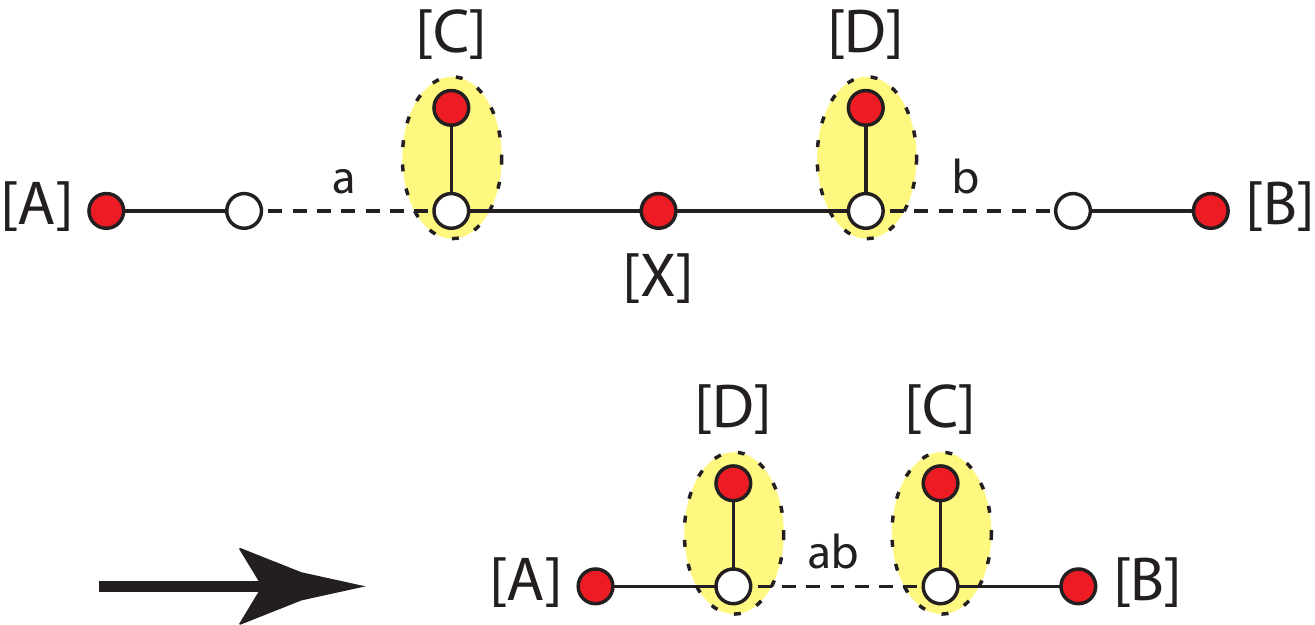}
  \caption{
  Representation of a composite graph with internal connections other
  than those to $\mbox{X}$ as an effective elementary graph.
  Highlights denote the absorption of other species into modifications
  of effective rate constants coupled to $\mbox{X}$ at $a$ and $b$.
  These are then used to define the elementary-form rate constants
  ${\hat{k}}_{ab}$ and ${\bar{\hat{k}}}_{ab}$ in the reduced graph.
    \label{fig:two_composite_reactions} 
  }
  \end{center}
\end{figure}

We first reduce Eq.~(\ref{eq:two_rate_const_comp_def}) to the base
case of the previous section, by absorbing the concentrations not to
be removed into a pair of effective rate constants 
\begin{eqnarray}
  \left[ \mbox{A} \right]
  k_a - 
  \left[ \mbox{X} \right]
  \left( 
    \left[ \mbox{C} \right]
    {\bar{k}}_a 
  \right)
& = & 
  J_a 
\nonumber \\
  \left[ \mbox{X} \right]
  \left( 
    \left[ \mbox{D} \right]
    k_b 
  \right) - 
  \left[ \mbox{B} \right]
  {\bar{k}}_b 
& = & 
  J_b . 
\label{eq:two_rate_const_comp_hat}
\end{eqnarray}
From these we derive a composition equation 
\begin{equation}
  \left[ \mbox{A} \right]
  {\hat{k}}_{ab} - 
  \left[ \mbox{B} \right]
  {\bar{\hat{k}}}_{ab} = 
  J_{ab} , 
\label{eq:bound_rate_const_comp_hat}
\end{equation}
corresponding to the graph representation in
Fig.~\ref{fig:two_composite_reactions}.  We may then define
${\hat{k}}_{ab}$ and ${\bar{\hat{k}}}_{ab}$ by the elementary
composition rule~(\ref{eq:composition_notation})
\begin{equation}
  \left( k_a , \left[ \mbox{C} \right] {\bar{k}}_a \right) \circ
  \left( \left[ \mbox{D} \right] k_b , {\bar{k}}_b \right) = 
  \left( {\hat{k}}_{ab} , {\bar{\hat{k}}}_{ab} \right) , 
\label{eq:composition_notation_hat}
\end{equation}
giving the transformation\footnote{Note that if $\left[ \mbox{C}
    \right]$ and $\left[ \mbox{D} \right]$ are the same species these
  cancel in the numerator and denominator of
  Eq.~(\ref{eq:elem_comp_rule_hat}), and the same applies to common
  factors in products $\prod_{k=1}^p \left[ {\mbox{C}}_k \right]$ and
  $\prod_{l=1}^q \left[ {\mbox{D}}_l \right]$.  Therefore these
  factors may simply be removed before the graph reduction if desired,
  because they encoded redundant constraints with the conservation law
  already implied by $\dot{\left[ \mbox{X} \right]} = 0$.  The
  irrelevance of redundant species in the graph reduction for removal
  of $\left[ \mbox{X} \right]$ is radically different from the
  graphically similar-looking role of a network catalyst which is both
  an input and an output of the same \emph{reaction}.  Network
  catalysts are essential to the determination of reaction rates.}
\begin{eqnarray}
  {\hat{k}}_{ab} 
& = & 
  \frac{
    k_a \left[ \mbox{D} \right] k_b
  }{
    \left[ \mbox{C} \right] {\bar{k}}_a + \left[ \mbox{D} \right] k_b
  }
\nonumber \\ 
  {\bar{\hat{k}}}_{ab} 
& = & 
  \frac{
    \left[ \mbox{C} \right] {\bar{k}}_a {\bar{k}}_b
  }{
    \left[ \mbox{C} \right] {\bar{k}}_a + \left[ \mbox{D} \right] k_b
  } . 
\label{eq:elem_comp_rule_hat}
\end{eqnarray}

\begin{figure}[ht]
  \begin{center} 
  \includegraphics[scale=0.5]{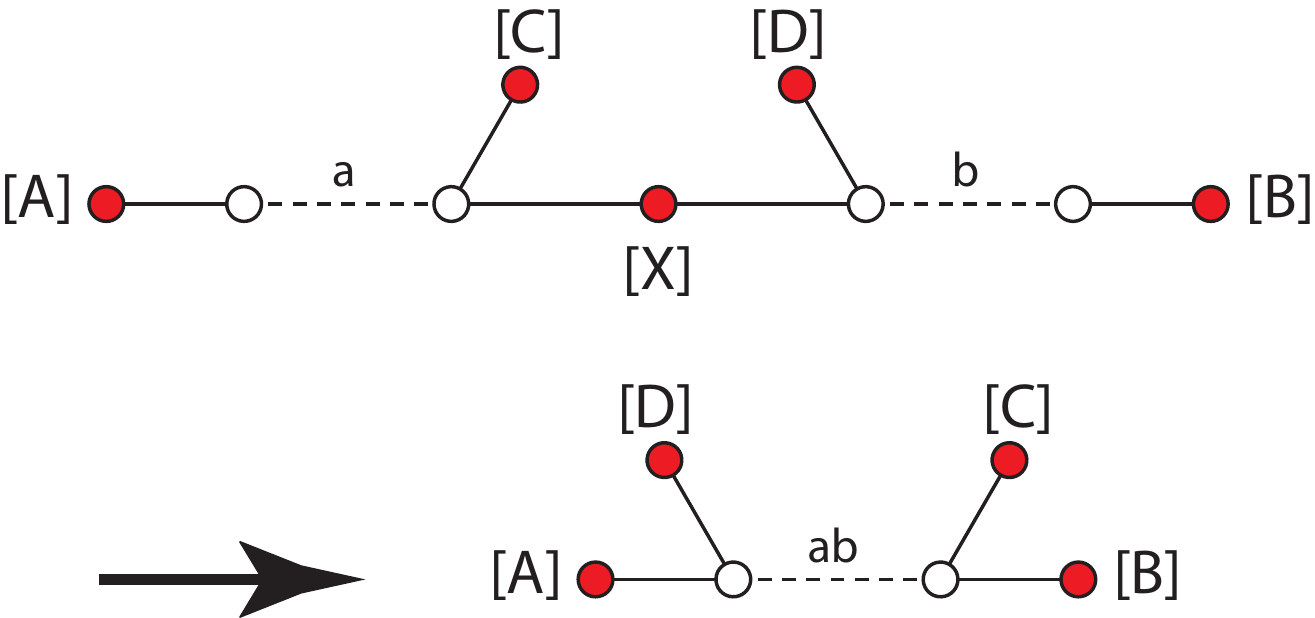}
  \caption{
  The composite graph corresponding to the reduction from
  Eq.~(\ref{eq:two_rate_const_comp_def}) to
  Eq.~(\ref{eq:bound_rate_const_comp_def}).  
    \label{fig:two_composite_direct} 
  }
  \end{center}
\end{figure}

Now removing the factors of $\left[ \mbox{C} \right]$ and $\left[
  \mbox{D} \right]$ used to define the hatted rate constants, 
\begin{eqnarray}
  {\hat{k}}_{ab}
& = & 
  \left[ \mbox{D} \right] k_{ab}
\nonumber \\
{\bar{\hat{k}}}_{ab} 
& = & 
  \left[ \mbox{C} \right]
  {\bar{k}}_{ab} , 
\label{eq:hat_nohat_reln}
\end{eqnarray}
we obtain a direct expression for the composition rule in
Eq.~(\ref{eq:bound_rate_const_comp_hat}), of
\begin{eqnarray}
  k_{ab} 
& = & 
  \frac{
    k_a k_b
  }{
    \left[ \mbox{C} \right] {\bar{k}}_a + \left[ \mbox{D} \right] k_b
  }
\nonumber \\ 
  {\bar{k}}_{ab} 
& = & 
  \frac{
    {\bar{k}}_a {\bar{k}}_b
  }{
    \left[ \mbox{C} \right] {\bar{k}}_a + \left[ \mbox{D} \right] k_b
  } , 
\label{eq:gen_comp_rule}
\end{eqnarray}
which is the interpretation of the graph reduction shown in
Fig.~\ref{fig:two_composite_direct}.

\subsubsection{Associativity for composite graphs}

\begin{figure}[ht]
  \begin{center} 
  \includegraphics[scale=0.5]{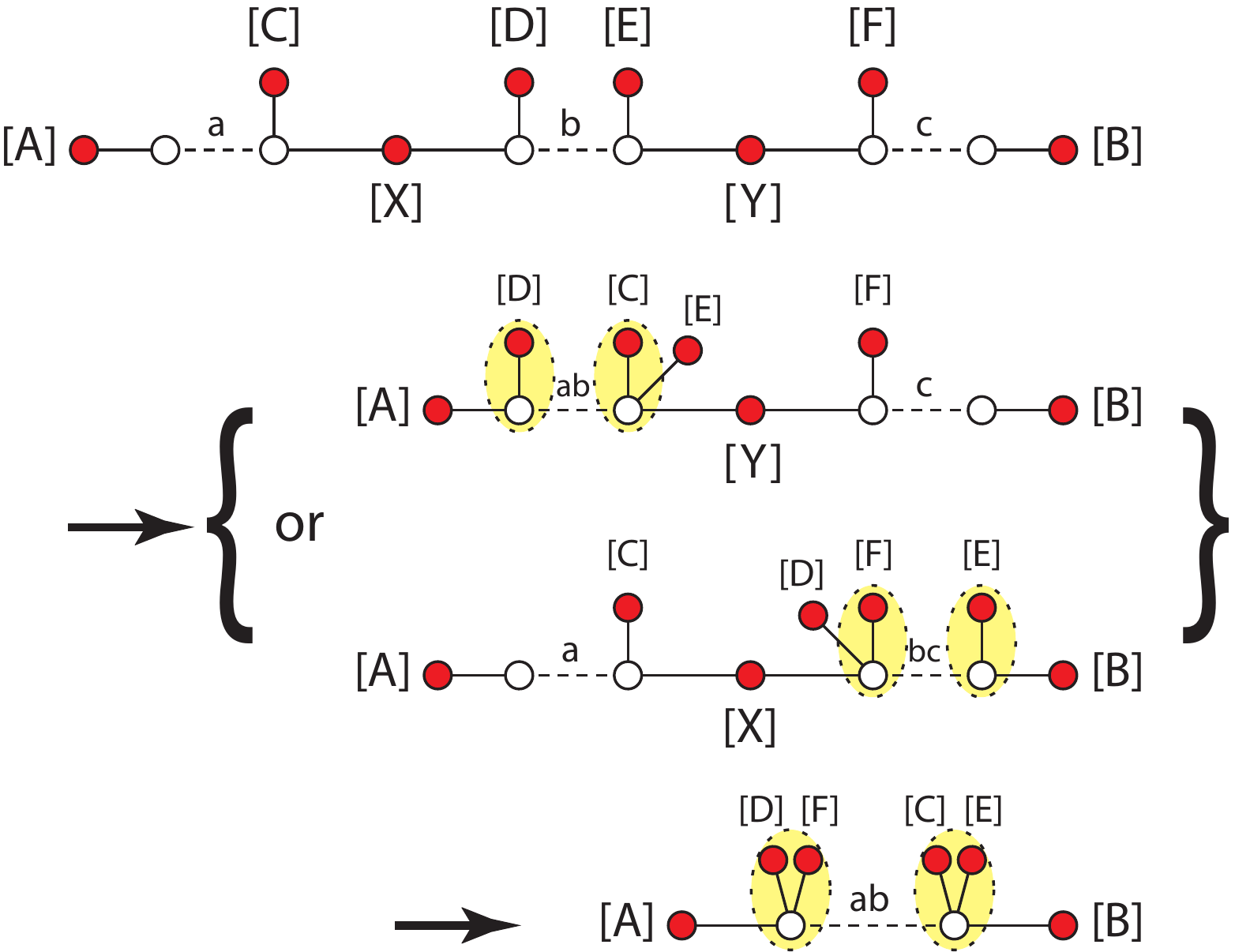}
  \caption{
  A two-step reduction with other internal connections, which may be
  performed by removing either $\mbox{X}$ or $\mbox{Y}$ first. 
    \label{fig:three_composite_reactions} 
  }
  \end{center}
\end{figure}

Associativity for composite graphs follows from the associativity of
the elementary composition rule~(\ref{eq:notation_comp_assoc}), via
the grouping~(\ref{eq:composition_notation_hat}).  To show how this
works, we demonstrate associativity for the minimal case shown in
Fig.~\ref{fig:three_composite_reactions}.  The important features are
that the graph ``re-wiring'' follows from composition of the rule
demonstrated in Fig.~\ref{fig:two_composite_direct}, and the
composition rule for rate constants permits consistent removal of the
necessary factors of reagent concentrations.

The application of the elementary reduction to remove $\mbox{X}$,
corresponding to the second line in
Fig.~\ref{fig:three_composite_reactions}, yields
Eq's.~(\ref{eq:composition_notation_hat},\ref{eq:elem_comp_rule_hat}).
An equivalent removal of $\mbox{Y}$ first (the third line of
Fig.~\ref{fig:three_composite_reactions}) gives
\begin{equation}
  \left( k_b , \left[ \mbox{E} \right] {\bar{k}}_b \right) \circ
  \left( \left[ \mbox{F} \right] k_c , {\bar{k}}_c \right) = 
  \left( {\hat{k}}_{bc} , {\bar{\hat{k}}}_{bc} \right) , 
\label{eq:composition_notation_Y}
\end{equation}
with rule 
\begin{eqnarray}
  {\hat{k}}_{bc} 
& = & 
  \frac{
    k_b \left[ \mbox{F} \right] k_c
  }{
    \left[ \mbox{E} \right] {\bar{k}}_b + \left[ \mbox{F} \right] k_c
  }
\nonumber \\ 
  {\bar{\hat{k}}}_{bc} 
& = & 
  \frac{
    \left[ \mbox{E} \right] {\bar{k}}_b {\bar{k}}_c
  }{
    \left[ \mbox{E} \right] {\bar{k}}_b + \left[ \mbox{F} \right] k_c
  } . 
\label{eq:elem_comp_rule_hat_bc}
\end{eqnarray}
The two equivalent rules for removing whichever internal node was not
removed in the first reduction are 
\begin{eqnarray}
  \left( 
    {\hat{k}}_{ab} , \left[ \mbox{E} \right] {\bar{\hat{k}}}_{ab} 
  \right) \circ
  \left( \left[ \mbox{F} \right] k_c , {\bar{k}}_c \right) 
& = & 
  \left( {\hat{k}}_{abc} , {\bar{\hat{k}}}_{abc} \right) , 
\nonumber \\
  \left( k_a , \left[ \mbox{C} \right] {\bar{k}}_a \right) \circ
  \left( 
    \left[ \mbox{D} \right]  {\hat{k}}_{bc} , {\bar{\hat{k}}}_{bc} 
  \right) 
& = & 
  \left( {\hat{k}}_{abc} , {\bar{\hat{k}}}_{abc} \right) . 
\label{eq:composition_notation_abc_hat}
\end{eqnarray}

Composing these rules for intermediate rate constants, we may check
that 
\begin{eqnarray}
  {\hat{k}}_{abc}
& = & 
  \frac{
    {\hat{k}}_{ab} \left[ \mbox{F} \right] k_c
  }{
    \left[ \mbox{E} \right] {\bar{\hat{k}}}_{ab} + 
    \left[ \mbox{F} \right] k_c
  }
\nonumber \\
& = & 
  \frac{
    \left( k_a \left[ \mbox{D} \right] k_b \right) 
    \left[ \mbox{F} \right] k_c
  }{
    \left[ \mbox{C} \right] {\bar{k}}_a 
    \left[ \mbox{E} \right] {\bar{k}}_b + 
    \left( 
      \left[ \mbox{C} \right] {\bar{k}}_a + 
      \left[ \mbox{D} \right] k_b 
    \right) 
    \left[ \mbox{F} \right] k_c
  }
\nonumber \\
& = & 
  \frac{
    k_a \left[ \mbox{D} \right] 
    \left( k_b \left[ \mbox{F} \right] k_c \right) 
  }{
    \left[ \mbox{C} \right] {\bar{k}}_a  
    \left( 
      \left[ \mbox{E} \right] {\bar{k}}_b + 
      \left[ \mbox{F} \right] k_c 
    \right) + 
    \left[ \mbox{D} \right] k_b \left[ \mbox{F} \right] k_c 
  }
\nonumber \\
& = & 
  \frac{
    k_a \left[ \mbox{D} \right]  {\hat{k}}_{bc}
  }{
    \left[ \mbox{C} \right] {\bar{k}}_a + 
    \left[ \mbox{D} \right] {\hat{k}}_{bc}
  } , 
\label{eq:elem_comp_assoc_abc_hat}
\end{eqnarray}
and a similar equation follows for ${\bar{\hat{k}}}_{abc}$.
Converting the hatted forms to the normal reaction form produces the
rate equation
\begin{equation}
  \left[ \mbox{A} \right]
  \left[ \mbox{D} \right]
  \left[ \mbox{F} \right]
  k_{abc} - 
  \left[ \mbox{C} \right]
  \left[ \mbox{E} \right]
  \left[ \mbox{B} \right]
  {\bar{k}}_{abc} = 
  J_{abc} . 
\label{eq:bound_rate_const_comp_abc}
\end{equation}

We may directly obtain the rate constants $k_{abc}$, ${\bar{k}}_{abc}$
with the composition rule
\begin{eqnarray}
  \left( k_{abc} , {\bar{k}}_{abc} \right) 
& = & 
  \left( 
    k_{ab} , {\bar{k}}_{ab} 
  \right) \circ
  \left( k_c , {\bar{k}}_c \right) 
\nonumber \\
& = & 
  \left( k_a , {\bar{k}}_a \right) \circ
  \left( 
    k_{bc} , {\bar{k}}_{bc} 
  \right) , 
\label{eq:composition_notation_abc}
\end{eqnarray}
using the appropriate version of the graph-dependent evaluation
rule~(\ref{eq:gen_comp_rule}) in each step.  The resulting
composition~(\ref{eq:composition_notation_abc}) is automatically
associative, because it satisfies the conversion
\begin{eqnarray}
  {\hat{k}}_{abc}
& = & 
  \left[ \mbox{D} \right] \left[ \mbox{F} \right] k_{abc}
\nonumber \\
{\bar{\hat{k}}}_{abc} 
& = & 
  \left[ \mbox{C} \right] \left[ \mbox{E} \right]
  {\bar{k}}_{abc} 
\label{eq:hat_nohat_reln_abc}
\end{eqnarray}
with Eq.~(\ref{eq:elem_comp_assoc_abc_hat}), which is associative.  As
a final check, the equilibrium constants in the normal reaction form
satisfy the necessary chain rule
\begin{equation}
  \frac{k_{abc}}{{\bar{k}}_{abc}} = 
  \frac{k_a}{{\bar{k}}_a} 
  \frac{k_b}{{\bar{k}}_b} 
  \frac{k_c}{{\bar{k}}_c} . 
\label{eq:eqm_const_compound_abc}
\end{equation}
Intermediate (hatted) rate constants have been used here to show how
associativity is inherited from the base case.  The examples below
work directly with the actual (un-hatted) rate constants, which keep
the network in its literal form at each reduction.

\subsection{Application to the citric-acid cycle reactions}

Using this graph representation and the associated graph reductions,
we may express the qualitative kinetics associated with network
autocatalysis in the rTCA cycle.  We use a minimal model network in
which only the cycle intermediates are represented explicitly, and
only the CHO stoichiometry is retained. As noted above,
  phosphorylated intermediates and thioesters, including the
  energetically important substrate-level phosphorylation of citrate
  and succinate, are not represented. External sources or sinks are
used to buffer only four compounds in the network, which are
${\mbox{CO}}_2$, ${\mbox{H}}_2$, ${\mbox{H}}_2 \mbox{O}$, and a pool
of reduced carbon which we take to be acetate (ACE, or ${\mbox{CH}}_3
\mbox{COOH}$) because it has the lowest free energy of formation of
cycle intermediates under reducing conditions (following
Ref.~\cite{Miller:TCA_Gibbs:90}) and is the natural drain
compound~\cite{Smith:universality:04}.

The purpose of network reduction in such a model is to produce a graph
in which each element corresponds to a specific control parameter for
the interaction of conservation laws with non-equilibrium boundary
conditions.  ${\mbox{CO}}_2$, ${\mbox{H}}_2$, and ${\mbox{H}}_2
\mbox{O}$ provide sources of carbon and reductant, and an output for
reduced oxygen atoms.  Because they comprise different ratios of three
elements, any set of concentrations is consistent with a Gibbs
equilibrium, and the chemical potentials corresponding to the elements
are preserved by the conservation laws of arbitrary reactions.  A
fourth boundary condition for acetate cannot be linearly independent
in equilibrium, and drives the steady-state reaction flux.

Such a model is limited in many ways.  The replacement of explicit
(and unknown) parasitic side reactions, from all cycle intermediates,
by a single loss rate for acetate may fail to capture
concentration-dependent losses, in a way that cannot simply be
absorbed into lumped rate constants.  Moreover, the rate constants
themselves depend on catalysts, and reasonable values for these in a
prebiotic or early-cellular context are unknown.  Therefore all
critical properties of the model are expressed relative to these rate
constants.  The reduction remains meaningful, however, because the
lumped-parameter rate constants are controlled by the three buffered
environmental compounds ${\mbox{CO}}_2$, ${\mbox{H}}_2$, and ${\mbox{H}}_2
\mbox{O}$, leaving the network flux to be controlled by the
disequilibrium concentration of acetate.

\subsubsection{The graph reduction sequence}

\begin{figure}[ht]
  \begin{center} 
  \includegraphics[scale=0.65]{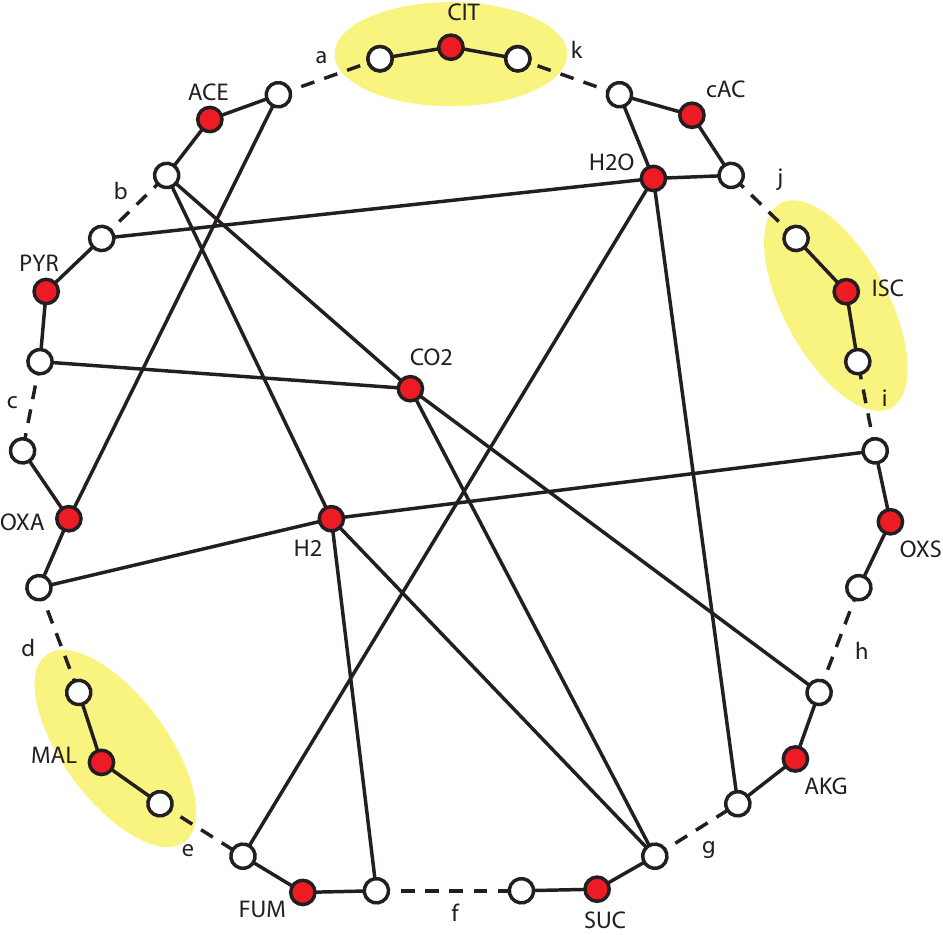}
  \caption{
  The projection of the TCA cycle onto CHO compounds.  Phosphates and
  thioesters are omitted, and the stoichiometry of all acids refers to
  the protonated forms, so that ${\mbox{H}}_2$ stands for general
  two-electron reductants.  Omission of explicit representations of
  substrate-level phosphorylation to form citryl-CoA and succinyl-CoA
  causes water elimination to accompany carboxylation of acetate and
  succinate in this graph, where in the actual cycle it would occur
  outside the graph, in the formation of pyrophosphates.  Highlighted
  species are sole outputs and sole inputs of their associated
  reactions, and can be removed with the elementary composition
  rule~(\ref{eq:elem_comp_rule}) of Sec.~\ref{sec:base_case}.  Legend:
  acetate (ACE), pyruvate (PYR), oxaloacetate (OXA), malate (MAL),
  fumarate (FUM), succinate (SUC), $\alpha$-ketoglutarate (AKG),
  oxalosuccinate (OXS), cis-aconitate (cAC), isocitrate (ISC),
  citrate (CIT).  
    \label{fig:rTCA_allnodes_del_triv} 
  }
  \end{center}
\end{figure}

\begin{figure}[ht]
  \begin{center} 
  \includegraphics[scale=0.65]{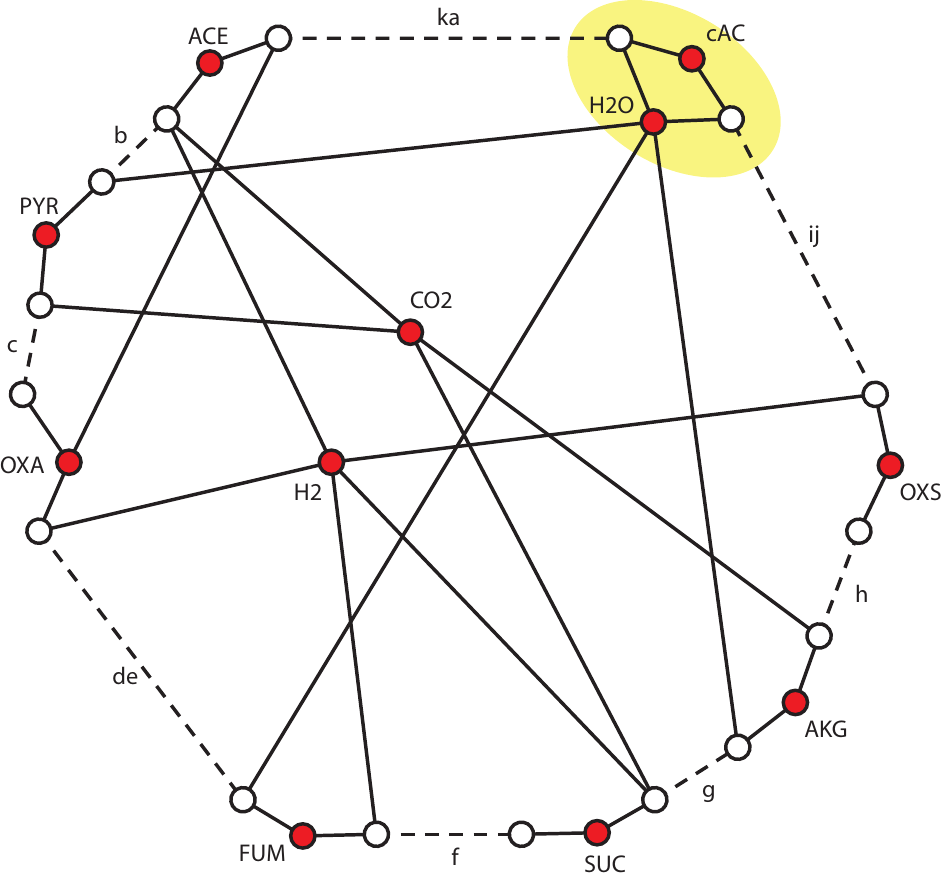}
  \caption{
  Graph of Fig.~\ref{fig:rTCA_allnodes_del_triv} with its highlighted
  species removed.  Cis-aconitate (cAC highlighted) has common factors
  of $\left[ {\mbox{H}}_2 \mbox{O} \right]$, and is the next internal
  node to be removed, by the rewrite rules of
  Sec.~\ref{sec:composite_case}, but with the simplifying feature that
  common factors cancel, so they resemble the base case.
    \label{fig:rTCA_notriv_del_cAC} 
  }
  \end{center}
\end{figure}

The bipartite graph for the minimal rTCA network in CHO compounds is
shown in Fig.~\ref{fig:rTCA_allnodes_del_triv}.  All networks in the
text are generated by equivalent methods.  Highlighted nodes are those
that can be removed by the base reduction in
Sec.~\ref{sec:base_case}.  Reactions are labeled with lowercase Roman
letters, and relative to the elementary half-reaction rate constants,
the lumped-parameter rate constants are given by
\begin{eqnarray}
  k_{de} 
& = & 
  \frac{
    k_d k_e
  }{
    {\bar{k}}_d + k_e
  }
\makebox[0.25in]{}
  {\bar{k}}_{de} = 
  \frac{
    {\bar{k}}_d {\bar{k}}_e
  }{
    {\bar{k}}_d + k_e
  }
\nonumber \\ 
  k_{ij} 
& = & 
  \frac{
    k_i k_j
  }{
    {\bar{k}}_i + k_j
  }
\makebox[0.25in]{}
  {\bar{k}}_{ij} = 
  \frac{
    {\bar{k}}_i {\bar{k}}_j
  }{
    {\bar{k}}_i + k_j
  }
\nonumber \\ 
  k_{ka} 
& = & 
  \frac{
    k_k k_a
  }{
    {\bar{k}}_k + k_a
  } 
\makebox[0.25in]{}
  {\bar{k}}_{ka} = 
  \frac{
    {\bar{k}}_k {\bar{k}}_a
  }{
    {\bar{k}}_k + k_a
  } , 
\label{eq:elem_comp_rule_rTCA}
\end{eqnarray}
with equivalent expressions for the $\bar{k}$s.  These define the
elementary reactions in the reduced graph of
Fig.~\ref{fig:rTCA_notriv_del_cAC}. Here and
  below, we give formulae only for the forward half-reaction rate
  constants $k$.  Formulae for the backward half-reaction rate
  constants $\bar{k}$ have corresponding forms as shown in the
  preceding sections.

\begin{figure}[ht]
  \begin{center} 
  \includegraphics[scale=0.65]{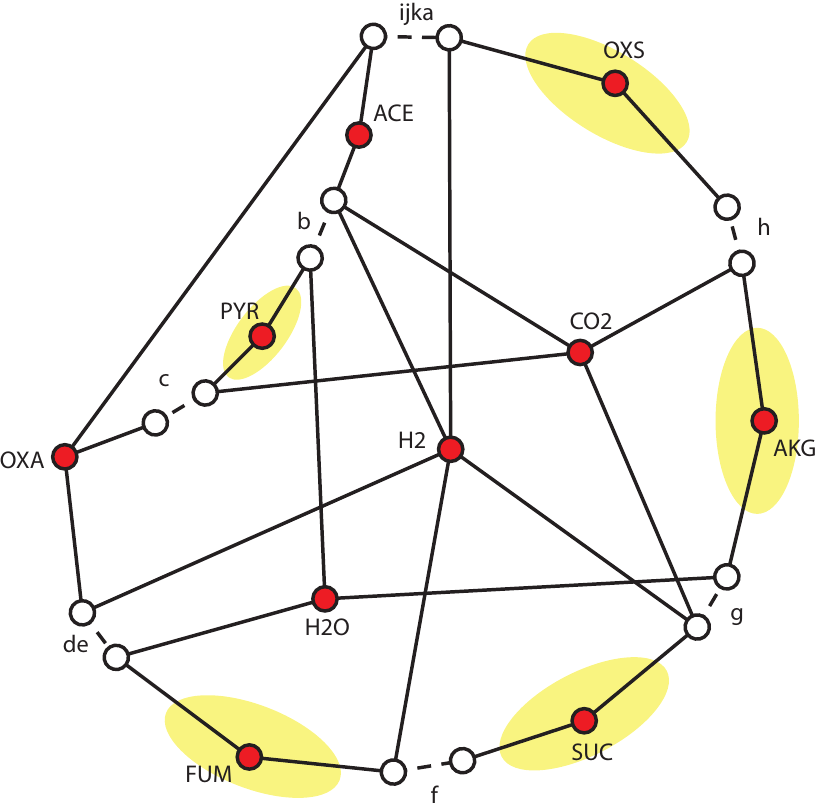}
  \caption{
  Graph of Fig.~\ref{fig:rTCA_notriv_del_cAC} with cAC and its
  parallel links to water removed.  For all remaining species except
  acetate (ACE), neither sources nor sinks are assumed, and these may
  be removed with non-trivial instances of the composition rule of
  Sec.~\ref{sec:composite_case}.  Each of these removals changes the
  degree of the remaining reactions, and thus changes the topology of
  the graph.
    \label{fig:rTCA_nocAC_del_nobound} 
  }
  \end{center}
\end{figure}

One further reduction that follows the elementary rule in
Fig.~\ref{fig:rTCA_notriv_del_cAC} is removal of cis-aconitate (cAC),
which involves a common factor of the solvent $\left[ {\mbox{H}}_2
  \mbox{O} \right]$.  The resulting lumped-parameter rate constants
are given by 
\begin{equation}
  k_{ijka} = 
  \frac{
    k_{ij} k_{ka}
  }{
    {\bar{k}}_{ij} + k_{ka}
  } 
\makebox[0.25in]{}
  {\bar{k}}_{ijka} = 
  \frac{
    {\bar{k}}_{ij} {\bar{k}}_{ka}
  }{
    {\bar{k}}_{ij} + k_{ka}
  } .
\label{eq:elem_comp_rule_cAC}
\end{equation}
These lead to the graph of Fig.~\ref{fig:rTCA_nocAC_del_nobound}. 

\begin{figure}[ht]
  \begin{center} 
  \includegraphics[scale=0.65]{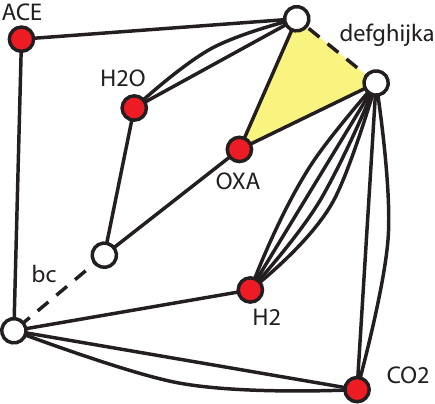}
  \caption{
  Graph of Fig.~\ref{fig:rTCA_nocAC_del_nobound} with all internal
  nodes from linear chains removed.  $\left[ {\mbox{H}}_2 \mbox{O}
    \right]$, $\left[ {\mbox{H}}_2 \right]$, $\left[ {\mbox{CO}}_2
    \right]$, and $\left[ \mbox{ACE} \right]$ are the four molecular
  concentrations to which boundary sources are coupled.  $\left[
  \mbox{OXA} \right]$ is retained as the last representation of the
  network catalysis of the loop, indicated by highlighting of the
  reaction in which OXA is input and output with equal stoichiometry.
  In steady state, OXA is in equilibrium with ACE, because it is not
  coupled to external currents. 
    \label{fig:rTCA_minimal_equiv} 
  }
  \end{center}
\end{figure}

All further graph reductions require the composition rules of
Sec.~\ref{sec:composite_case}, and result in changes of the input or
output stoichiometries of the unreduced nodes.  All highlighted
compounds in Fig.~\ref{fig:rTCA_nocAC_del_nobound} may be removed, and
the resulting lumped-parameter rate constants are given by 
\begin{eqnarray}
  k_{bc} 
& = & 
  \frac{
    k_b k_c
  }{
    \left[ {\mbox{H}}_2 \mbox{O} \right] {\bar{k}}_b + 
    \left[ {\mbox{CO}}_2 \right] k_c
  }
\nonumber \\ 
  k_{def} 
& = & 
  \frac{
    k_{de} k_f
  }{
    \left[ {\mbox{H}}_2 \mbox{O} \right] {\bar{k}}_{de} + 
    \left[ {\mbox{H}}_2 \right] k_f
  }
\nonumber \\ 
  k_{defg} 
& = & 
  \frac{
    k_{def} k_g
  }{
    \left[ {\mbox{H}}_2 \mbox{O} \right] {\bar{k}}_{def} + 
    \left[ {\mbox{H}}_2 \right] \left[ {\mbox{CO}}_2 \right] k_g
  }
\nonumber \\ 
  k_{defgh} 
& = & 
  \frac{
    k_{defg} k_h
  }{
    {\left[ {\mbox{H}}_2 \mbox{O} \right]}^2 {\bar{k}}_{defg} + 
    \left[ {\mbox{CO}}_2 \right] k_h
  }
\nonumber \\ 
  k_{defghijka} 
& = & 
  \frac{
    k_{defgh} k_{ijka}
  }{
    {\left[ {\mbox{H}}_2 \mbox{O} \right]}^2 {\bar{k}}_{defgh} + 
    \left[ {\mbox{H}}_2 \right] k_{ijka}
  } . 
\label{eq:gen_comp_rule_rTCA}
\end{eqnarray}
These define the maximal reduction of the original rTCA graph, to the
graph shown in Fig.~\ref{fig:rTCA_minimal_equiv}.  

The lumped-parameter rate equations for
Fig.~\ref{fig:rTCA_minimal_equiv}, parametrized by lumped-parameter
rate constants, are
\begin{eqnarray}
  J_{bc}
& = & 
  \left[ \mbox{ACE} \right]
  \left[ {\mbox{H}}_2 \right]
  {\left[ {\mbox{CO}}_2 \right]}^2
  k_{bc} 
\nonumber \\
& & 
  \mbox{} - 
  \left[ \mbox{OXA} \right]
  \left[ {\mbox{H}}_2 \mbox{O} \right]
  {\bar{k}}_{bc}
\nonumber \\
  J_{defghijka}
& = & 
  \left[ \mbox{OXA} \right]
  {\left[ {\mbox{H}}_2 \right]}^4
  {\left[ {\mbox{CO}}_2 \right]}^2
  k_{defghijka} 
\nonumber \\
& & 
  \mbox{} - 
  \left[ \mbox{OXA} \right]
  \left[ \mbox{ACE} \right]
  {\left[ {\mbox{H}}_2 \mbox{O} \right]}^2
  {\bar{k}}_{defghijka} . 
\label{eq:rTCA_min_rate_eqns}
\end{eqnarray}
In steady state $J_{bc} = 0$ and $\left[ \mbox{OXA} \right]$ may be
replaced with the equilibrium function 
\begin{equation}
  \left[ \mbox{OXA} \right] = 
  \frac{k_{bc}}{{\bar{k}}_{bc}}
  \frac{
    \left[ {\mbox{H}}_2 \right] {\left[ {\mbox{CO}}_2 \right]}^2
  }{
    \left[ {\mbox{H}}_2 \mbox{O} \right]
  }
  \left[ \mbox{ACE} \right] . 
\label{eq:OXA_eqm}
\end{equation}

\subsubsection{Network reaction fluxes and their control parameters}

For the remainder of the appendix we replace the subscript
${}_{defghijka}$ with designation ${}_{\mbox{\scriptsize rTCA}}$ in
currents $J$, half-reaction rate constants $k$, $\bar{k}$, and
equilibrium constants $K$.  Dimensionally, the rate constants require
the concentration of OXA in the mass-action law, and so presume that
the anaplerotic segment ${}_{bc}$ has been handled. 

Plugging Eq.~(\ref{eq:OXA_eqm}) into the second rate equation of
Eq.~(\ref{eq:rTCA_min_rate_eqns}), and supposing $\left[ \mbox{OXA}
\right]$ is in equilibrium with $\left[ \mbox{ACE} \right]$ at a
(non-equilibrium) steady state for the network as a whole, we obtain
the only independent mass-action rate equation for the reduced
network.  This is the current producing acetate:
\begin{widetext}
\begin{equation}
  J_{\mbox{\scriptsize rTCA}} = 
  {\bar{k}}_{\mbox{\scriptsize rTCA}}
  \frac{k_{bc}}{{\bar{k}}_{bc}}
  \left[ {\mbox{H}}_2 \right] 
  {\left[ {\mbox{CO}}_2 \right]}^2
  \left[ {\mbox{H}}_2 \mbox{O} \right]
  \left[ \mbox{ACE} \right]
  \left(
    \frac{
      k_{\mbox{\scriptsize rTCA}}
    }{
      {\bar{k}}_{\mbox{\scriptsize rTCA}}
    }
    \frac{
      {\left[ {\mbox{H}}_2 \right]}^4 {\left[ {\mbox{CO}}_2 \right]}^2
    }{
      {\left[ {\mbox{H}}_2 \mbox{O} \right]}^2
    } - 
    \left[ \mbox{ACE} \right]
  \right) . 
\label{eq:ACE_OXA_eqm}
\end{equation}
\end{widetext}
The first term in parenthesis in Eq.~(\ref{eq:ACE_OXA_eqm}) is the
concentration at which acetate would be in equilibrium with the
inorganic inputs, which we denote
\begin{equation}
  {\left[ \mbox{ACE} \right]}_G \equiv 
    \frac{
      k_{\mbox{\scriptsize rTCA}}
    }{
      {\bar{k}}_{\mbox{\scriptsize rTCA}}
    }
    \frac{
      {\left[ {\mbox{H}}_2 \right]}^4 {\left[ {\mbox{CO}}_2 \right]}^2
    }{
      {\left[ {\mbox{H}}_2 \mbox{O} \right]}^2
    } . 
\label{eq:Gibbs_ace_notation}
\end{equation}
Therefore the network response is proportional to the offset of
$\left[ \mbox{ACE} \right]$ from its equilibrium value, with a rate
constant that depends on the particular contributions of chemical
potential from $\left[ {\mbox{CO}}_2 \right]$ and reductant. Although the
lumped-parameter rate constant in this relation appears complex, the
consistency conditions with single-reaction equilibrium constants
ensure that $k_{\mbox{\scriptsize rTCA}} /
{\bar{k}}_{\mbox{\scriptsize rTCA}}$ is independent of synthetic
pathway and equal to the exponential of the Gibbs free energy of
formation.

\subsection{Interaction of Wood-Ljungdahl with rTCA}

We may envision an early Wood-Ljungdahl ``feeder'' pathway to
acetyl-CoA as a reaction with the same stoichiometry as rTCA for the
creation of acetate, but fixed half-reaction rate constants that do
not depend on the internal concentrations in the network.  This may be
a pre-pterin mineral pathway~\cite{Huber:Ace_CoA:00}, in which rate
constants are determined by the abiotic environment, or an early
pathway using pterin-like cofactors, if the concentrations of these
are somehow buffered from the instantaneous flows through the
reductive pathway.  Labeling this ``linear'' effective reaction WL,
the rate equation becomes
\begin{equation}
  J_{\mbox{\scriptsize WL}} = 
  {\bar{k}}_{\mbox{\scriptsize WL}}
  {\left[ {\mbox{H}}_2 \mbox{O} \right]}^2
  \left(
    \frac{k_{\mbox{\scriptsize WL}}}{{\bar{k}}_{\mbox{\scriptsize WL}}}
    \frac{
      {\left[ {\mbox{H}}_2 \right]}^4 {\left[ {\mbox{CO}}_2 \right]}^2
    }{
      {\left[ {\mbox{H}}_2 \mbox{O} \right]}^2
    } - 
    \left[ \mbox{ACE} \right]
  \right) . 
\label{eq:ACE_OXA_WL_feeder}
\end{equation}
Note that $k_{\mbox{\scriptsize WL}} / {\bar{k}}_{\mbox{\scriptsize
    WL}} = k_{\mbox{\scriptsize rTCA}} / {\bar{k}}_{\mbox{\scriptsize
    rTCA}}$ because both are expressions for the equilibrium constant
which depends only on the free energy of reaction.

To understand the performance of a joint network in the presence of
losses, as the simplest case introduce a reaction labeled Env standing
for dilution of acetate to an environment at zero concentration.  The
dilution current becomes
\begin{equation}
  J_{\mbox{\scriptsize Env}} =
  k_{\mbox{\scriptsize Env}} 
  \left[ \mbox{ACE} \right] . 
\label{eq:dilution_curr_gen}
\end{equation}
At a non-equilibrium steady state the total losses must equal the
total supply currents, so that 
\begin{equation}
  J_{\mbox{\scriptsize Env}} = 
  J_{\mbox{\scriptsize rTCA}} + 
  J_{\mbox{\scriptsize WL}} . 
\label{eq:dilute_equals_feed}
\end{equation}

\begin{figure}[ht]
  \begin{center} 
  \includegraphics[scale=0.75]{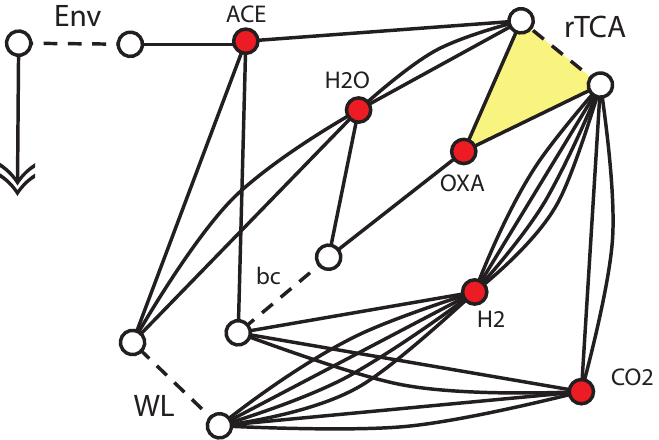}
  \caption{
  Hypergraph model for parallel reactions through the rTCA and WL
  pathways, coupled to a linear drain reaction representing dilution
  of acetate by the environment. 
    \label{fig:rTCA_minimal_equiv_WL_drain} 
  }
  \end{center}
\end{figure}

The un-reduced equation for steady-state currents can be written
\begin{widetext}
\begin{eqnarray}
  J_{\mbox{\scriptsize rTCA}} + 
  J_{\mbox{\scriptsize WL}} 
& = & 
  {\left[ {\mbox{H}}_2 \mbox{O} \right]}^2  
  \left\{
    \sqrt{k_{\mbox{\scriptsize rTCA}} {\bar{k}}_{\mbox{\scriptsize rTCA}}} 
    \frac{K_{bc}}{K_{\mbox{\scriptsize rTCA}}}
    \frac{
      \left[ {\mbox{CO}}_2 \right]
    }{
      \left[ {\mbox{H}}_2 \right]
    }
    {\left[ \mbox{ACE} \right]}_G^{1/2}
    \left[ \mbox{ACE} \right] + 
    {\bar{k}}_{\mbox{\scriptsize WL}}
  \right\}
  \left(
    {\left[ \mbox{ACE} \right]}_G - 
    \left[ \mbox{ACE} \right]
  \right) 
\nonumber \\
  = J_{\mbox{\scriptsize D}}
& = & 
  k_{\mbox{\scriptsize D}}
  \left[ \mbox{ACE} \right]
\label{eq:unred_SS_currs}
\end{eqnarray}
\end{widetext}
The graph corresponding to this model for rate laws is shown in
Fig.~\ref{fig:rTCA_minimal_equiv_WL_drain}. 

The variable that characterizes the ``impedance'' of a chemical
reaction network, and displays thresholds for autocatalysis when these
exist, is the ratio of the output acetate concentration to the value
that would exist in a Gibbs equilibrium with the inputs: 
\begin{equation}
  x \equiv 
  \frac{
    \left[ \mbox{ACE} \right]
  }{
    {\left[ \mbox{ACE} \right]}_G
  } . 
\label{eq:x_def}
\end{equation}
For a network with no reaction barriers (either in rate constants or
due to limitations of network catalysts, the output $x \rightarrow
1$.  

The two control parameters that govern the relative contributions of
the rTCA loop and the direct WL feeder are 
\begin{eqnarray}
  z_{\mbox{\scriptsize rTCA}}
& = & 
  \frac{
    \sqrt{
      k_{\mbox{\scriptsize rTCA}} 
      {\bar{k}}_{\mbox{\scriptsize rTCA}}
    } 
  }{
    k_{\mbox{\scriptsize Env}}
  }
  \frac{K_{bc}}{K_{\mbox{\scriptsize rTCA}}}
  \frac{
    \left[ {\mbox{CO}}_2 \right]
    {\left[ {\mbox{H}}_2 \mbox{O} \right]}^2  
  }{
    \left[ {\mbox{H}}_2 \right]
  }
  {\left[ \mbox{ACE} \right]}_G^{3/2}  
\nonumber \\
  z_{\mbox{\scriptsize WL}}
& = & 
  \frac{
    {\bar{k}}_{\mbox{\scriptsize WL}}
    {\left[ {\mbox{H}}_2 \mbox{O} \right]}^2  
  }{
    k_{\mbox{\scriptsize Env}}
  } . 
\label{eq:control_parms_def}
\end{eqnarray}
Each control parameter is a ratio of lumped half-reaction rates that
feed $\left[ \mbox{ACE} \right]$ to the environment dilution constant
$k_{\mbox{\scriptsize Env}}$ through which it drains.

In terms of $z_{\mbox{\scriptsize WL}}$ and $z_{\mbox{\scriptsize
    rTCA}}$, the normalized concentration $x$ -- which is proportional
by $k_{\mbox{\scriptsize Env}}$ to the total current through the system
-- satisfies
\begin{equation}
  x = 
  \frac{1}{2}
  \left( 
    1 - 
    \frac{1 + z_{\mbox{\scriptsize WL}}}{z_{\mbox{\scriptsize rTCA}}}
  \right) + 
  \sqrt{
    \frac{z_{\mbox{\scriptsize WL}}}{z_{\mbox{\scriptsize rTCA}}} + 
    \frac{1}{4}
    {
      \left( 
        1 - 
        \frac{1 + z_{\mbox{\scriptsize WL}}}{z_{\mbox{\scriptsize rTCA}}}
      \right) 
    }^2
  } . 
\label{eq:x_quad_sol}
\end{equation}

The solution to Eq.~(\ref{eq:x_quad_sol}) is shown versus base-10
logarithms of $z_{\mbox{\scriptsize rTCA}}$ and $z_{\mbox{\scriptsize
    WL}}$ in Fig.~\ref{fig:rTCA_WL_conc_bw_surf} in the main text.
The critical (unsupported) response of the rTCA loop occurs at
$z_{\mbox{\scriptsize WL}} \rightarrow 0$ and $z_{\mbox{\scriptsize
    rTCA}} = 1$.  It is identified with the discontinuity in the
derivative $\partial x / \partial z_{\mbox{\scriptsize rTCA}}$ at
$z_{\mbox{\scriptsize rTCA}} = 1$ and the exactly zero value of $x$
for $z_{\mbox{\scriptsize rTCA}} < 1$.  As $z_{\mbox{\scriptsize WL}}$
increases from zero, the transition becomes smooth, and a nonzero
concentration $x$ is maintained against dilution at \emph{all} values
of $z_{\mbox{\scriptsize rTCA}}$.

\bibliographystyle{unsrt} 
\bibliography{DES}

\end{document}